\documentclass[epj]{svjour}
\pdfoutput=1
\usepackage{graphicx}
\usepackage{float}
\input{epsf}
\usepackage{amssymb}
\usepackage{amsmath}

\begin{document}

\title{Freed by interaction kinetic states in the Harper model}

\author{
Klaus M. Frahm
\and
Dima L. Shepelyansky
}
\institute{
Laboratoire de Physique Th\'eorique du CNRS, IRSAMC, 
Universit\'e de Toulouse, UPS, 31062 Toulouse, France
}

\titlerunning{Freed by interaction kinetic states in the Harper model}
\authorrunning{K.M.Frahm  and D.L.Shepelyansky}

\abstract{We study the problem of two interacting 
particles in a one-dimensional quasiperiodic lattice
of the Harper model. We show that a short or  long range interaction
between particles leads to emergence of delocalized pairs
in the non-interacting localized phase. The properties of these
Freed by Interaction Kinetic States (FIKS) are analyzed numerically
including the advanced Arnoldi method. We find that
the number of sites populated by FIKS pairs
grows algebraically with the system size with the maximal exponent $b=1$,
up to a largest lattice size $N=10946$ reached in our numerical simulations, 
thus corresponding to a complete delocalization of pairs.
For delocalized FIKS pairs 
the spectral properties of such quasiperiodic operators
represent a deep mathematical problem.
We argue that FIKS pairs can be detected in the framework of
recent cold atom experiments [M.~Schreiber {\it et al.}
Science {\bf 349}, 842 (2015)] by a simple setup modification.
We also discuss possible implications of FIKS pairs for
electron transport in the regime of charge-density wave and high $T_c$
superconductivity.
}


\date{Dated: September 9, 2015}

\maketitle

\section{ Introduction}
\label{sec1}
The Harper model \cite{harper} describes the quantum evolution of an electron
in a two-dimensional periodic potential in a magnetic field.
Due to periodicity it can be reduced to a one-dimensional
Schr\"odinger equation on a quasiperiodic lattice
known as the almost Mathieu operator.
This equation is characterized by a dimensional Planck constant
determined by the magnetic flux through the lattice cell.
The complex structure of the spectrum of this model 
was discussed in \cite{azbel} and was directly
demonstrated in \cite{hofstadter}.
As shown by Aubry and Andr\'e \cite{aubry},
for irrational flux values $\alpha/2\pi$ this
one-dimensional (1D) system
has a metal-insulator transition 
with ballistic states for $\lambda <2$ (large hopping)
and localized states for $\lambda>2$ (small hopping).
The rigorous proof is given in \cite{lana1}.
The review on this model can be found in \cite{sokoloff}
and more recent results are reported in \cite{geisel,austin}.

It is interesting to study the case of 
Two Interacting Particles (TIP) in the Harper model.
The model with Hubbard interaction between two particles 
was introduced in \cite{dlsharper} and it was shown that 
an interaction of moderate strength leads to the appearance of a localized 
component in the metallic non-interacting phase
at $\lambda <2$ while in the localized phase $\lambda >2$ 
such an interaction
does not significantly affect the properties of localized states.
Further studies also showed that the interactions provide only an 
enhancement of localization properties \cite{barelli,orso}.

These results for the Harper model show an opposite tendency compared to the 
case of TIP in the 1D Anderson model with disorder where moderate 
Hubbard interaction leads to an increase of the localization length for TIP
comparing to the non-interacting case 
\cite{dlstip,imry,pichard,frahm1995,vonoppen,frahm_tip_green}.

Thus the result of Flach, Ivanchenko, Khomeriki \cite{flach}
on appearance of delocalized TIP states at certain large
interactions in the localized phase of the Harper model at $\lambda>2$
is surprising and very interesting. In a certain way one has in this
TIP Harper model the appearance of Freed by Interaction Kinetic States (FIKS).
In this work we investigate the properties of these FIKS pairs
in more detail using numerical simulations for the time evolution
of wave functions and a new approach which allows to determine accurate eigenvectors 
for large system sizes up to $\sim 10^4$ (corresponding to a two-particle 
Hilbert space of dimension $\sim 10^8$). This approach is based on a combination 
of the Arnoldi method with a new, highly efficient, algorithm for 
Green's function evaluations. 

We note that the delocalization transition in the Harper model
has been realized recently in experiments with non-interacting
cold atoms in optical lattices \cite{roati}. Experiments with
interacting atoms have been reported in \cite{modugno}
and more recently in \cite{bloch} showing delocalization
features of interactions. Thus the investigations of the 
properties of FIKS pairs are of actual interest due to 
the recent experimental progress with cold atoms. 
We will discuss the possible implications of FIKS pairs 
to cold atom and solid state experiments after presentation of our results.

The paper is composed as follows: we describe the model in Section~\ref{sec2},
the new Green
function Arnoldi method is introduced in Section~\ref{sec3}, the analysis of 
time evolution of wave functions is presented in Section~\ref{sec4},
the properties of FIKS eigenstates for the Hubbard interaction
are described in Section~\ref{sec5} and for 
the long rang interactions in Section~\ref{sec6},
properties of FIKS eigenstates in momentum and energy representations are
analyzed in Section~\ref{sec7}, possible implications for 
the cold atom experiments \cite{modugno,bloch}
are discussed in Section~\ref{sec8}, 
the dependence on the flux parameter is studied in 
Section~\ref{sec9} and the discussion of 
the results is presented in Section~\ref{sec10}.

\section{Model description}
\label{sec2}

We consider particles in a one-dimensional lattice of size $N$. 
The one-particle Hamiltonian $h^{(j)}$ for particle $j$ is given by:
\begin{eqnarray}
\label{eq_h1}
h^{(j)}&=&T^{(j)}+V^{(j)},\\
\label{eq_T1}
T^{(j)}&=&-\sum_{x} \Bigl(|x\!>_j\,<\!x+1|_j+h.~c.\Bigr),\\
V^{(j)}&=&\sum_{x} V_1(x)\,|x\!>_j\,<\!x|_j.
\label{eq_V1}
\end{eqnarray}
The kinetic energy $T^{(j)}$ is given by the standard tight-binding 
model in one dimension with hopping elements $t=-1$ linking nearest neighbor 
sites with periodic boundary conditions. We consider a quasiperiodic 
potential of the form $V_1(x)=\lambda\cos(\alpha x+\beta)$ which leads 
for $\lambda>2$ to localized eigenfunctions with localization length 
$\ell=1/\log(\lambda/2)$ \cite{aubry}. Usually one chooses 
$\alpha=2\pi(\sqrt{5}-1)/2$ such that $\alpha/(2\pi)\approx 0.61803$ is the 
golden ratio, the ``most'' irrational number. 
For time evolution we manly use the golden mean value 
(together with the choice $\beta=0$) while for the eigestates we mainly use
the rational Fibonacci approximant  
$\alpha \rightarrow 2\pi f_{n-1}/f_n$ where $f_n$ 
is a certain Fibonacci number and where the 
system size is just $N=f_n$.  
Furthermore, in order to avoid the parity symmetry with respect to $x\to N-x$ 
at $\beta=0$ (that leads to an artificial eigenvalue degeneracy)
we choose for this case $\beta=(\sqrt{5}-1)/2$. We will see later that 
this Fibonacci approximant of $\alpha$ is very natural and 
useful in the interpretation at finite system sizes 
(especially with respect to Fourier transformation). 
In our main numerical studies for the eigenvectors we consider
system sizes/Fibonacci numbers in the range $55\le f_n\le 10946$ and the 
parameter $\lambda$ is always fixed at 
$\lambda=2.5$ with a one-particle localization length
$\ell=1/\log(\lambda/2)\approx 4.48$ \cite{aubry,sokoloff}. 

In Secs.~\ref{sec8} and \ref{sec9} we also consider different 
irrational values of $\alpha/(2\pi)$ (or suitable rational approximants 
for finite system size). This is
motivated by the recent experiments of 
Ref.~\cite{bloch} and interest  to  the overall dependence of the 
FIKS properties on the flux parameter $\alpha$. 

We now consider the TIP case, when each particle is described by the 
one-particle Hamiltonian $h^{(j)}$, and is coupled by an interaction potential 
$U(x_1-x_2)$ with another particle. 
Here we use $U(x)=U/(1+w|x|)$ for $|x|<U_R$ \cite{U_boundary} 
and $U(x)=0$ if $|x|\ge U_R$ with 
$U_R$ being the interaction range, $U$ is the global interaction strength 
and $w$ is a parameter describing the decay of the interaction. We choose
mostly $w=0$ but in certain cases also $w=1$. The case $U_R=1$ corresponds 
to the case of the on-site Hubbard interaction studied in \cite{dlsharper,flach}. 
Here we consider both symmetric two-particle states (bosons) and (for 
$U_R\ge 2$) also anti-symmetric two-particle states (fermions). 

The total two-particle Hamiltonian is 
given by 
\begin{equation}
\label{ham_tot}
H=h^{(1)}+h^{(2)}+\hat U
\end{equation}
where 
\begin{equation}
\label{U_operator}
\hat U=\sum_{x_1,x_2} U(x_1-x_2)\,|x_1,x_2\!><\!x_1,x_2|
\end{equation} 
is the interaction operator in the two-particle 
Hilbert space and with the notation $|x_1,x_2\!>=|x_1\!>_1|x_2\!>_2$ 
for the non-symmetrized two-particle states. 

Our aim is to determine if the interaction may induce at least 
partial delocalization, i.~e. at least for some eigenstates at certain 
energies. This can be done by a time evolution calculation from the 
Schr\"odinger equation using a Trotter formula approximation (see 
Sec. \ref{sec4}) or by a numerical computation of (some) eigenfunctions 
of $H$. The size of the (anti-)symmetrized Hilbert space 
is $N_2=N(N+s)/2\approx N^2/2$ 
with $s=1$ ($s=-1$) for the boson (fermion) case and 
therefore a direct full numerical diagonalization of $H$ is limited to $N$ 
smaller than a few hundred, e.g. $N\le 250$ \cite{flach}. 

Since the Hamiltonian $H$ corresponds to a sparse matrix one can in 
principle apply the Arnoldi method \cite{arnoldi,stewart,frahmulam} or 
more precisely, since $H$ is a Hermitian matrix, 
the Lanczos method \cite{arnoldi_comment},
to determine certain eigenvalues and eigenvectors.
In the next Section, we will present a new method based on the 
particular structure of $H$, the {\em Green function Arnoldi method}, 
which is even more efficient than the standard 
implicitely restarted Arnoldi method. Thus, it
allows to study larger system 
sizes, to obtain more eigenvalues, for much more parameter values and with 
virtually exact eigenvalues and eigenvectors, i.~e. 
$\delta^2 E(\psi)\sim 10^{-28}$-$10^{-20}$ implying that there are 
only numerical rounding errors due to the limited precision of standard 
double precision numbers. The description of the Arnoldi method and 
definition of $\delta^2 E(\psi)$ are given in Appendix \ref{appendix1}.

\section{Green's function Arnoldi method}
\label{sec3}

Let $E$ be some energy value for which we want to determine numerically 
eigenvalues of $H$ close to $E$ and the corresponding eigenvectors. 
Furthermore let $G=(E-H)^{-1}$ be 
the Green function or resolvent of $H$ at energy $E$. 
The idea of the Green function Arnoldi method is to apply 
the Arnoldi method to the resolvent $G$ and 
not to $H$ which is sufficient since 
the eigenvectors of $G$ are identical to those of $H$ and the 
eigenvalues $E_j$ of $H$ can be obtained from the eigenvalues 
$\gamma_j$ of $G$ simply by $E_j=E-1/\gamma_j$. 
The important point is that the largest eigenvalues $\gamma_j$ 
of $G$, which result from the simple Arnoldi 
method, provide exactly the eigenvalues $E_j$ close to a given value $E$ 
which we may choose arbitrarily. Therefore it is not necessary to 
apply the quite complicated (and rather expensive) implicitly restarted 
Arnoldi method in order to focus on a given energy interval. 

For this we need an efficient method to evaluate the product 
$G|\varphi\!>$ of $G$ to an arbitrary vector $|\varphi\!>$ and 
an arbitrary value of $E$. We have developped 
a new, highly efficient, numerical algorithm
to determine $G|\varphi\!>$ with a complexity 
${\cal O}(U_R^3 N^3)$ for an initial preparation step at a given 
value of $E$ and 
${\cal O}(N^3)$ for the matrix vector multiplication, provided the value 
of $E$ is kept fixed. For larger system sizes, when localization 
of one-particle eigenstates can be better exploited, the complexity of the 
matrix vector multiplication can even be reduced to 
${\cal O}(c\,N^2)$ with $c\sim 10^2$ being a rather large constant. 
For comparison we remind that a naive matrix vector multiplication has a 
complexity of ${\cal O}(N_2^2)={\cal O}(N^4)$ assuming that the full 
matrix $G$ has been calculated and stored previously. 

Our algorithm is based on the following ``magic'' exact formula:
\begin{equation}
\label{eq_Green}
G=G_0+G_0 ({\bf 1}-\hat U\bar G_0)^{-1} \hat U G_0
\end{equation}
where $G_0$ is the resolvent at vanishing interaction and $\bar G_0$ 
is its projection on the smaller subspace of dimension $\approx U_R\,N$ 
of sites in two-particle space where the interaction operator has a 
non-vanishing action. The computation of $\bar G_0$ and the matrix inverse 
in (\ref{eq_Green}) can therefore be done with ${\cal O}(U_R^3 N^3)$ 
operations 
and has to be done only once for a given value of the Green function 
energy $E$. The full matrix $G_0$ does not need to be computed since 
we can efficiently compute the product $G_0|\varphi\!>$ on a given 
vector $|\varphi\!>$ using a transformation of $|\varphi\!>$
from position to energy representation 
(in the basis of non-interacting two-particle product eigenstates) where 
$G_0$ is diagonal and a further transformation back to position 
representation. Both 
transformations can be done with complexity ${\cal O}(N^3)$ due to the 
product property of non-interacting two-particle eigenstates. Therefore 
(\ref{eq_Green}) allows to compute the product $G|\varphi\!>$ 
also for the full resolvent $G$ with ${\cal O}(N^3)$ operations which 
is exactly what we need to apply the Arnoldi method to $G$. 
A second, even more efficient, variant of the Green function Arnoldi 
method actually uses directly 
vectors in energy representation thus reducing the number of necessary 
transformation steps by a factor of two and also provides certain other 
advantages. These and other details of this approach 
are described in Appendix \ref{appendix2} while Appendix \ref{appendix3} 
provides the proof of (\ref{eq_Green}).

\section{Time evolution} 
\label{sec4}

We start our numerical study with a calculation for 
the time evolution with respect to the Hamiltonian (\ref{ham_tot}) 
using a Trotter formula approximation:
\begin{equation}
\label{eq_evolution1}
|\psi(t+\Delta t)\!>=\exp(-iH_p\Delta t)\,\exp(-iH_x\Delta t)\,|\psi(t)\!>
\end{equation}
with $H_p=T^{(1)}+T^{(2)}$ and $H_x=V^{(1)}+V^{(2)}+\hat U$. 
The time evolution step (\ref{eq_evolution1}) is valid for the limit of 
small $\Delta t$ and allows for an efficient evaluation by first applying 
$\exp(-iH_x\Delta t)$ 
(diagonal in position representation) to the vector $|\psi(t)\!>$, 
then transforming the resulting vector to momentum representation 
by Fast Fourier Transform using the library FFTW \cite{fftw}, applying 
$\exp(-iH_p\Delta t)$ (diagonal in momentum representation) and finally 
retransforming the vector back to position representation. 
For a finite value of $\Delta t$ (\ref{eq_evolution1}) can be viewed as the 
``exact'' time evolution of a ``modified'' Hamiltonian with 
$H$ corrected by a sum of (higher order) commutators of $H_p$ and 
$H_x$. We have chosen $\Delta t=0.1$ and verified that it provides 
quantitatively correct results for the delocalization properties and 
its parameter dependence (this was done by comparison
with data at smaller $\Delta t$ values). This integration method for the time evolution
already demonstrated its efficiency for TIP in a disordered 
potential \cite{dlstip}. 

In all our numerical studies we fix $\lambda=2.5$
which has a modest one-particle localization length \cite{dlsharper,flach}.
The main part of studies is done for the irrational 
golden value of flux or rotation number $\alpha/(2\pi)=(\sqrt{5}-1)/2$ 
(all Sections except Secs.~\ref{sec8},\ref{sec9}). 
For the time evolution we choose the quasimomentum at $\beta=0$ and use 
the system size $N=512$ with an 
initial state with both particles localized at the 
center point $x_0=N/2$ with $|\psi(0)\!>=|x_0,x_0\!>$ for the boson case 
or an anti-symmetrized state with one-particle at position $x_0$ and the 
other one at position $x_0-1$, i.~e. 
$|\psi(0)\!>=(|x_0,(x_0-1)\!>-|(x_0-1),x_0\!>)/\sqrt{2}$, for 
the fermion case.

To study the localization properties we use the one-particle density 
of states:
\begin{equation}
\label{eq_onepart_dens}
\rho_1(x)=\sum_{x_2} |<\!x,x_2\,|\,\psi\!>|^2
\end{equation}
representing the probability of finding one-particle at position $x$. 
We are  interested in the case where only a small 
weight of density is delocalized from the initial state. Thus,
we introduce an effective one-particle density
without the 20\% center box by using $\rho_{\rm eff}(x)=C\,\rho_1(x)$ 
for $0\le x<0.4N$ or $0.6N\le x<N$ and $\rho_{\rm eff}(x)=0$ 
for $0.4N\le x<0.6N$. Here $C$ is a constant that assures the 
proper normalization $\sum_x \rho_{\rm eff}(x)=1$. 
Using this effective density we define two length scales 
to characterize the (low weight) delocalization which are the 
inverse participation ratio 
\begin{equation}
\label{eq_ipr_def}
\xi_{\rm IPR}=\left(\sum_x \rho_{\rm eff}^2(x)\right)^{-1} \; , 
\end{equation}
which gives the approximate number of sites over which the density 
(outside the 20\% center box) extends and the variance length 
$\langle(x-x_0)^2\rangle^{1/2}$ with 
\begin{equation}
\label{eq_var_len}
\langle(x-x_0)^2\rangle=\sum_x\,(x-x_0)^2\,\rho_{\rm eff}(x) \; .
\end{equation}

\begin{figure}
\begin{center}
\includegraphics[width=0.48\textwidth]{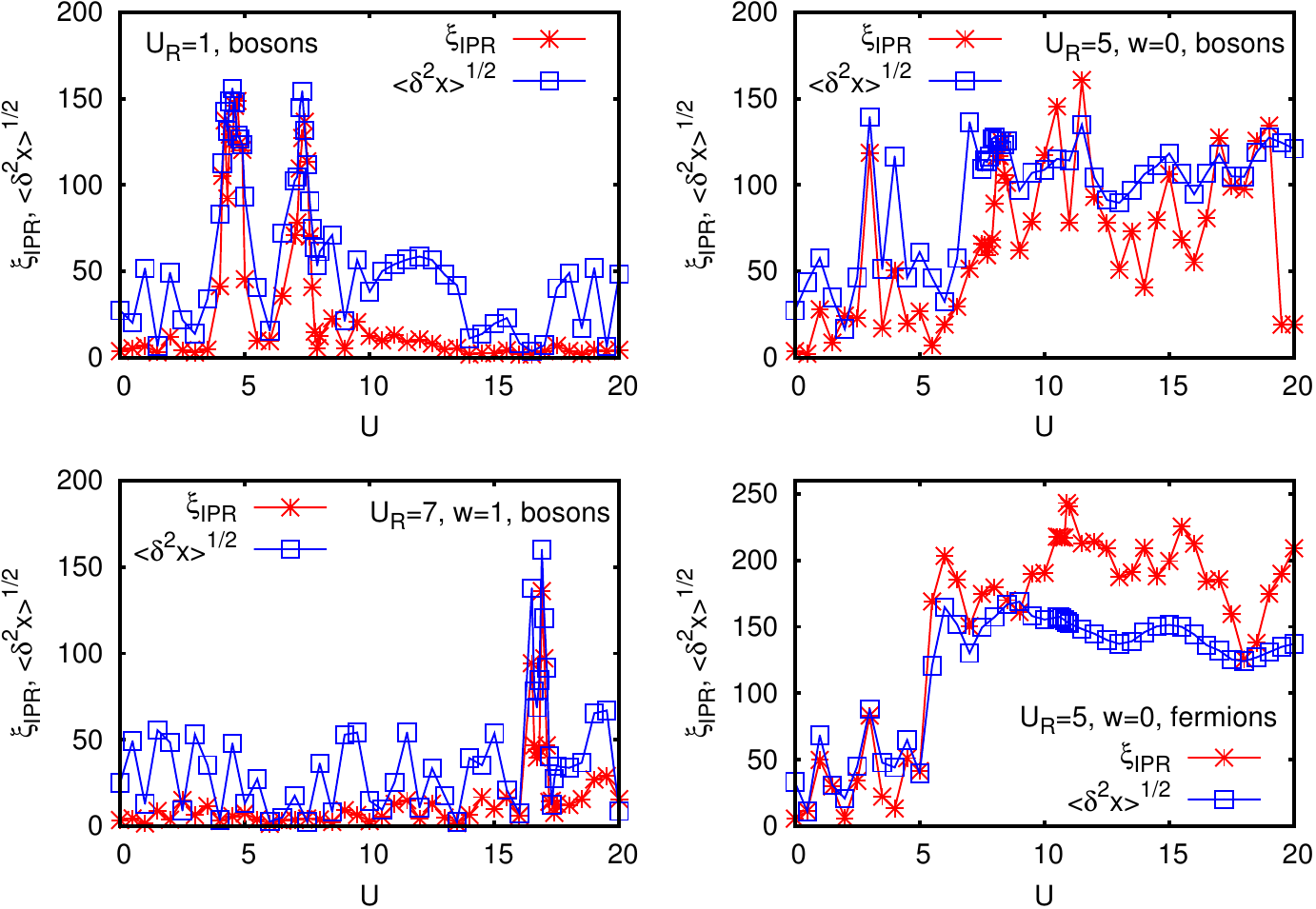}
\caption{Inverse participation ratio $\xi_{\rm IPR}$ and 
variance length 
$\langle\delta^2x\rangle^{1/2}=\langle (x-x_0)^2\rangle^{1/2}$ 
of the time evolution two-particle state for system size $N=512$ and 
iteration time $t=5120$ (or $t=20480$ for bottom left panel) 
versus interaction 
strength $U$. The initial state at $t=0$ is localized either with 
both particles in the center position $x_0=N/2$ (boson case) 
or antisymmetrized with one-particle in position $x_0$ 
and the other particle in position $x_0-1$ (fermion case).
Both quantities have been calculated from an effective one-particle density 
without a center box of size 20\% (with respect to system size). 
The different panels correspond to different cases of interaction range
$U_R$, decay parameter $w$ and boson/fermion case. 
Here $\alpha/(2\pi)=(\sqrt{5}-1)/2$ and $\beta=0$.
}
\label{fig1}
\end{center}
\end{figure}

Fig. \ref{fig1} shows the dependence of both length scales on 
the interaction strength $U$ for values up to $U\le 20$ and different cases 
of interaction range $U_R$ and decay parameter $w$ at iteration 
time $t=5120$ (or $t=20480$ for the boson case with $U_R=7$ and $w=1$). 
For each case there are a few values of interaction strength where 
the delocalization is rather strong, even if the weight of the delocalized
component is relatively small. For the Hubbard interaction 
case $U_R$ we find the two interesting values $U=4.5$ and $U=7.4$ in a rather
good agreement with the results of Ref. \cite{flach}. However, a closer 
inspection of the one-particle density reveals that there is still a strong 
localized main peak close to initial point $x_0$ and the delocalization 
only applies to a small weight of the initial state. We also note that 
the quantity 
(\ref{eq_ipr_def}) captures peaks in $U$ in a more clear way compared to 
(\ref{eq_var_len}). We attribute this to additional fluctuations
added by a large distance from $x_0$ to $x$ values outside of the central box.

The localized main peak can be understood by the 
assumption that only a small fraction of (two-particle) eigenvectors with 
specific energy eigenvalues are delocalized while the other eigenvectors 
remain strongly localized. Indeed, the initial vector $|\psi(0)\!>$, 
localized at $x_0$ and expanded in a basis of two-particle energy 
eigenstates, contains contributions from all possible energy eigenvalues.
The time evolution from the Schr\"odinger equation only 
modifies the phases of the energy expansions coefficients but not the 
amplitudes and therefore the wave packet 
at arbitrary time $|\psi(t)\!>$ contains rather uniform 
contributions from the same 
energy values. Obviously the delocalization effect in the wave packet 
only happens for the small weight corresponding to the limited fraction of 
delocalized eigenvectors while the other contributions 
form the central peak close to the initial position. 

We have therefore computed a {\em tail state} $|\psi_{\rm tail}(t)\!>$ 
from the wave packet $|\psi(t)\!>$ by removing (putting to zero) 
a big 60\% center box in a similar way as for $\rho_{\rm eff}(x)$ (but 
in the two-particle space and using a larger center box). 
The energy eigenvectors who contribute to $|\psi_{\rm tail}(t)\!>$ 
obviously only cover the delocalized eigenvectors and 
assuming that the latter exist 
only for certain specific energies we can try to determine this energy range 
(for delocalization) by computing the expectation value $\langle H\rangle$ 
of $H$ and 
its energy variance [see Eq. (\ref{deltaE2})] with respect to 
$|\psi_{\rm tail}(t)\!>$ (after 
proper renormalization of $|\psi_{\rm tail}(t)\!>$). Furthermore 
the square norm $\|\psi_{\rm tail}(t)\|^2$, which is the probability 
of propagating outside the 60\% centerbox, gives also a good measure 
for the delocalization effect. 

\begin{table}
\caption{Time evolution parameters for certain cases 
of short and long range interactions for interaction 
values with strong delocalization. All rows except the last one correspond 
to the boson case and the last row to the fermion case. 
The iteration time is $t=5120$ except for the case with
$U=16.9$, $U_R=7$ and $w=1$ where $t=20480$.
Here $\alpha/(2\pi)=(\sqrt{5}-1)/2$ and $\beta=0$.
}
\begin{center}
\begin{tabular}{|r|r|r|r|r|r|r|}
\hline
$U$ & $U_R$ & $w$ & $\xi_{\rm IPR}$ & 
$\langle H\rangle$ & $\delta^2E$ & $\|\psi_{\rm tail}(t)\|^{2^{\phantom 1}}$ \\
\hline
\hline
4.4 & 1 & 0 & 129.16 & -3.0756 & 0.2257 & 0.04175 \\ 
\hline
4.5 & 1 & 0 & 125.22 & -3.0645 & 0.2454 & 0.0383 \\ 
\hline
4.7 & 1 & 0 & 148.56 & -3.0347 & 0.2594 & 0.02596 \\ 
\hline
7.2 & 1 & 0 & 109.53 & 1.8072 & 0.4891 & 0.05801 \\ 
\hline
7.4 & 1 & 0 & 136.60 & 1.1369 & 3.0897 & 0.04102 \\ 
\hline
7.8 & 1 & 0 & 15.13 & 1.8151 & 0.6851 & 0.0001974 \\ 
\hline
8.0 & 5 & 0 & 89.26 & 8.7256 & 0.3260 & 0.01406 \\ 
\hline
16.9 & 7 & 1 & 136.06 & 10.1893 & 0.5026 & 0.03268 \\ 
\hline
10.9 & 5 & 0 & 243.17 & 10.8879 & 0.4431 & 0.0795 \\ 
\hline
\end{tabular}
\label{table1}
\end{center}
\end{table}

In Table~\ref{table1} we show for certain cases with strong delocalization 
the values of the quantities $\xi_{\rm IPR}$, $\langle H\rangle$, 
$\delta^2E$ and $\|\psi_{\rm tail}(t)\|^2$. For $U_R=1$ and the 
first peak at $U\approx 4.5$ the maximum for $\xi_{\rm IPR}$ corresponds 
to $U=4.7$ while the maximum of $\|\psi_{\rm tail}(t)\|^2$ corresponds 
to $U=4.4$. Therefore the intermediate value $U=4.5$ used in Ref.~\cite{flach} 
is indeed promising. For all these three values of $U$ the average energy 
$\langle H\rangle\approx -3.05$ of the tail state 
corresponds rather well to the 
approximate eigenvalue region $E\approx -3.1$ at 
$U=4.5$ for delocalized eigenstates found in  \cite{flach} and confirmed 
by our detailed eigenvector analysis presented in the next Section. 
Furthermore, the corresponding energy variance is indeed rather small. 

For $U_R=1$ there is also a second local maximum of $\xi_{\rm IPR}$ 
at $U=7.4$ and close to this value there is also a local maximum of 
$\|\psi_{\rm tail}(t)\|^2$ at $U=7.2$. We have also included in 
Table~\ref{table1} the value $U=7.8$ which is close to the second 
interaction value $U=7.9$ used in Ref.~\cite{flach}. The value $U=7.8$ seems 
less optimal but our eigenvector analysis shows that this value 
is quite optimal for {\em two} different energy ranges $E\approx 1.8$ 
and $E\approx -2.8$ with well delocalized eigenstates for both energies. 
According to Table~\ref{table1} the average energy of the tail state is  
$\langle H\rangle\approx 1.8$ for $U=7.2$ and $U=7.8$ but with a 
somewhat larger value of the variance (in comparison to the case $U=4.5$) 
indicating that the main contributions in the tail 
state arise from the first energy range $E\approx 1.8$ but the second value 
$E\approx -2.8$ provides also some smaller contributions therefore increasing 
the variance. For $U=7.4$ the average energy of the tail state 
is even reduced to $\langle H\rangle\approx 1.1$ and the 
variance $\delta^2E\approx 3.1$ is quite large 
which indicates clearly that for this case both energy ranges have more 
comparable contributions in the tail state. 
In Fig.~1(a) of Ref.~\cite{flach} these two energy values can be roughly 
identified with a somewhat stronger delocalization at $E\approx -2.8$. 
Our eigenvector calculations (see next Section) 
for larger system sizes confirm that for modest values of 
system sizes the delocalization is stronger at $E\approx -2.8$ but at larger 
sizes it is considerable stronger at $E\approx 1.8$.

The values of $\|\psi_{\rm tail}(t)\|^2$ between 
$10^{-4}$ and $5.8 \times 10^{-2}$ represent the weight of the delocalized eigenstates 
in the wave packets. These values are significantly smaller than unity showing 
that the 
main contribution still corresponds to the central peak at $x_0$ and the 
localized eigenstates at other energy values but they are also considerably 
larger than 
the values $\sim 10^{-14}$ for $U$ values with minimal (or absent) 
small weight delocalization. In general, the maximal values of $U$ for the 
two length scales shown in Fig. \ref{fig1} correspond rather well also to 
the local maximal values for $\|\psi_{\rm tail}(t)\|^2$. 
For the other three cases of Fig. \ref{fig1}, with long range interaction 
we can also identify certain values of $U$ with rather strong delocalization 
(for both length scales and the squared tail norm). According to 
Table~\ref{table1} we find for these three cases 
$\xi_{\rm IPR}\sim 10^2$, $\|\psi_{\rm tail}(t)\|^2\sim 10^{-2}$ and 
rather sharp average energy values of the tail state with a small variance. 

We have repeated this type of analysis also for many other 
long range interaction cases and in certain cases we have been able to 
identify optimal values of $U$ and $E$ for strong delocalization
where the approximate energy obtained from the time evolution 
tail state was used 
as initial value of $E$ for the Green function Arnoldi 
method to compute eigenstates (see Sec. \ref{sec5}).

We also computed the inverse participation ratio and the 
variance length using the full one-particle density of states 
(including the center box) and also these quantities have somewhat 
maximal values at the optimum $U$ values for delocalization found 
above but their maximum values are much smaller than the length scales 
shown in Fig. \ref{fig1}. Therefore it would be more difficult (or 
impossible) to distinguish between small weight long range delocalization 
and high weight small or medium range delocalization (i.e. where the full 
wave packet delocalizes but for a much smaller length scale). For this reason 
we prefer to compute the inverse participation ratio and the 
variance length using the effective one-particle density without center box 
and with the results shown in Fig. \ref{fig1}. 

\begin{figure}
\begin{center}
\includegraphics[width=0.48\textwidth]{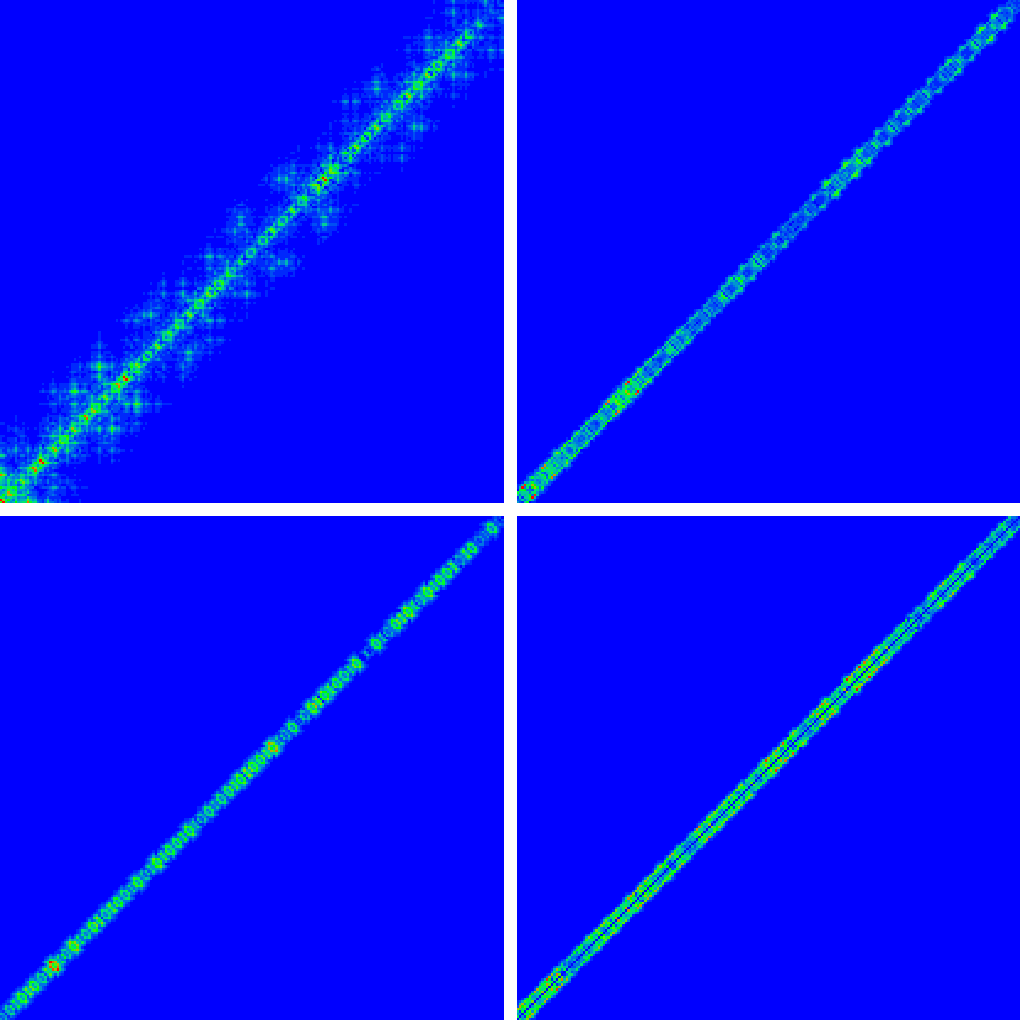}
\caption{Density plot of time evolution state for 
$t=5120$ (or $t=20480$ for bottom left panel), 
system size $N=512$, 
the four cases of Fig. \ref{fig1} and with a value of 
$U$ corresponding to strongest delocalization: 
$U=4.5$, $U_R=1$, boson case (top left panel), 
$U=8$, $U_R=5$, $w=0$, boson case (top right panel), 
$U=16.9$, $U_R=7$, $w=1$, boson case (bottom left panel), 
$U=10.9$, $U_R=5$, $w=0$, fermion case (bottom right panel). 
We show only  zoomed region of size $205\times 205$ 
with left bottom corner at position $x_1=x_2=307$ which 
corresponds to the right/top boundary of the 20\% center box. 
The colors indicate red for maximum, green for medium and blue for 
minimum values (same distribution of colors in other figures
of density plots).
}
\label{fig2}
\end{center}
\end{figure}

In Fig. \ref{fig2} we show the density plots of a zoomed region 
of the time evolution state for the four cases of Fig. \ref{fig1} and 
the optimal delocalization values for $U$ 
($U=4.5$ for $U_R=1$ and the three values given in Table~\ref{table1} 
for the cases with 
$U_R>1$ and also mentioned in the figure caption of Fig. \ref{fig2}). 
The zoomed region correspond to a box of size $205\times 205$ 
with left bottom corner at position $x_1=x_2=307$. This value corresponds 
exactly to the right/top boundary of the 
20\% center box which has been removed when determining the effective 
one-particle density of states $\rho_{\rm eff}(x)$. For positions 
inside the center box between $205$ and $306$ 
the time evolution state has a strong peaked structure 
with considerably larger values of the amplitude than the right/top part 
shown in Fig. \ref{fig2}. The left/lower part (between 
$0$ and $204$) is similar in structure with similar amplitudes 
to the right/top part. 
Fig. \ref{fig2} clearly confirms the complete 
small weight delocalization along the diagonal $x_1\approx x_2$ 
of the wave packet at sufficiently long 
iterations times $t=5120$ (or $t=20480$ for the case with $U_R=7$ and $w=1$).

\begin{figure}
\begin{center}
\includegraphics[width=0.48\textwidth]{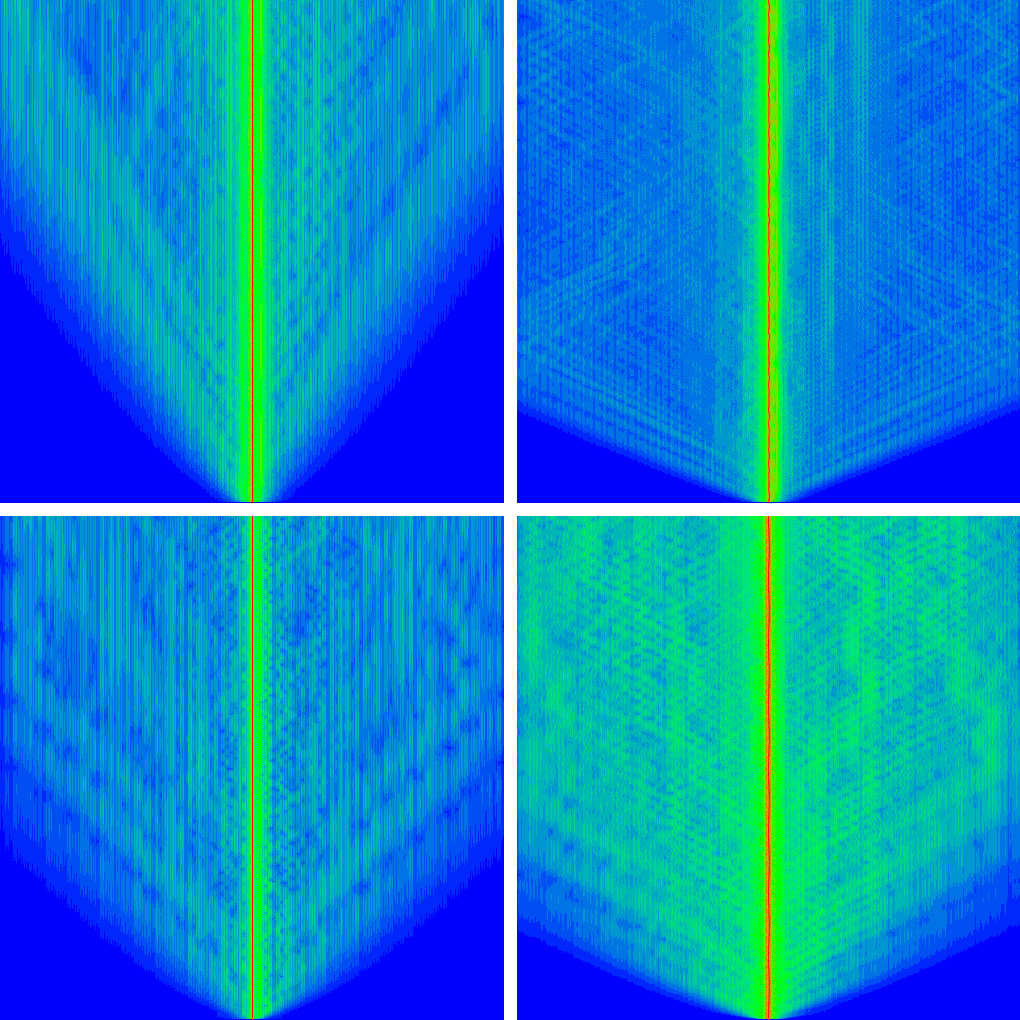}
\caption{Density plot for the time dependence of one-particle density from 
the time evolution state with $x$-position ($0\le x<512$) corresponding to 
the horizontal axis and time $t$ ($0\le t \le 5120$ or 
$0\le t \le 20480$ for bottom left panel) corresponding to 
the vertical axis. The four panels correspond to the same 
parameter values of $U$, $U_R$, $w$ and boson/fermion cases 
as in four panels of Fig. \ref{fig2}. 
}
\label{fig3}
\end{center}
\end{figure}

The time evolution of the one-particle density of states can be 
seen in Fig. \ref{fig3} with its time dependence corresponding to 
the vertical axis and position dependence corresponding to the 
horizontal axis for the same cases and parameters of Fig.~\ref{fig2}. 
In all cases 
one can identify a strong central peak at $x_0$ and a low weight 
delocalization with a characteristic length scale increasing 
linearly in time, thus corresponding to a ballistic dynamics already 
observed for the Hubbard interaction case in \cite{flach}. 
One can also observe in Figs. \ref{fig2} and \ref{fig3} that for 
$U=8.0$, $U_R=5$, $w=0$, boson case, the weight of the delocalized part 
of the wave packet is minimal of the four shown cases which is in 
agreement with the lowest value of $\|\psi_{\rm tail}(t)\|^2$ 
for the same case.

\section{Eigenstates for Hubbard interaction}
\label{sec5}

In this Section we present our results for the 
two-particle eigenstates for the case of the Hubbard interaction with 
$U_R=1$. In order to characterize the delocalization properties of eigenstates we 
use two quantities. One is the inverse participation ratio in 
position representation $\xi_x$,
obtained from the one-particle density of states (\ref{eq_onepart_dens}) 
of eigenstate  $|\psi\!>$,  by 
\begin{equation}
\label{eq_IPRxdef}
\xi_x=\left(\sum_x \rho_{\rm 1}^2(x)\right)^{-1}  \; .
\end{equation}
Another one is the inverse participation ratio in energy representation 
$\xi_E$ obtained from an expansion of a two-particle eigenstate $|\psi\!>$ 
of $H$ in the basis of non-interacting energy 
product eigenstates $|\phi_\nu,\phi_\mu\!>$ (of $H_0$) by
\begin{equation}
\label{eq_IPREdef}
\xi_E=\left(\sum_{\nu,\mu} \Bigl|<\!\phi_\mu,\phi_\nu\,|
\,\psi\!>\Bigr|^4\right)^{-1}.
\end{equation}
The quantity $\xi_x$ is identical to the ``participation number'' 
used in Ref. \cite{flach}. It is similar (but different) to the quantity 
(\ref{eq_ipr_def}), used in the previous Section, but for the 
full one-particle density and not the effective density without the 
20\% center box. Thus $\xi_x$ counts the number of $x$-positions over which 
the one-particle density extends and obeys the exact inequality $\xi_x\le N$.
It is not to be confused with the inverse participation ratio in the 
two particle $(x_1,x_2)$-space, a quantity we did not study. Instead we use 
the other quantity $\xi_E$ that counts the number of non-interacting energy 
product eigenstates of $H_0$ which contribute in the eigenstate. This quantity 
may be larger than $N$ as we will see for the case of long range interactions 
in the next Section. It is very convenient to determine  $\xi_E$ with the second 
variant of the Green function Arnoldi method where the main computations 
are done in the energy representation using the non-interacting energy 
product eigenstates $|\phi_\nu,\phi_\mu\!>$ as basis states. For the case 
of two particles localized far away from each other,
the quantity $\xi_E$ is very close to 
unity while $\xi_x$ is closer to $3$-$4$ due to the finite localization 
length of the one-particle Harper problem. For a ballistic 
delocalized state along the 
diagonal $x_1=x_2$ we expect that both $\xi_x$ and $\xi_E$ are $\sim CN$ 
with some constant $C$ of order or a bit smaller than unity. 

In this and the next Sections we choose the system size to be a 
Fibonacci number $N=f_n$, the rational case $\alpha/(2\pi)=f_{n-1}/f_n$ 
and $\beta=(\sqrt{5}-1)/2$. However, we have verified that 
the strong delocalization of eigenstates for certain values of $U$ and $E$ 
is also valid for the irrational case for arbitrary $N$ with 
$\alpha/(2\pi)=(\sqrt{5}-1)/2$ and $\beta=0$. 
For example for $U_R=1$, $U=4.5$, $E\approx -3.1$ ($U=7.8$, $E\approx -2.8$) 
we find for the rational case with $N=4181$ that the eigenstate 
with maximal $\xi_E$ corresponds to $E=-3.09901$, $\xi_E=795.960$ and 
$\xi_x=1172.887$ ($E=-2.78600$, $\xi_E=501.321$ and $\xi_x=475.573$) 
while for the irrational case with $N=4000$ we have 
$E=-3.09963$, $\xi_E=763.440$ and $\xi_x=889.854$ 
($E=-2.78716$, $\xi_E=559.130$ and $\xi_x=588.186$). 

We consider as system size $N$ all Fibonacci numbers between 55 
and 10946. For each system size we apply the Green function Arnoldi method 
with a typical Arnoldi dimension $n_A\approx 0.7N$-$0.8N$ 
slightly smaller than $N$ except for the largest case $N=10946$ for 
which we choose $n_A=2000$ or $n_A=3000$ and the smallest cases $N=55$ or 
$N=89$ where we choose $n_A\sim 300$-$400$. From all $n_A$ Ritz eigenvalues we 
retain only those with a minimal quality requirement of 
$\delta^2E(\psi)<10^{-8}$ which corresponds roughly to $2/3$ of all $n_A$ 
eigenvalues. It turns out that among these ``acceptable'' eigenvalues most of 
them are virtually exact with $\delta^2E(\psi)<10^{-20}$ 
(or even better), especially for the eigenvalues closest to the Green function 
energy $E$ or with rather large values of $\xi_E$ or $\xi_x$. Only some 
eigenvalues at the boundaries $E\pm \Delta E$ (with $\Delta E$ depending 
on $N$ and $n_A$)  of the obtained energy band were of 
modest quality with $\delta^2E(\psi)$ between $10^{-20}$ and $10^{-8}$.

Concerning the interaction strength $U$ and the approximate energy range $E$ 
we present here the detailed results for the eigenvectors of four cases which 
are $U=4.5$ combined with $E=-3.1$, $U=7.2$ combined with $E=1.8$ 
and also the less optimal interaction strength $U=7.8$ with two 
possible energy values $E=-2.8$ and $E=1.8$. For three of theses cases 
($U=4.5$, $U=7.2$ and $U=7.8$ with $E=1.8$) the approximate energy range can 
be obtained as the average energy $\langle H\rangle$ of the tail state 
computed 
from the time evolution and given in Table~\ref{table1}. For the last case 
the second interesting energy value $E=-2.8$ for $U=7.8$ 
can be found by exact diagonalization for small system sizes ($N=55$ and 
$N=89$) and was also identified in Fig.~1(a) of Ref.~\cite{flach}. (Actually, 
the Green function Arnoldi method 
is for small system sizes also suitable for a full matrix 
diagonalization by choosing $n_A=N(N+1)/2$ identical to the dimension 
of the symmetrized two-particle Hilbert space.) 

The Green function Arnoldi method requires to fix a preferential 
energy for the Green function which determines the approximate 
energy range of computed eigenvalues and eigenvectors. 
For this we use a refinement procedure 
where at each system size $N$ this energy is either chosen as 
the eigenvalue of the eigenstate with maximum $\xi_E$
obtained from the last smaller system size or, for the smallest system 
size $N=55$, as one of the above given approximate energy values 
essentially obtained as the average energy of the time 
evolution tail state. This systematic refinement is indeed necessary 
if one does not want to miss the strongest delocalized states since 
the typical energy width of ``good'' eigenvalues 
provided by the method decreases rather strongly 
with increasing system size, e.~g. $\Delta E\sim 10^{-3}$ for $N=10946$. 

In this way we obtained indeed the strongest delocalized states up to the 
largest 
considered system size. However, for $N=10946$ we added one or two additional 
runs at some suitable neighbor values for $E$ which allowed us to obtain 
a more complete set of delocalized states. We also made an 
additional verification that overlapping states, obtained by two 
different runs at different $E$ values, were indeed identical for both 
runs and did not depend 
on the precise value of $E$ used in the Green function Arnoldi 
method provided that the eigenvalue of the overlapping eigenstate 
was sufficiently close to both $E$ values. 
In general, if one is interested in an eigenstate which by accident 
is close to the boundary of the good energy interval and is therefore of 
limited quality, one can easily improve its quality 
by starting a new run with a Green function energy closer to the eigenvalue 
of this state.

\begin{figure}
\begin{center}
\includegraphics[width=0.48\textwidth]{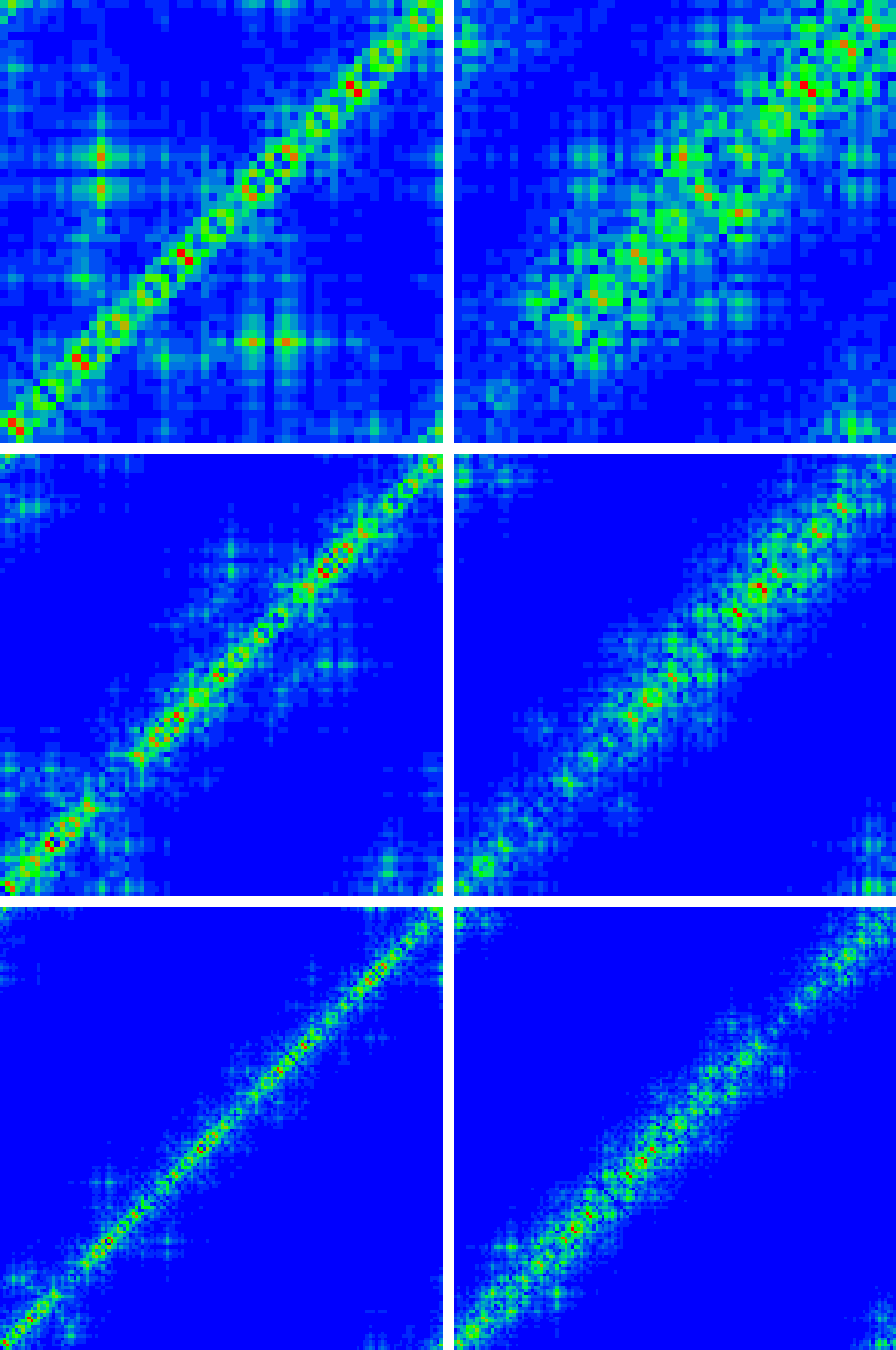}
\caption{Density plot of FIKS eigenstates 
with maximal value $\xi_E$ for system size $N=55$ (top panels), 
$N=89$ (center panels), $N=144$ (bottom panels), $U_R=1$ 
and interaction strength $U=4.5$ (left column) or $U=7.8$ (right column). 
The corresponding energy eigenvalues and values for both types 
of inverse participation ratios are: 
{\em Top left:} $E=-3.10334$, $\xi_E=22.756$, $\xi_x=30.794$.
{\em Top right:} $E=-2.75868$, $\xi_E=33.274$, $\xi_x=24.901$.
{\em Center left:} $E=-3.09588$, $\xi_E=50.742$, $\xi_x=49.867$.
{\em Center right:} $E=-2.78575$, $\xi_E=35.139$, $\xi_x=28.198$.
{\em Bottom left:} $E=3.09966$, $\xi_E=61.373$, $\xi_x=63.353$.
{\em Bottom right:} $E=-2.78596$, $\xi_E=56.210$, $\xi_x=47.958$.
}
\label{fig4}
\end{center}
\end{figure}

In Fig. \ref{fig4} we show density plots for 
the strongest delocalized eigenstates 
(in $\xi_E$) for the two cases $U=4.5$, $E\approx -3.1$ and $U=7.8$, 
$E\approx -2.8$ and the three smallest system sizes $N=55$, $N=89$ and 
$N=144$. In all cases the eigenstate extends to the full diagonal along 
$x_1\approx x_2$ with a width of about 7 sites ($U=4.5$) or about 15 sites 
($U=7.8$) with a quasiperiodic structure of holes or strong peaks. 
One can also identify some additional peaks with $|x_1-x_2|\sim 20$-$30$ which 
can be interpreted as a resonant coupling of the main state 
with some product state of non-interacting one-particle eigenstates with 
both particles localized at some modest distance a bit larger than 
the one-particle localization length $\ell\approx 4.48$ and where the 
eigenvalue of the main state is very close to the total energy of the product 
state. 

\begin{figure}
\begin{center}
\includegraphics[width=0.48\textwidth]{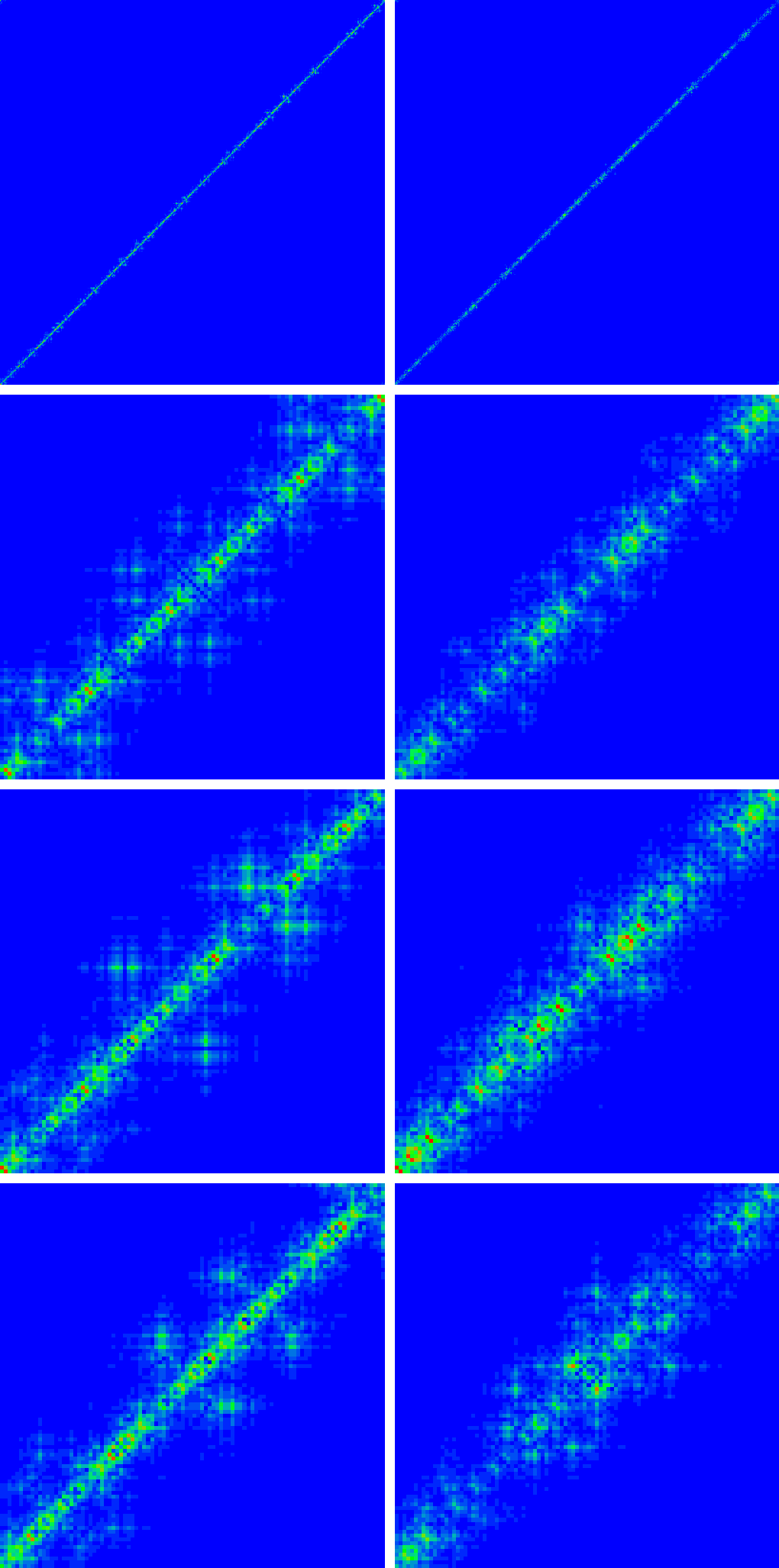}
\caption{Density plot of FIKS eigenstates 
with maximal value of $\xi_E$ for system size $N=1597$, $U_R=1$ 
(both columns) and interaction strength $U=4.5$, energy eigenvalue 
$E= -3.09644$, $\xi_E=616.638$, $\xi_x=716.050$ 
 (left column) or $U=7.8$, $E=-2.78777$, $\xi_E=330.269$, $\xi_x=355.236$ 
 (right column). 
The first row corresponds to the full eigenstates and 
the other rows correspond to zoomed regions of size $100\times 100$ 
with bottom left corner at position $x_1=x_2=0$ (second row), 
$x_1=x_2=700$ (third row) and $x_1=x_2=1400$ (fourth row). 
}
\label{fig5}
\end{center}
\end{figure}

\begin{figure}
\begin{center}
\includegraphics[width=0.48\textwidth]{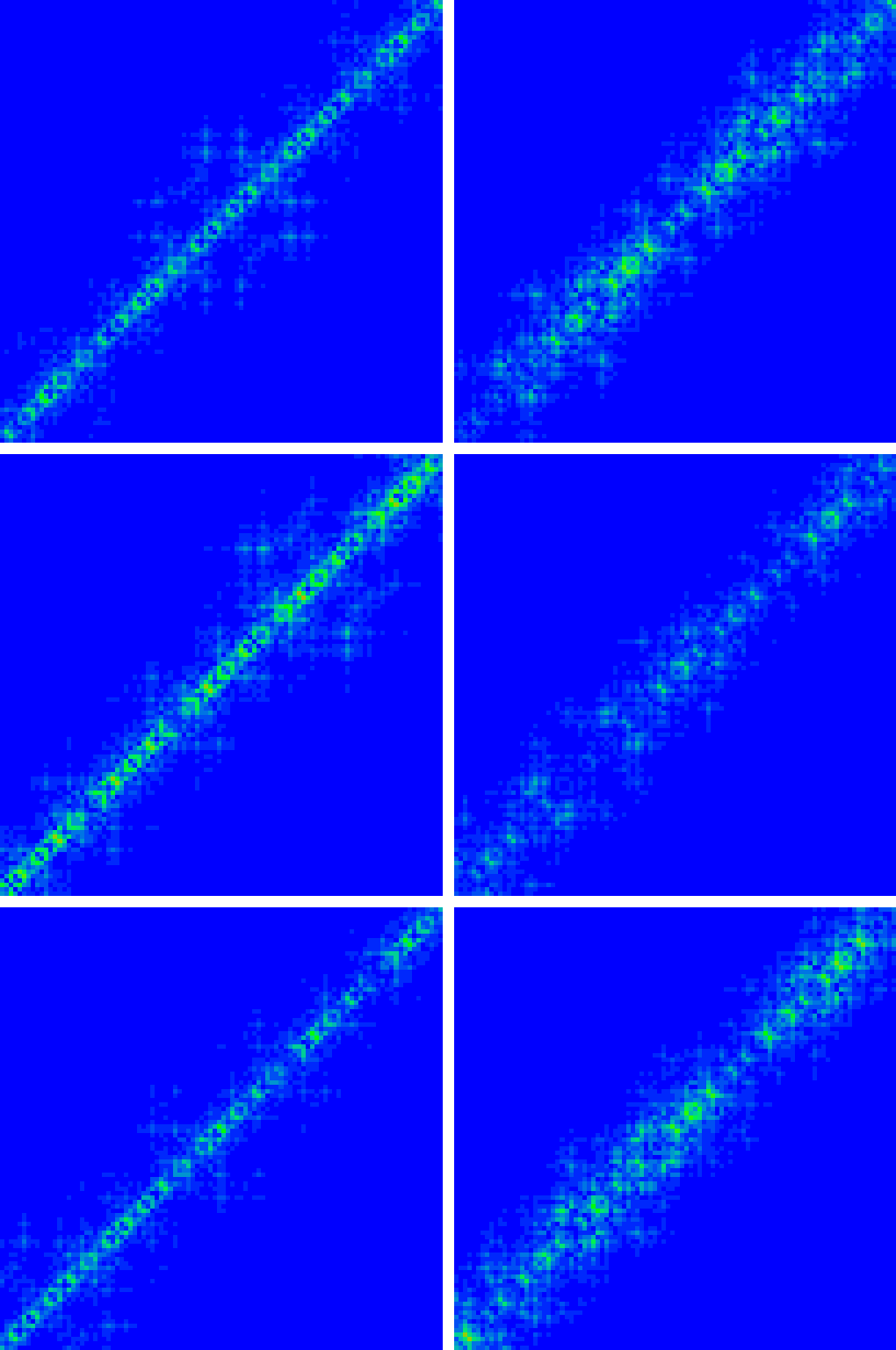}
\caption{Density plot of FIKS eigenstates 
with maximal value of $\xi_E$ for system size $N=10946$, $U_R=1$ 
(both columns) and interaction strength $U=4.5$, energy eigenvalue 
$E= -3.09749$, $\xi_E=2099.806$, $\xi_x=3105.529$ 
 (left column) or $U=7.8$, $E=-2.78707$, $\xi_E=952.498$, $\xi_x=1147.965$ 
 (right column). 
All panels correspond to zoomed regions of size $100\times 100$ 
with bottom left corner at position $x_1=x_2=0$ (first row), 
$x_1=x_2=5000$ (second row) and $x_1=x_2=10000$ (third row). 
}
\label{fig6}
\end{center}
\end{figure}

In Fig. \ref{fig5} and Fig. \ref{fig6}, the strongest delocalized states for 
$N=1597$ ($N=10946$) and the same values of $U$ and approximate energy 
as in Fig. \ref{fig4} are shown as full states (only 
for $N=1597$) and with three zoomed regions of size $100\times 100$ 
at three different positions on the diagonal (for $N=1597$ and $N=10946$). 
Again the 
eigenstates extend to the full diagonal size with a certain width 
and one can identify a a quasiperiodic structure of holes and peaks and 
some resonant couplings to product states of non-interacting 
one-particle eigenstates. Higher quality gif files for the 
full eigenstate of these (and some other) cases are available 
for download at \cite{webpage}. 

Figs. \ref{fig4}-\ref{fig6} also show that, 
apart from the common features, with increasing system size the 
eigenstates seem to become ``thinner'', i.~e. the weight of the hole parts 
seems to increase and the strength of peaks seems to decrease, especially 
for the case $U=7.8$ and approximate energy $E=-2.8$.

\begin{figure}
\begin{center}
\includegraphics[width=0.48\textwidth]{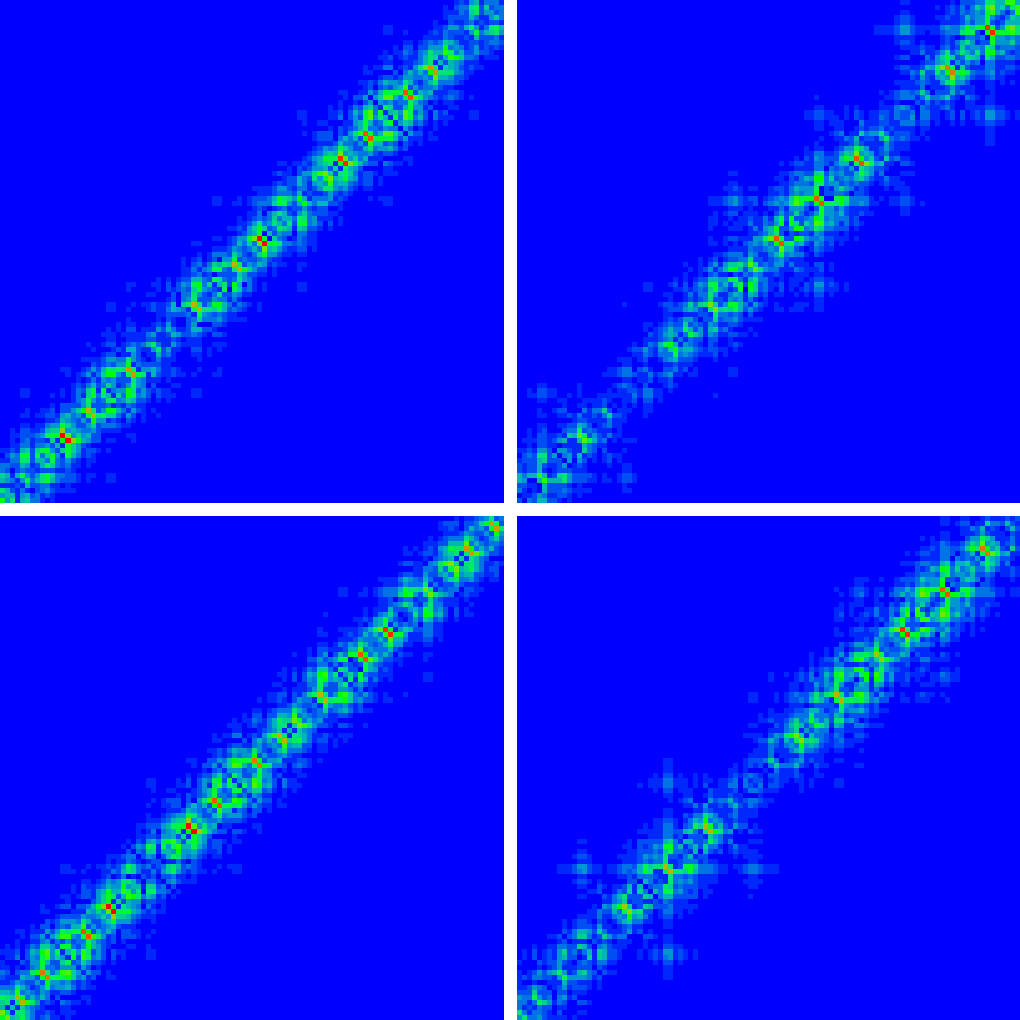}
\caption{Density plot of FIKS eigenstates 
with maximal value of $\xi_E$ for $E\approx 1.8$, 
$U_R=1$, $U=7.2$ (left column) 
or $U=7.8$ (right column) and 
system size $N=1597$ (top panels) or $N=10946$ (bottom panels). 
All panels correspond to zoomed regions of size $100\times 100$ 
with bottom left corner at position 
$x_1=x_2=700$ (top panels) or $x_1=x_2=5000$ (bottom panels). 
The corresponding energy eigenvalues and values for both types 
of inverse participation ratios are: 
{\em Top left:} $E=1.79597$, $\xi_E=638.916$, $\xi_x=506.113$,
{\em Top right:} $E=1.81744$, $\xi_E=475.972$, $\xi_x=359.239$, 
{\em Bottom left:} $E=1.79652$, $\xi_E=5694.610$, $\xi_x=4834.890$,
{\em Bottom right:} $E=1.81741$, $\xi_E=2086.088$, $\xi_x=1843.227$.
The figures obtained for other zoomed regions (on the diagonal) 
for these states 
are very similar and these four eigenstates extend clearly to the 
full diagonal $x_1\approx x_2$ and all values with $0\le x_1<N$. 
}
\label{fig7}
\end{center}
\end{figure}

Fig. \ref{fig7} shows a zoomed region of size $100\times 100$ 
roughly in the middle of the diagonal for 
strongest delocalized eigenstates for $N=1597$ and $N=10946$ and 
the two cases $U=7.2$ and $U=7.8$, both with the 
approximate energy $E=1.8$. 
Globally one observes in Fig.~\ref{fig7} the same features as in the 
Figs.~\ref{fig5}-\ref{fig6} for the previous two cases 
but with a detail structure on the diagonal which is significantly different, 
i.~e. quite large width and different pattern for the quasiperiodic peak-hole 
structure. One observes that the eigenstates for $U=7.2$ are very compact 
while for $U=7.8$ they are a bit less compact, with more holes, but also 
with additional small satellite contributions from product pair-states 
at distance $\approx 20$ from the diagonal. These satellite contributions 
are absent at $U=7.2$. Apart from this the pattern for both cases in 
Fig.~\ref{fig7} is 
rather similar, i.e. the FIKS eigenstates for $E\approx 1.8$, and 
$U=7.2$ or $U=7.8$ belong to the same family but obviously the value 
$U=7.2$ is more optimal with a compacter structure, larger values of 
$\xi_E$ and $\xi_x$. This is also in agreement with the discussion of the 
time evolution states in the previous Section. It is interesting to note 
that even for the case $U=7.8$ with a modest 
squared tail norm $\approx 2\times 10^{-4}$ (instead of $5\times 10^{-2}$ 
for $U=7.2$, see Table~\ref{table1}) there are very clear FIKS eigenstates 
and even at two different energy regions. 

We have also calculated eigenstates up to system sizes $N=2584$ 
for the additional case $U=7.2$ and $E\approx -2.8$ in order to verify 
if the second energy value is also interesting for $U=7.2$. 
Here one finds also some FIKS eigenstates but 
of reduced quality if compared to $U=7.8$ and $E\approx -2.8$, i.~e. 
smaller values of $\xi_E$ and $\xi_x$ and for larger system sizes the 
eigenstates do not extend to the full diagonal, i.~e. about 20-40\% of 
the diagonal is occupied for $N=2584$. For this additional case we do 
not present any figures.

\begin{figure}
\begin{center}
\includegraphics[width=0.48\textwidth]{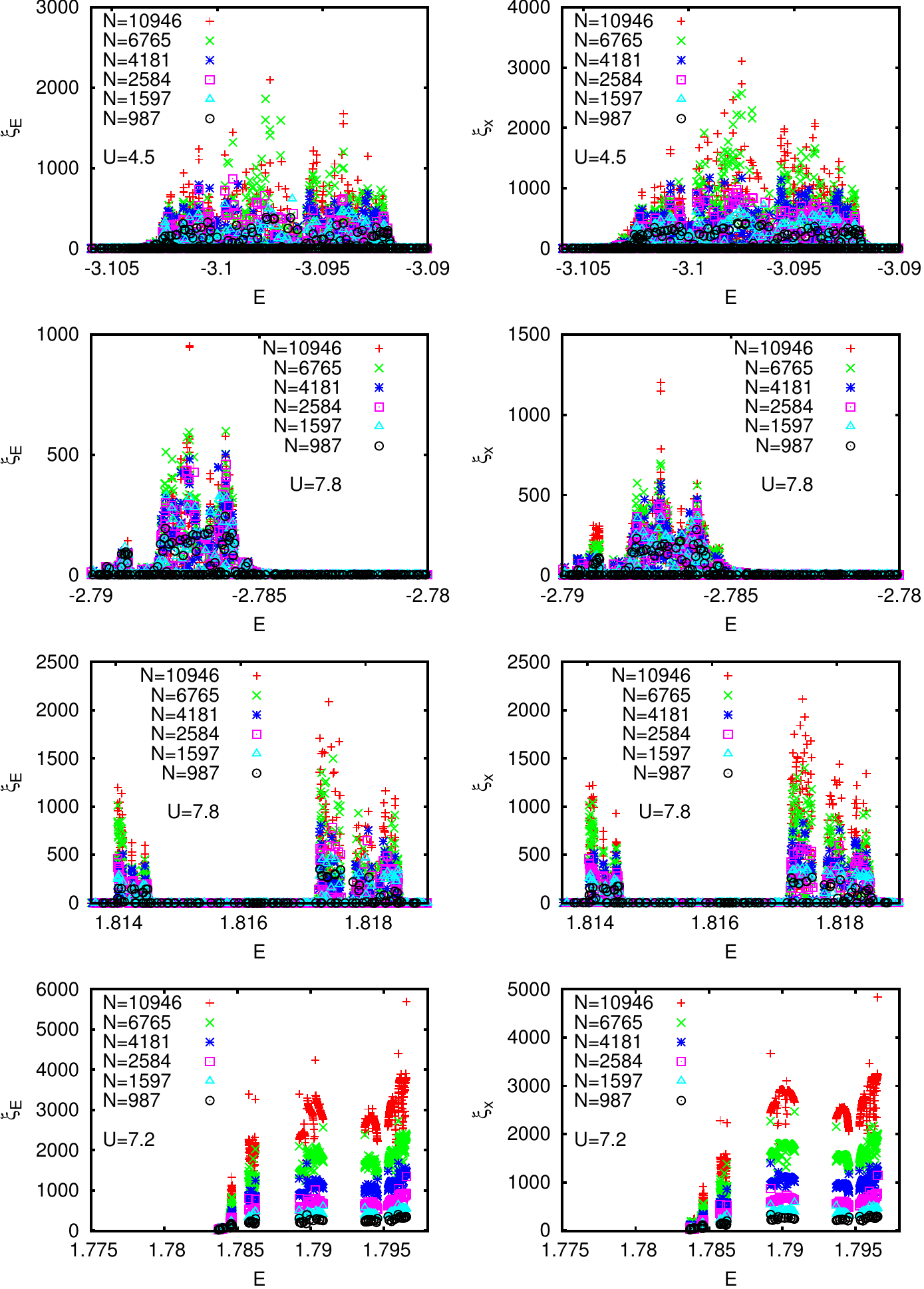}
\caption{Inverse participation ratio of eigenstates 
versus eigenvalue energy for the system sizes 
$N=987$, 1597, 2584, 4181, 6765, 10946 and $U_R=1$. 
The left column of panels correspond 
to the inverse participation ratio $\xi_E$ in energy representation 
and the right column to the inverse participation 
ratio $\xi_x$ in position representation. 
First row of panels correspond to $U=4.5$ and 
the energy region $E\approx -3.098$, second (third) row of panels 
correspond to $U=7.8$ and $E\approx -2.787$ ($E\approx 1.817$)
and fourth row of panels 
correspond to $U=7.2$ and $E\approx 1.79$. }
\label{fig8}
\end{center}
\end{figure}

In Fig. \ref{fig8} both types of inverse participation ratios 
$\xi_E$ and $\xi_x$ of eigenstates are shown as a function of the 
energy eigenvalue 
for all four cases (corresponding to Figs.~\ref{fig5}-\ref{fig7}) 
with energies in the interesting regions and 
for the six largest values of the system size between 987 and 10496. 
Both quantities increase considerably with system size and the 
overall shape of the cloud of points seems to be similar for each value 
of $N$ but with a vertical scaling factor increasing with $N$. The 
figures for $\xi_E$ and $\xi_x$ are rather similar with somewhat larger 
(maximum) values for $\xi_x$ (except for $U=7.2$ where the maximum value 
of $\xi_E$ is larger). For $U=4.5$ the energy region of 
delocalized states extends from $E\approx -3.103$ to $E\approx -3.092$ 
and for $N=10496$ two supplementary runs with Green's function energy values 
shifted to the left ($E=-3.104$) and right ($E=-3.094$) from the center 
($E=-3.0977$) were necessary to obtain 
a complete cloud of data points. 
For $U=7.8$ and approximate energy 
$E=-2.8$ the main region of delocalized eigenstates extends from 
$E\approx -2.788$ to $E\approx -2.786$ with a secondary small region 
at $E\approx -2.789$. For the secondary region and $N=10946$ also 
an additional run with a shifted Green function energy was necessary. 
For $U=7.8$ and approximate energy 
$E=1.8$ the main region of delocalized eigenstates extends from 
$E\approx 1.8172$ to $E\approx 1.8186$ also with a secondary small region 
at $E\approx 1.814$ and for this secondary region and $N=10946$ also 
an additional run with a shifted Green function energy was necessary. 

For $U=7.2$ and approximate energy $E=1.8$ the main region of 
delocalized eigenstates extends from $E\approx 1.793$ to $E\approx 1.797$. 
For this particular case one observes the absence of eigenstates 
with very small values of $\xi_E\approx 1$ and $\xi_x\approx$3-4. 
We have verified, by choosing different values of the Arnoldi dimension $n_A$ 
and the Green function energy, that the absence of such states is stable 
with respect to different parameters of the numerical method. 
Apparently in this energy region there are no strongly localized product states 
(of one-particle energy eigenstates)  with a modest distance 
between the two particles such that there would be some contribution of 
them in the initial state used for the Arnoldi method. There may still 
be other 
product states in this energy region but with the two particles localized 
further away such that the Arnoldi method cannot detect them.

\begin{figure}
\begin{center}
\includegraphics[width=0.48\textwidth]{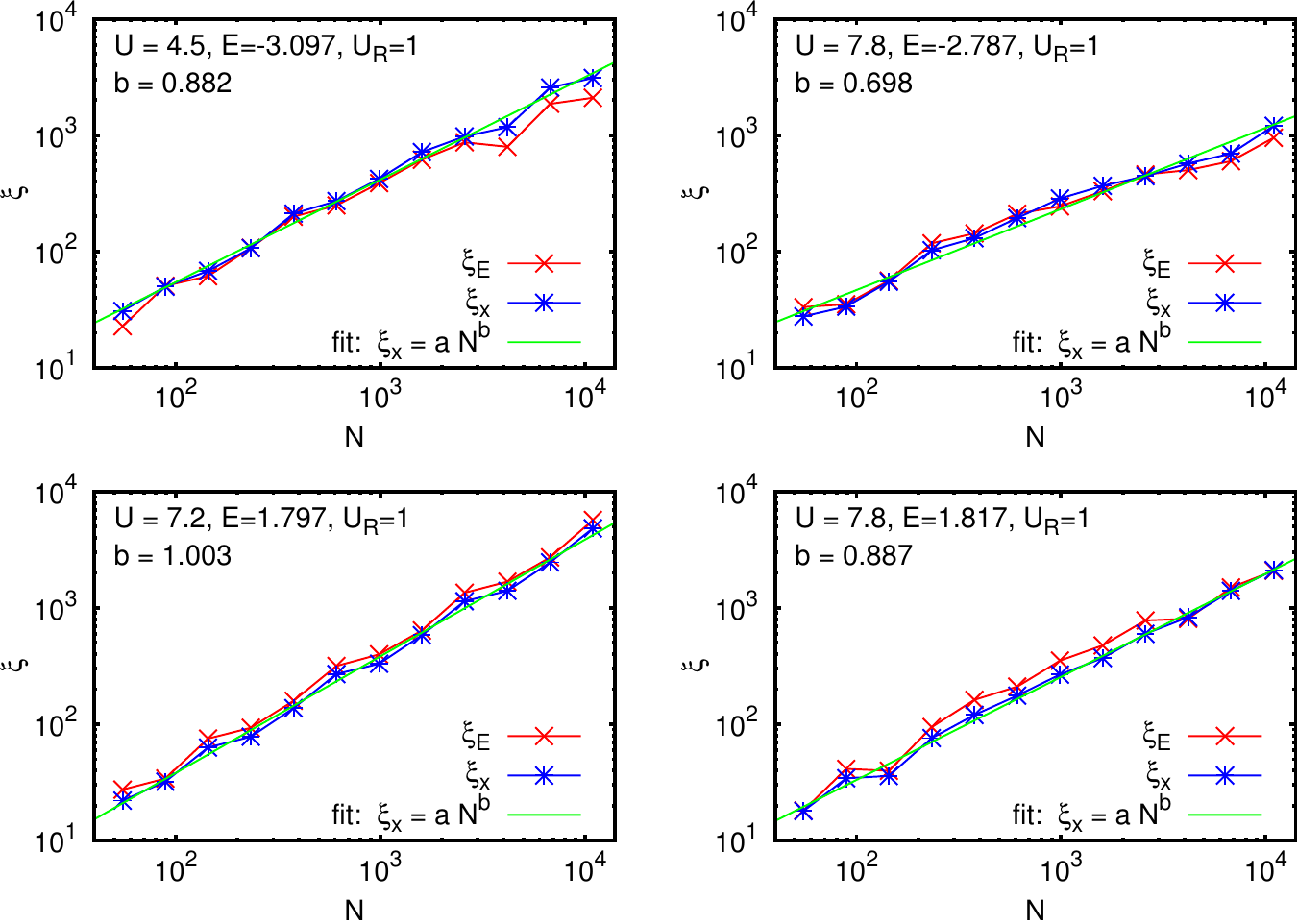}
\caption{Largest inverse participation ratio (for given values of $N$
and approximate energy) of FIKS eigenstates 
versus system size $N$ in a double logarithmic scale 
using all Fibonacci numbers between 55 and 10946. 
Top (bottom) left panel corresponds to $U=4.5$ ($U=7.2$) and 
the energy region $E\approx -3.1$ ($E\approx 1.8$). 
Top (bottom) right panel corresponds to $U=7.8$ and $E\approx -2.8$ 
($E\approx 1.8$). The blue line with stars 
corresponds to the inverse participation ratio 
$\xi_x$ in position representation, the red line 
with crosses to the inverse participation ratio 
$\xi_E$ in energy representation and 
the green line to the power law fit $\xi_x=a\,N^b$ with 
fit results shown in Table~\ref{table2}. The shown energy values 
in the panels refer to the eigenvector with maximal $\xi_E$ 
for the largest system size. 
Note that for given values of $N$ and approximate energy 
the eigenstates with maximal $\xi_x$ and 
maximal $\xi_E$ may be different. 
}
\label{fig9}
\end{center}
\end{figure}

\begin{table}
\caption{Approximate energy $E$ (for largest system size) and results 
of the power law fit $\xi_x=a\,N^b$ for the same cases 
and data sets as in Fig. \ref{fig9}. }
\begin{center}
\begin{tabular}{|r|r|r|l|l|}
\hline
$U$ & $U_R$ & $E$ & $a$ & $b$ \\
\hline
\hline
$4.5$ & $ 1$ & $-3.097$  & $0.940 \pm 0.137$ & $0.882 \pm 0.021$  \\
\hline
$7.2$ & $ 1$ & $ 1.797$  & $0.375 \pm 0.054$ & $1.003 \pm 0.021$  \\
\hline
$7.8$ & $ 1$ & $-2.787$  & $1.878 \pm 0.380$ & $0.698 \pm 0.029$  \\
\hline
$7.8$ & $ 1$ & $ 1.817$  & $0.559 \pm 0.073$ & $0.887 \pm 0.019$  \\
\hline
\end{tabular}
\label{table2}
\end{center}
\end{table}

The scenario of strongly delocalized eigenstates for certain narrow energy 
bands found in Ref. \cite{flach} is clearly confirmed also for larger 
system sizes up to $N=10946$. However, the maximum values of $\xi_E$ and 
$\xi_x$ do not scale always linearly with $N$ as can be seen in 
Fig. \ref{fig9} 
which shows the dependence of maximum values of $\xi_E$ and $\xi_N$ 
for all four cases (of interaction strength and approximate energy) 
as a function of the system size $N$ in a double logarithmic scale. 
Note that in Fig. \ref{fig9} the 
data points for maximum $\xi_E$ (for given values of $N$, $U$ and 
approximate energy) may correspond to other eigenstates than for the 
data points for maximum $\xi_x$, i.~e. the maximum values for the two 
quantities are obtained at two different eigenstates. 
For example for $U=4.5$ and $N=6765$ the eigenstate with maximum 
$\xi_E$ corresponds to $E=-3.09771$, $\xi_E=1861.131$, $\xi_x=2538.299$ 
while the eigenstate with maximum $\xi_x$ corresponds 
to $E=-3.09749$, $\xi_E=1406.560$, $\xi_x=2573.484$, a state which 
ranks on the 5th position in the list of states with maximum values 
for $\xi_E$. However, despite such particular cases the appearance 
of large values for $\xi_E$ (strong delocalization in one-particle energy 
representation) or $\xi_x$ (strong delocalization in position representation) 
are rather well correlated which is obvious since 
the transformation from energy to position representation corresponds somehow 
to a ``smoothing'' on the length scale of the one-particle localization 
length $\ell\approx 4.48$. 

The results of the power law fit $\xi_x=a\,N^b$ using the data sets of 
Fig. \ref{fig9} are shown in Table~\ref{table2}. 
For $U=4.5$ or $U=7.8$ (both energy ranges) the fit values of the exponent 
$b$, which are either close to $0.9$ or $0.7$, seem 
to indicate a kind of fractal structure of the eigenstates since even for 
the largest system sizes the corresponding eigenstates extend to the full 
length of the diagonal $x_1\approx x_2$. Therefore the reduction of $\xi_x$ 
with respect to a linear behavior in $N$ is due to 
the internal structure (appearance of more ``holes''). This is also in 
agreement with our above observation 
that delocalized eigenstates seem to become thinner for larger systems sizes 
and this effect is strongest for the case $U=7.8$, $E\approx -2.8$ which 
also corresponds to the smallest value of the exponent 
$b=0.698$ among the three cases. However, for $U=7.2$ the exponent is rather 
precisely unity and no fractal or increasing hole structure (with 
increasing system size) is visible in the FIKS eigenstates (see also 
Fig.~\ref{fig7}).

\section{Eigenstates for long range interaction} 
\label{sec6}

\begin{figure}
\begin{center}
\includegraphics[width=0.48\textwidth]{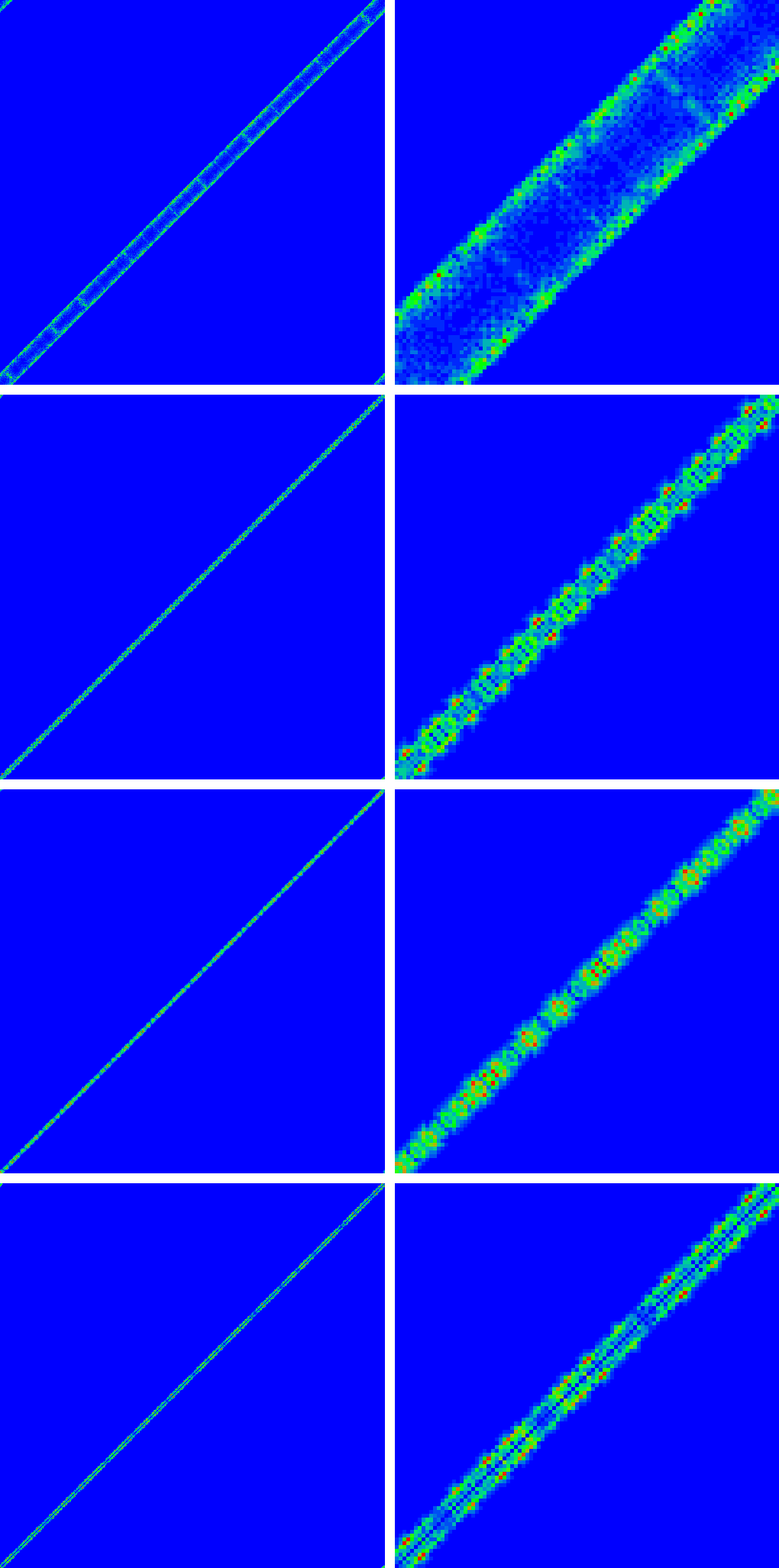}
\caption{Density plot of FIKS eigenstates for different cases of 
long range interaction $U_R>1$ 
with maximal inverse participation 
ratio $\xi_E$ in energy representation for $N=610$. 
The left column corresponds to the full eigenstate and 
the right column to a zoomed region of size $100\times 100$ 
with bottom left corner at position $x_1=x_2=200$. 
{\em First row:} $U_R=20$, $w=0$, $U=14$, boson case, energy eigenvalue 
$E=14.00502$, $\xi_E=263.410$, $\xi_x=350.519$.
{\em Second row:} $U_R=5$, $w=0$, $U=8$, boson case, $E=8.79607$, 
$\xi_E=787.137$, $\xi_x=397.779$.
{\em Third row:} $U_R=7$, $w=1$, $U=17$, boson case, $E=10.22864$,
$\xi_E=635.918$, $\xi_x=307.585$.
{\em Fourth row:} $U_R=5$, $w=0$, $U=10.9$, fermion case, $E=11.53294$,
$\xi_E=535.618$, $\xi_x=360.478$.
}
\label{fig10}
\end{center}
\end{figure}

We now turn to the case of long range interactions with $U_R>1$. 
We remind that we consider a model where the particles are 
coupled by the interaction potential 
$U(x_1-x_2)$ with $U(x)=U/(1+w|x|)$ for $|x|<U_R$ \cite{U_boundary} 
and $U(x)=0$ if $|x|\ge U_R$. For the decay parameter $w$ we mostly 
choose $w=0$ (i.~e. ``no decay'') or for the boson case also $w=1$ 
(decay $\sim |x_1-x_2|^{-1}$ provided that $|x_1-x_2|<U_R$). 

We considered many different cases with $2\le U_R\le 7$ and one case 
with $U_R=20$ and performed for each case a time evolution analysis as 
described in Sec. \ref{sec4} to find good candidates of the interaction 
strength $U$ for strong delocalization. Using the tail state analysis 
we also obtained suitable approximate energy values to start the 
Green function Arnoldi method for the smallest system size $N=55$ we 
considered. Then we refined the Green function energy for 
larger system sizes in the same way as described above.
In many cases (but not always) 
this procedure leads to a nice data set of well delocalized two-particle 
eigenstates for a given narrow energy band. In certain cases the 
refinement procedure gets trapped at a ``wrong'' energy, 
i.e. which is promising for a particular small system size but where the 
localization saturates at some medium value for $\xi_E$ for larger system 
sizes or is simply less optimal than some other energy. In these cases 
it might be useful to manually select a different 
eigenvalue obtained from the last smaller system (e.~g. for $N=55$ or $N=89$) 
to force the refinement of energies into a direction of stronger delocalized 
states. 

We mention that for the larger 
values of $U_R$ the computational cost [$\sim (NU_R)^3$] 
and the memory requirement  [$\sim (NU_R)^2$] of the initial preparation 
part of the Green function Arnoldi method is considerably increased and 
therefore we have limited for these cases the maximal 
considered system size to $N\le 1597$. 

In Fig. \ref{fig10} we show the strongest delocalized state (in $\xi_E$) 
for $N=610$ and the case $U=14.0$, $U_R=20$, $w=0$, boson case (top panels) 
and the three cases with $U_R>1$ already presented in Figs. 
\ref{fig1}-\ref{fig3} of Sec. \ref{sec4} (second to fourth row of panels). 
Concerning the case $U_R=7$, $U=16.9$, $w=1$, bosons (of Sec. \ref{sec4}), 
it turns out that 
for the eigenstate analysis the interaction strength $U=17.0$ is somewhat 
more optimal than the case of $U=16.9$. Therefore we show in Fig. \ref{fig10} 
(and other figures in this Section) the case of $U=17.0$ instead of 
$U=16.9$. For each case the left column panel of Fig. \ref{fig10} 
shows the full state and the right column panel a zoomed region 
of size $100\times 100$ with bottom left corner at position $x_1=x_2=200$ 
for a better visibility. 

The energy eigenvalues of the three boson states in Fig. \ref{fig10}: 
$E=14.00502$, $E=8.79607$ or $E=10.22864$ (top three rows of panels) 
correspond quite well to the approximate energies obtained from the tail 
state analysis of the time evolution wave packet for 
the same (or very similar) parameters: 
$\langle H\rangle=14.00247$, $\langle H\rangle=8.72561$ or 
$\langle H\rangle=10.18926$ (see also Table~\ref{table1}). 
However for the fermion case (fourth row of panels with $U=10.9$, $U_R=5$, 
$w=0$) the energy eigenvalue of the strongest delocalized state at $N=610$ is 
$E=11.53294$ while the approximate energy obtained from the tail 
state analysis $\langle H\rangle=10.88786$ is somewhat different. 
Here the refinement procedure to optimize $\xi_E$ leads already at the 
first Green function Arnoldi calculation for $N=55$ and $n_A=400$ 
to an energy shift from 
$10.9$ (as initial Green's function energy) to $11.5$ (as eigenvalue 
of the eigenstate with maximum $\xi_E$). However, optimizing for $\xi_x$ 
(instead of $\xi_E$) or fixing manually the value $E=10.9$ for $N=144$ 
results in a different set of strongly delocalized eigenstates close to the 
energy $E=10.84$ with somewhat smaller values for $\xi_E$ but larger 
values for $\xi_x$ than the first set of delocalized eigenstates at 
$E=11.53$. 

The eigenstates shown in  Fig. \ref{fig10} have the same common features 
as the eigenstates shown in Figs. \ref{fig4}-\ref{fig7} for the Hubbard 
short range interaction discussed previously such 
as extension to the full diagonal at $x_1\approx x_2$, a certain width of 
$\sim 10$-$20$ sites, quasiperiodic structure of holes and peaks etc. but the 
detail pattern is specific for each case. For the very long interaction 
range $U_R=20$ one observes more a double diagonal structure with main 
contributions for positions such that $x_2\approx x_1\pm 20$.

\begin{figure}
\begin{center}
\includegraphics[width=0.48\textwidth]{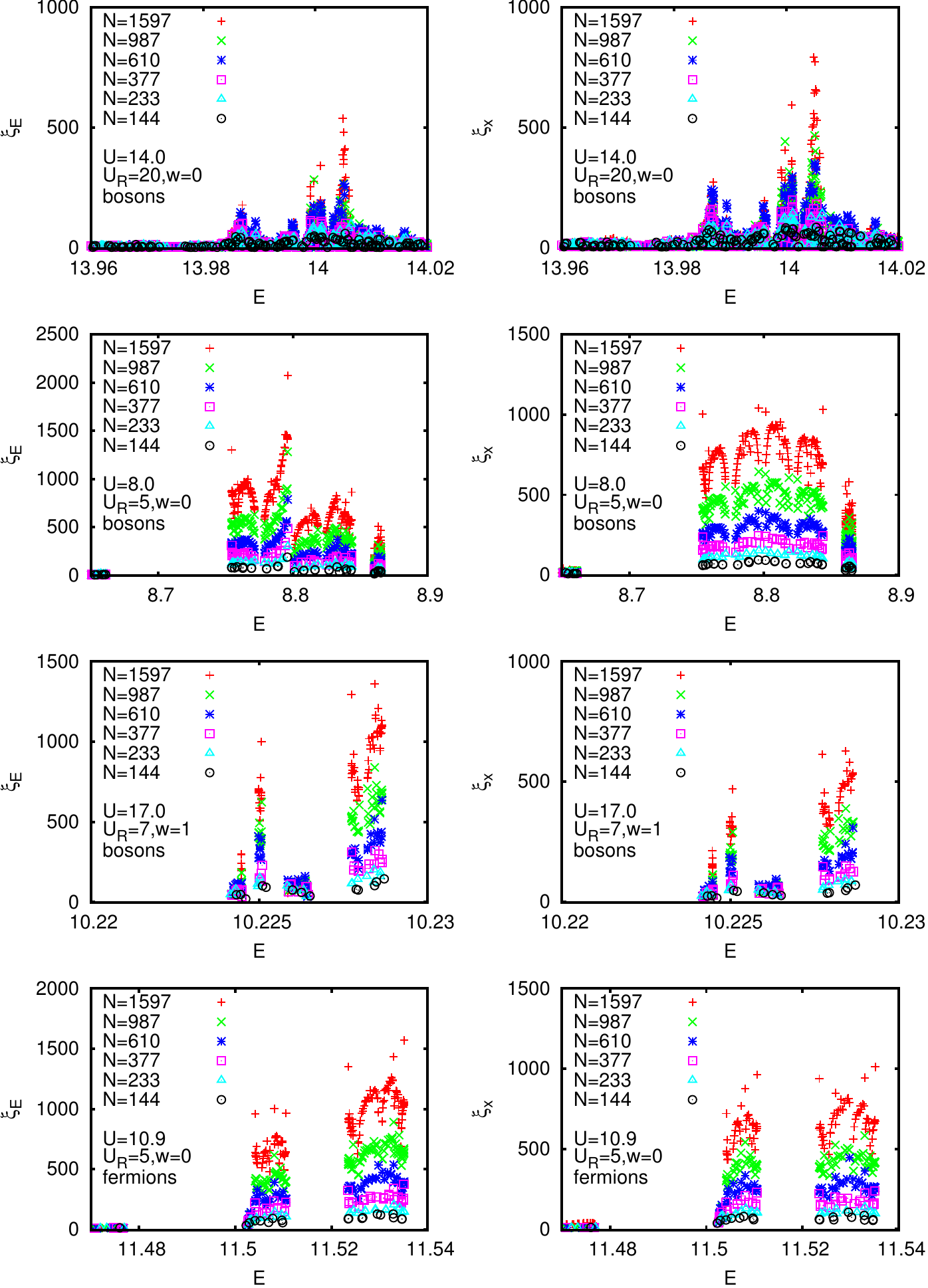}
\caption{Inverse participation ratio of eigenstates 
versus eigenvalue energy for the system sizes 
$N=144$, 233, 377, 610, 987, 1597 and the same four cases 
with $U_R>1$ as in Fig. \ref{fig10} (see labels in panels 
for the values of the parameters $U$, $U_R$, $w$ and boson or fermion case). 
The left column of panels correspond 
to the inverse participation ratio $\xi_E$ in energy representation 
and the right column to the inverse participation 
ratio $\xi_x$ in position representation. }
\label{fig11}
\end{center}
\end{figure}

The energy dependence of both $\xi_E$ and $\xi_x$ for all four cases of 
Fig. \ref{fig10} and all system sizes between $144$ and $1597$ 
is shown in Fig. \ref{fig11}. As in the Hubbard interaction case 
(see Fig. \ref{fig8}) the typical values of $\xi_E$ and $\xi_x$ 
increase systematically with the system size and for each case there 
is a certain narrow, quite well defined, energy band for strongly 
delocalized eigenstates. 

In addition to this, for the three cases presented in the three lower 
rows of panels in Fig. \ref{fig11} one does not see many data points for 
strongly localized states (with $\xi_E\sim 1$) 
inside or close to this narrow energy band in contrast to 
Fig. \ref{fig8} where a lot of eigenstates with very small values 
of $\xi_E$ or $\xi_x$ are visible (for three out of four cases). 
The reason for this is that the total 
energy for these three cases is outside the interval $|E|<6$ 
for non-interacting product states (at $\lambda=2.5$) where the two 
particles are localized more or less far away with only small (or absent) 
effects due to the interaction. Therefore contributions of such products 
state cannot be seen for the particular narrow energy bands visible in 
Fig. \ref{fig11}. 

In principle this argument also applies to the first row of panels 
in Fig. \ref{fig11} 
(with $U_R=20$ and $U=14.0$), i.e. here products states with 
particles localized {\em far away} cannot be not seen as well. 
However, for the long interaction range $U_R=20$ and due to the fact that 
the interaction is uniform in this range there are {\em other} 
products states where 
both particles are localized at a {\em distance smaller than $U_R$} which is 
possible due to the small one-particle localization length  $\ell=4.48<20$. 
The spatial structure of these kind of product states is not modified by 
the uniform interaction. Therefore they are strongly localized, but 
obviously the energy eigenvalue of such a short range product range 
is shifted by the mean value of the uniform interaction $U=14.0$ 
(with respect to the sum of the two one-particle energies) 
therefore explaining that it is possible to find such states for energies 
close to $E\approx 14$. This explains also that more complicated effects 
of the interaction, such as the creation of strongly delocalized two-particle 
states, happen if both particles are at an approximate distance $\sim 20$ 
such that the interaction coupling matrix elements (between 
non-interacting product states with both particles at critical distance 
$\sim U_R$) have a more complicated and subtle structure 
due to complicated boundary effects. 
One may note that this particular type of interaction is similar to 
the {\em bag model} studied in \cite{dlstip,frahm1995}.

\begin{figure}
\begin{center}
\includegraphics[width=0.48\textwidth]{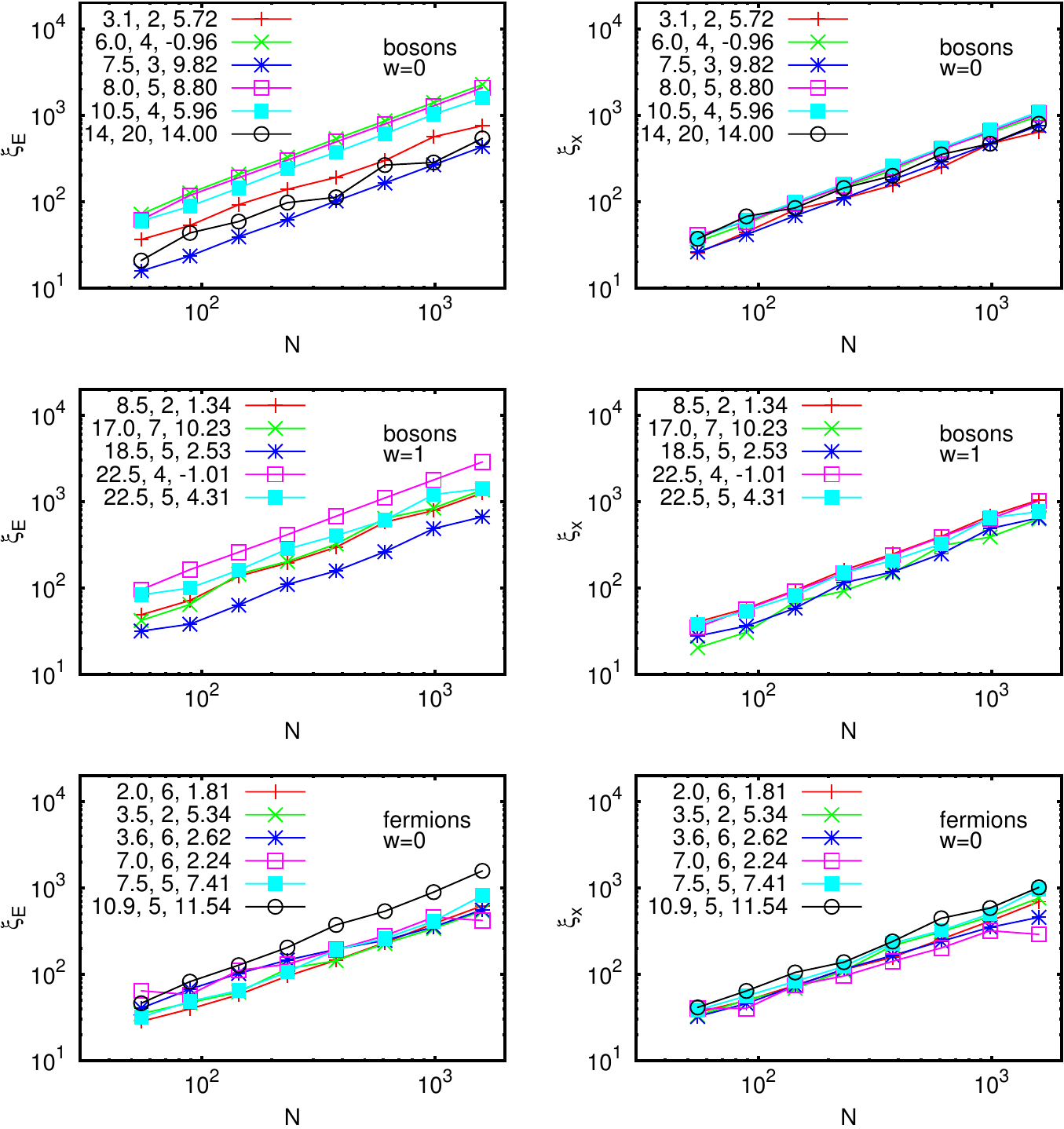}
\caption{Largest inverse participation ratio (for a given value of $N$) 
of FIKS eigenstates versus system size $N$ using all Fibonacci 
numbers between 55 and 1597 for selected cases of long range interactions 
(same data sets as in Table \ref{table3}). 
The left column corresponds to the inverse participation ratio 
$\xi_E$ in energy representation and the right column corresponds to 
the inverse participation ratio $\xi_x$ in position representation.
Top (center) panels correspond to the boson case with the decay parameter 
$w=0$ ($w=1$). Bottom panels correspond to the fermion case with the decay 
parameter $w=0$. 
The three numbers in the color labels in the top left corner represent 
the interaction strength $U$, the interaction range $U_R$ and 
the approximate energy eigenvalue $E$ (for $N=1597$ and the state 
with largest $\xi_E$). 
Note that for a given value of $N$ and set of interaction parameters 
the eigenstates with maximal $\xi_x$ and maximal $\xi_E$ may be different. 
}
\label{fig12}
\end{center}
\end{figure}

\begin{table}
\caption{Approximate energy $E$ and  results 
of the power law fit $\xi_x=a\,N^b$ for selected cases of long range 
interactions (same data sets as in Fig. \ref{fig12}). 
The bottom six rows of the table correspond to the fermion case and the 
other top rows to the boson case. }
\begin{center}
\begin{tabular}{|r|r|r|r|l|l|}
\hline
$U$ & $U_R$ & $w$ & $E$ & $a$ & $b$ \\
\hline
\hline
$ 3.1$ & $ 2$ & 0 & $  5.72$  & $0.633 \pm 0.117$ & $0.942 \pm 0.032$  \\
\hline
$ 6.0$ & $ 4$ & 0 & $ -0.96$  & $0.637 \pm 0.046$ & $0.999 \pm 0.013$  \\
\hline
$ 7.5$ & $ 3$ & 0 & $  9.82$  & $0.465 \pm 0.007$ & $1.002 \pm 0.003$  \\
\hline
$ 8.0$ & $ 5$ & 0 & $  8.80$  & $0.753 \pm 0.055$ & $0.978 \pm 0.013$  \\
\hline
$10.5$ & $ 4$ & 0 & $  5.96$  & $0.684 \pm 0.001$ & $1.000 \pm 0.000$  \\
\hline
$14.0$ & $20$ & 0 & $ 14.00$  & $1.118 \pm 0.172$ & $0.885 \pm 0.027$  \\
\hline
$ 8.5$ & $ 2$ & 1 & $  1.34$  & $0.727 \pm 0.062$ & $0.986 \pm 0.015$  \\
\hline
$17.0$ & $ 7$ & 1 & $ 10.23$  & $0.339 \pm 0.084$ & $1.032 \pm 0.043$  \\
\hline
$18.5$ & $ 5$ & 1 & $  2.53$  & $0.485 \pm 0.100$ & $0.981 \pm 0.036$  \\
\hline
$22.5$ & $ 4$ & 1 & $ -1.01$  & $0.635 \pm 0.001$ & $1.000 \pm 0.000$  \\
\hline
$22.5$ & $ 5$ & 1 & $  4.31$  & $0.842 \pm 0.168$ & $0.936 \pm 0.034$  \\
\hline
$ 2.0$ & $ 6$ & 0 & $  1.81$  & $0.991 \pm 0.147$ & $0.873 \pm 0.026$  \\
\hline
$ 3.5$ & $ 2$ & 0 & $  5.34$  & $0.696 \pm 0.111$ & $0.947 \pm 0.027$  \\
\hline
$ 3.6$ & $ 6$ & 0 & $  2.62$  & $1.308 \pm 0.158$ & $0.807 \pm 0.021$  \\
\hline
$ 7.0$ & $ 6$ & 0 & $  2.24$  & $2.349 \pm 0.718$ & $0.683 \pm 0.053$  \\
\hline
$ 7.5$ & $ 5$ & 0 & $  7.41$  & $0.793 \pm 0.138$ & $0.945 \pm 0.030$  \\
\hline
$10.9$ & $ 5$ & 0 & $ 11.54$  & $0.896 \pm 0.141$ & $0.949 \pm 0.027$  \\
\hline
\end{tabular}
\label{table3}
\end{center}
\end{table}

Fig. \ref{fig12} shows in a double logarithmic scale
the size dependence of the maximal inverse participation ratios $\xi_E$ 
(left column) or $\xi_x$ (right column)
for the above and many other selected cases, 
with different values of $U$, $U_R$, $w$ and boson/fermion case. 
The typical values of $\xi_E$ and $\xi_x$ clearly increase 
strongly with system size $N$ with typical exponents $b\sim 0.7$-$1$ 
obtained from the power low fit $\xi_x=a\,N^b$ as can be seen 
in Table \ref{table3}. For two particular cases the behavior is even linear 
with high precision with $b=1$ and a fit error below $0.03$\% 
(the two data sets shown with $b=1.000\pm 0.000$ in Table \ref{table3}). 

Actually, these two cases are also characterized by the absence 
of strongly localized states with $\xi_E\approx 1$ in the narrow energy band 
(and accessible by the Arnoldi method) 
in a similar way as the case $U=7.2$ for $U_R=1$ discussed previously 
and one may conjecture that the presence of strongly 
localized products states (accessible by the Arnoldi method and with a modest 
distance between both particles) at the same energies as the FIKS 
eigenstates might be a necessary condition to lower the exponent from the 
linear behavior $b=1$ to a fractal value $b<1$, eventually due to some 
weak coupling of FIKS states to strongly localized pairs. Such localized pairs 
with modest distance 
would also be reasonable for the appearance of satellite peaks visible 
in many (but not all) FIKS eigenstates (see discussion in 
Sec.~\ref{sec5}). 

For certain other cases of Table \ref{table3} the exponents are 
clearly below 1, e.g. $b\approx 0.7$ or $b\approx 0.8$ indicating 
a kind of modest fractal structure of the eigenstates in a similar way 
as for the Hubbard case with $U=7.8$ and $E\approx -2.8$. 

Furthermore, both the figure labels of Fig. \ref{fig12} 
and also Table \ref{table3} provide the approximate energy values for 
the narrow energy delocalization band and in many cases these energy values 
also lie inside the interval $|E|<6$ of non-interacting product states 
with both particles localized far away, confirming that the strong 
delocalization effect may happen for both cases $|E|<6$ and $|E|>6$.

\section{Momentum and energy representation of eigenstates} 
\label{sec7}

It is illustrative to present the FIKS eigenstates which are 
delocalized along the 
diagonal $x_1\approx x_2$ in other representations such as a momentum 
representation using discrete Fourier transform or in the 
energy representation in terms of non-interacting product one-particle 
eigenstates, a representation already used for the algorithm of the 
Green function Arnoldi method described in Sec. \ref{sec3} 
and Appendix \ref{appendix2}. 

We first write a two-particle eigenstate with wave function 
$\psi(x_1,x_2)$ for $x_1,\,x_2\in\{0,\,\ldots,\,N-1\}$ 
in momentum representation by discrete Fourier transform:
\begin{equation}
\label{eq_FT1}
\bar\psi(p_1,p_2)=\frac1N
\sum_{x_1,x_2}\,\exp(i\,k_{p_1}\,x_1+i\,k_{p_2}\,x_2)\,
\psi(x_1,x_2)
\end{equation}
with $k_{p_j}=2\pi p_j/N$ for $p_j=0,\,\ldots,\,N-1$ and $j=1,2$. 
The momentum eigenfunction (\ref{eq_FT1}) can be efficiently 
evaluated using Fast Fourier Transform using the library fftw3 \cite{fftw}
which also works very well with optimal complexity 
${\cal O}(N^2\,\log(N))$ (for a two-dimensional discrete Fourier transform)
for arbitrary values of $N$, even for prime numbers and not only 
for powers of two. However, it turns out that the density plot of 
the momentum eigenfunction (\ref{eq_FT1}) 
has typically a quite complicated or bizarre 
structure and does not reveal much useful insight in the delocalization 
effect visible in position representation. Actually, the momentum 
representation with the simple ordering of momenta $k_p$ with 
$p=0,\,\ldots,\,N-1$ is not appropriate to study the quasiperiodic 
potential $V_1(x)=\lambda\,\cos(\alpha x+\beta)$. 

To understand this more clearly let us revisit the eigenvalue equation 
of an eigenfunction $\phi(x)$ with eigenvalue $\epsilon$ for 
the one-particle Hamiltonian with this quasiperiodic potential:
\begin{equation}
\label{eq_phi1_eigenvalue}
\epsilon\, \phi(x)=t[\phi(x+1)+\phi(x-1)]+\lambda\cos(\alpha x+\beta)\phi(x)
\end{equation}
where we have used a generalized hopping matrix element $t$ and 
where for simplicity $x$ may take arbitrary integer values for 
an infinite system and $\alpha/(2\pi)$ is an irrational number such as 
the golden ratio $\alpha/(2\pi)=(\sqrt{5}-1)/2$. In Ref. \cite{aubry} 
a duality transformation was introduced by expanding the 
eigenfunction in the form:
\begin{equation}
\label{eq_duality1}
\phi(x)=\sum_{p} 
\exp[i(\tilde\beta x+\alpha p\, x + \beta p)]\,\bar\phi(p)
\end{equation}
where the sum runs over all integer values of $p$, 
$\tilde\beta$ is some arbitrary parameter and for convenience 
we have taken out a phase factor $\exp(i\beta p)$ from the precise 
definition of $\bar\phi(p)$. This expansion defines unique 
coefficients $\bar\phi(p)$ only for irrational values of $\alpha$. 
Inserting (\ref{eq_duality1}) into (\ref{eq_phi1_eigenvalue}) one finds that 
the function $\bar\phi(p)$ obeys a similar eigenvalue equation of the form:
\begin{equation}
\label{eq_phi2_eigenvalue}
\tilde\epsilon\, \bar\phi(p)=
t[\bar\phi(p+1)+\bar\phi(p-1)]+\tilde\lambda\cos(\alpha p+\tilde \beta)
\bar\phi(p)
\end{equation}
with $\tilde\epsilon=2t\epsilon/\lambda$, $\tilde\lambda=4t^2/\lambda$ 
and $\tilde\beta$ is the parameter used in (\ref{eq_duality1}). 
For $|t|=1$ this transformation maps the case $\lambda>2$ to the 
case $\tilde\lambda=4/\lambda<2$. In Ref. \cite{aubry}, using this transformation 
together with Thouless formula (and some technical complications related 
to a finite size and rational approximation limit of $\alpha$), 
it was argued that for $\lambda>2$ the eigenfunctions $\phi(x)$ are 
localized with a localization length $\ell=1/\log(\lambda/2)$ and for 
the dual case (with $\tilde\lambda<2$) the functions $\bar\phi(p)$ 
are delocalized. 

The important lesson we can take from the duality transformation 
(\ref{eq_duality1}) is that it uses only a sum over discrete momentum values 
$q_p=(\tilde\beta+\alpha\,p){\rm \,mod\,}(2\pi)$, i.~e.
\begin{equation}
\label{eq_duality2}
\phi(x)=\sum_{p} 
\exp(iq_p\, x+i\beta p)\,\bar\phi(p), 
\end{equation}
instead of a 
continuous integration over $q\in[0,2\pi[$ which would normally be the proper 
way to perform a Fourier transform from the discrete infinite 
one-dimensional integer lattice space for $x$ to the continuous variable 
$q\in[0,2\pi[$. However, the quasiperiodic potential only couples (in the 
dual equation) momenta $q$ and $\tilde q$ such that 
$\tilde q=(q\pm\alpha){\rm \,mod\,}(2\pi)$ and therefore 
the discrete sum in (\ref{eq_duality1}) is sufficient. 
Furthermore, two momentum values obeying this relation have to be considered 
as ``neighbor'' values in dual space, i.e. the {\em 
natural proper ordering}
of momentum values is given by the discrete series 
$q_p=(\tilde\beta+\alpha\,p){\rm \,mod\,}(2\pi)$ with increasing 
integer values for $p$. 

Let us now consider the case of finite system size $N$ with periodic 
boundary conditions $\phi(0)=\phi(N)$ in (\ref{eq_phi1_eigenvalue}). If we 
want to construct a proper dual transformation for this case  we have to 
chose a 
rational value for $\alpha/(2\pi)=M/N$ where $0<M<N$ and the integer numbers 
$M$ and $N$ are relatively prime (if $M$ and $N$ are not relatively prime 
we would have a periodic potential with a non-trivial period being shorter 
than the system size requiring an analysis by Bloch theorem etc.). 
In this case 
we may directly use (\ref{eq_duality1}) to define the duality transformation 
provided that the sum is limited to the finite set $p=0,\,\ldots,\,N-1$ 
[and not infinite as for the case of infinite system size with irrational 
$\alpha/(2\pi)$]. Furthermore, for convenience
we chose the parameter $\tilde\beta=0$. 
Then the discrete momentum values $q_p$ become 
\begin{equation}
\label{eq_momentum}
q_p=(\alpha\,p){\rm \,mod\,}(2\pi)=
2\pi\frac{(pM){\rm \,mod\,}N}{N}=k_{\sigma(p)}
\end{equation}
where $k_p=2\pi p/N$ is the momentum value for the discrete 
Fourier Transform [see also below (\ref{eq_FT1})] and 
with $\sigma(p)=(pM){\rm \,mod\,}N$ being a permutation 
of the set $\{0,\,\ldots,\,N-1\}$ because $M$ and $N$ are relatively 
prime. We remind that for the eigenstate analysis in the previous Sections 
we had used the choice $M=f_{n-1}$ and $N=f_n$ where $f_n$ is the $n$-th 
Fibonacci number and we note that two subsequent Fibonacci numbers are 
indeed always relatively prime. For this particular choice we call the 
permutation $\sigma(p)$ the {\em golden permutation}. 
The permutation property of $\sigma(p)$ and Eq. (\ref{eq_momentum}) ensure 
that the discrete momentum values $q_p$ of the dual transformation 
(\ref{eq_duality1}) coincide {\em exactly} with the discrete momentum values 
used for the discrete Fourier Transform for a finite lattice of size $N$. 
However, there is a modified ordering between $q_p$ and $k_p$ because 
of the permutation and 
``neighbor'' momenta $k_p$ and $k_{p+1}$ of the discrete 
Fourier Transform are not neighbor values for the dual transformation and 
therefore the direct naive momentum representation 
(\ref{eq_FT1}) is not appropriate. The proper dual transformed 
representation corresponds to the golden permutation Fourier representation 
defined by 
\begin{eqnarray}
\label{eq_FT2}
\bar\psi_g(p_1,p_2)&=&
\bar\psi(\sigma(p_1),\sigma(p_2))\\
\nonumber
&=&\sum_{x_1,x_2}\,\exp(i\,q_{p_1}\,x_1+i\,q_{p_2}\,x_2)\,
\psi(x_1,x_2)
\end{eqnarray}
where the second identity with $q_p$ (instead of $k_p$) is valid 
due to (\ref{eq_momentum}). For $\bar\psi_g(p_1,p_2)$ 
neighbor values in $p_1$ or $p_2$ correspond indeed to neighbor 
values in the dual transformation. 

We mention that for a finite system size $N$ and an irrational choice of 
$\alpha/(2\pi)$ the momenta, $q_p=(\alpha\,p){\rm \,mod\,}(2\pi)$, used for 
the duality transformation do not coincide exactly with 
the discrete momenta of the discrete Fourier 
transform, in particular the quantity 
\begin{equation}
\label{eq_sigmatry}
\sigma(p)=\left(\frac{Np\alpha}{2\pi}\right){\rm \,mod\,}N
\end{equation}
would typically not be an integer number. At best one could try to define an 
approximate duality transformation with a modified permutation by 
rounding (\ref{eq_sigmatry}) to the next integer number but even in this case 
one would typically not obtain a permutation and it would be necessary to 
correct or modify certain $\sigma(p)$ values in order to avoid identical 
$\sigma(p)$ values for different integers $p$. 

If we want to choose a finite system size $N$ which is not a Fibonacci number 
we could try for the choice of 
$\alpha/(2\pi)$ a rational approximation $M/N$ of 
the golden ratio $(\sqrt{5}-1)/2$ with $M$ 
being the closest integer to $N(\sqrt{5}-1)/2$ and the denominator fixed 
by the given system size. However, in this case one might 
obtain a value of $M$ such that $M$ and $N$ are not relatively prime and 
(if we want to keep the same denominator) it would necessary to chose 
a different value of $M$ relatively prime to $N$ and still rather 
close to $N(\sqrt{5}-1)/2$ 
therefore reducing the quality of the rational approximation. For this 
reason we have in the preceding Sections mostly concentrated on the choice of 
Fibonacci numbers for the system size such that 
we can use the best rational approximation for the golden number 
and where we can always define in a simple and clear way the golden 
permutation by $\sigma(p)=(pf_{n-1}){\rm \,mod\,}f_n$. 

\begin{figure}
\begin{center}
\includegraphics[width=0.48\textwidth]{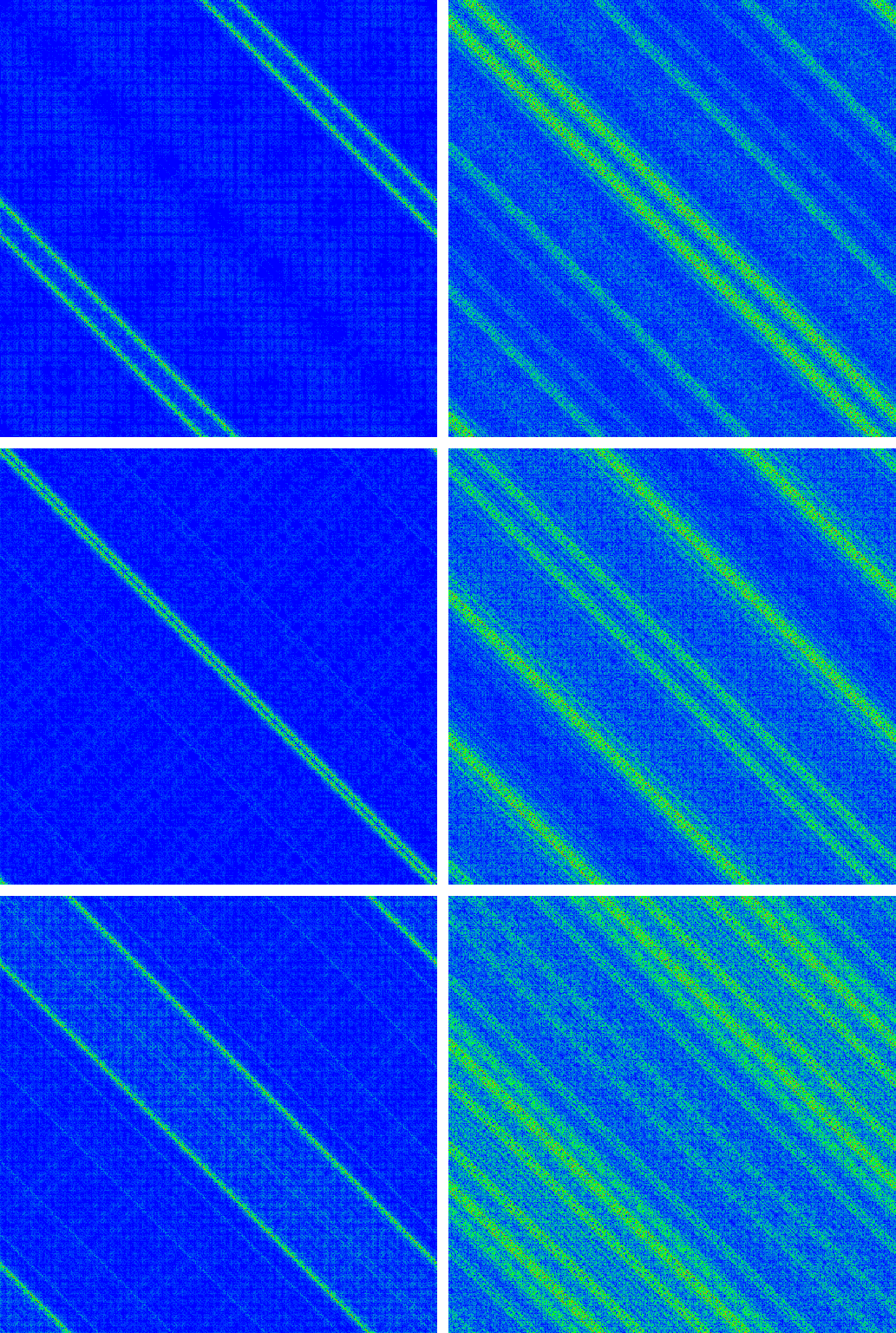}
\caption{Density plot of the three FIKS eigenstates in golden 
permutation Fourier representation with largest values 
of $\xi_E$ for $N=610$, $U_R=1$, $U=4.5$ (left 
column) or $U=7.8$ (right column). 
The corresponding energy eigenvalues and values for both types 
of inverse participation ratios are: 
{\em Top left:} $E=-3.09750$, $\xi_E=249.137$, $\xi_x=271.208$.
{\em Top right:} $E=-2.78586$, $\xi_E=211.058$, $\xi_x=194.241$.
{\em Center left:} $E=-3.09964$, $\xi_E=239.312$, $\xi_x=265.885$.
{\em Center right:} $E=-2.78599$, $\xi_E=200.958$, $\xi_x=176.454$.
{\em Bottom left:} $E=-3.09815$, $\xi_E=233.773$, $\xi_x=250.700$.
{\em Bottom right:} $E=-2.78593$, $\xi_E=190.171$, $\xi_x=193.885$.
}
\label{fig13}
\end{center}
\end{figure}

In Fig. \ref{fig13} the three eigenstates with 
maximum $\xi_E$ for $N=610$, $U_R=1$ and the two cases $U=4.5$ and 
$U=7.8$ (and $E\approx -2.8$) 
are shown in the golden permutation Fourier representation. 
One sees clearly that for the center of mass coordinate there is a 
strong momentum localization around a few typical values while 
for the relative coordinate all momentum values seem to contribute to the 
eigenstate leading to momentum delocalization in this direction. 
This is just dual to the typical behavior of such eigenstates 
in position representation with delocalization in the center of mass 
coordinate and localization in the relative coordinate. 
However, the precise detailed structure, in momentum space on a length 
scale of a few pixels and well inside the stripes seen in Fig. \ref{fig13}, 
is still quite complicated and subtle. 

The ``localization length'' in momentum space for the center of mass 
coordinate is considerably shorter for the case $U=4.5E$ with about 
10 pixels (i.~e. discrete momentum values) 
than for the other case $U=7.8$ (and $E\approx -2.8$) 
with about 30 pixels. This 
observation relates to the stronger quasiperiodic hole-peak structure 
in the eigenstates seen in Figs. \ref{fig4}-\ref{fig6} for the 
case $U=7.8$ (and $E\approx -2.8$). 

We have also tried for the irrational case and non-Fibonacci numbers 
for $N$ to define an approximate golden permutation which in principle 
provides similar figures as in Fig. \ref{fig13} 
but with a considerable amount of additional irregularities 
concerning the momentum structure etc.

\begin{figure}
\begin{center}
\includegraphics[width=0.48\textwidth]{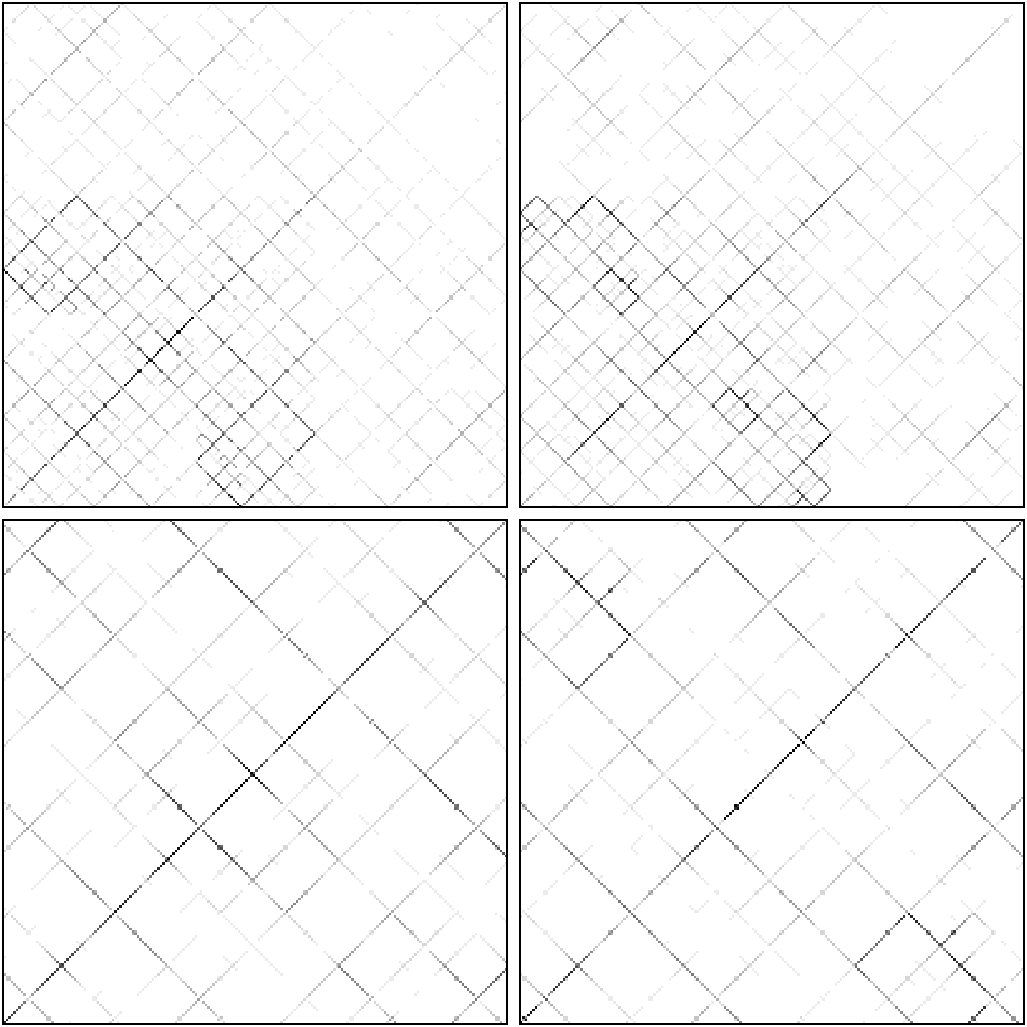}
\caption{Density plot of the FIKS eigenstates in non-interacting 
energy representation with the largest value 
of $\xi_E$ for $N=233$ (top panels) or $N=610$ (bottom panels), 
$U_R=1$, $U=4.5$ (left column) or $U=7.8$ (right column). 
Black represents maximum, grey medium and white minimum values of 
the expansion amplitudes of the shown eigenstate with respect to 
non-interacting energy product eigenstates $|\phi_\nu,\phi_\mu\!>$. 
The horizontal (vertical) axis corresponds to the index $\nu$ ($\mu$) 
ordered with respect 
to increasing values of the corresponding one-particle energy $\epsilon_\nu$ 
($\epsilon_\mu$) of the first (second) particle. 
Top panels for $N=233$ correspond to full eigenstates and 
bottom panels for $N=610$ correspond to a zoomed region of size 
$200\times 200$ with left bottom corner at position 
$x_1=x_2=100$. 
The corresponding energy eigenvalues and values for both types 
of inverse participation ratios are: 
{\em Top left:} $E=-3.09669$, $\xi_E=107.409$, $\xi_x=106.818$.
{\em Top right:} $E=-2.78569$, $\xi_E=117.697$, $\xi_x=102.577$.
{\em Bottom left:} $E=-3.09750$, $\xi_E=249.137$, $\xi_x=271.208$.
{\em Bottom right:} $E=-2.78586$, $\xi_E=211.058$, $\xi_x=194.241$.
}
\label{fig14}
\end{center}
\end{figure}

Another type of interesting eigenvector representation is obtained by an 
expansion of a two-particle eigenstate in the basis $|\phi_\nu,\phi_\mu\!>$ 
of non-interacting one-particle product eigenstates. Fig. \ref{fig14} 
shows black and white density plots for the amplitudes of certain 
eigenstates in such a representation for the two sizes $N=233$ and 
$N=610$ and the two values of the interaction $U=4.5$ and $U=7.8$ (both 
for $U_R=1$). Both axis correspond to the one-particle index ordered with 
respect to increasing values of the corresponding one-particle energy. 
We remind that in the second variant of the Green function Arnoldi method 
the main calculations are actually done in this energy representation, 
which is therefore more easily accessible than the standard position 
representation. 

One observes a kind of self-similar structure with (approximate) golden 
ratio rectangles of different sizes along the diagonals. 
The inverse participation ratio $\xi_E$ in energy representation corresponds 
approximately to the number of black dots in the black and white density plots 
of Fig. \ref{fig14}. 

We mention that when the one-particle eigenstate ordering in the energy 
representation is done with respect to the maximum positions of the 
one-particle eigenstates (instead of the one-particle energy) one obtains 
a clear banded structure with main values/peaks for $\nu\approx \mu\pm 5$
(Figure not shown). 

\section{Implications for cold atom experiments}
\label{sec8}

Motivated by recent experiments on cold atoms \cite{bloch} we present 
also some results for a modified value of the flux parameter $\alpha$ 
used in the quasiperiodic potential $V_1(x)$. 
In the experiment of Ref. \cite{bloch} the rational value for 
$\alpha/(2\pi)\approx 532/738=266/369$ was used. 
This value has the finite continued fraction expansion 
$[0;1,2,1,1,2,1,1,8]$ with 
\begin{equation}
\label{cf_expansion}
[a_0;a_1,a_2,a_3,\ldots]=
a_0+\cfrac{1}{a_1+\cfrac{1}{a_2+\cfrac{1}{a_3+\cdots}}}.
\end{equation}
We define two numbers $\alpha_j$, $j=1,2$ such that 
$\alpha_j/(2\pi)$ is irrational and close to the experimental rational value by
\begin{equation}
\label{eq_alpha1}
\frac{\alpha_1}{2\pi}=[0;1,2,1,\ldots]=\frac{\sqrt{10}-1}{3}
=0.7207592200561264\ldots
\end{equation}
and
\begin{eqnarray}
\label{eq_alpha2}
\frac{\alpha_2}{2\pi}&=&[0;1,2,1,1,2,1,1,8,\ldots]\\
\nonumber
&=&\frac{\sqrt{39999}-169}{43}
=0.7208720926598791\ldots
\end{eqnarray}
where the initial pattern of shown coefficients in the continued fraction 
expansion (except the leading zero) 
repeats indefinitely with a period of 3 (or 8) for the case of 
$\alpha_1$ (or $\alpha_2$). The first choice provides a ``stronger'' 
irrational number for $\alpha_1/(2\pi)$ while the second choice is 
closer to the experimental value. In this Section we choose for all 
numerical computations one of these two values (or rational approximations of 
them for the eigenvector calculations) and furthermore we fix the 
phase offset and the interaction range by $\beta=(\sqrt{5}-1)/2$ and $U_R=1$. 

\begin{figure}
\begin{center}
\includegraphics[width=0.48\textwidth]{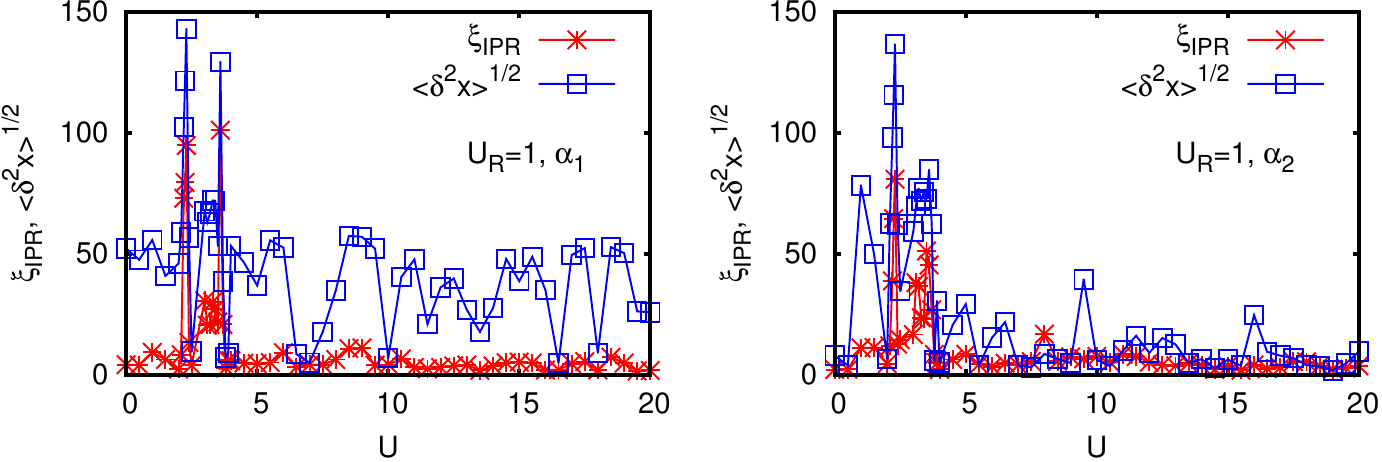}
\caption{Inverse participation ratio $\xi_{\rm IPR}$ and 
variance length 
$\langle\delta^2x\rangle^{1/2}=\langle (x-x_0)^2\rangle^{1/2}$ 
of the time evolution two-particle state for system size $N=512$, 
iteration time $t=5120$ and $\alpha_1$ (left panel) or $\alpha_2$ 
(right panel) versus interaction strength $U$. 
The initial state at $t=0$ is localized with 
both particles in the center position $x_0=N/2$.
Inverse participation ratio and variance  length 
have been calculated from an effective one-particle density 
without a center box of size 20\% (with respect to system size). 
The data points for $2<U<2.5$ have been calculated with a doubled 
iteration time $t=10240$. The values of $\alpha_1, \alpha_2$ from
(\ref{eq_alpha1}), (\ref{eq_alpha2})
correspond to experimental conditions of \cite{bloch}.
}
\label{fig15}
\end{center}
\end{figure}

First, we performed the time-evolution analysis already described in 
Sec.~\ref{sec4} using either $\alpha_1$ or $\alpha_2$. 
Fig.~\ref{fig15} shows the dependence 
of the inverse participation ratio $\xi_{\rm IPR}$ 
and the variance length  [both computed without the 
20\% center box, see  (\ref{eq_ipr_def}) and (\ref{eq_var_len})] 
on the interaction strength $U$ 
($0\le U\le 20$) for a system size 
$N=512$ and an iteration time $t=5120$. As in Sec. \ref{sec4} we chose 
for $t=0$ an initial state with both particles localized at the 
center point $x_0=N/2$. 
For both $\alpha$ values we observe strong peaks for both 
length scales at values $U=2.25$-$2.3$ and $U=3.6$ indicating the 
possible existence of FIKS states at these interaction values (or very close).
A closer inspection reveals that the first peak close to $U=2.25$ requires 
a longer iteration time $t=10240$ in order to provide saturation of
the two length scales and therefore in Fig.~\ref{fig15} 
the data points for $2<U<2.5$ are computed with 
this increased iteration time. 

\begin{table}
\caption{Time evolution parameters for the interaction values $U=2.25$ 
and $U=3.6$ and both values of $\alpha$ using the data sets of 
Fig.~\ref{fig15}. These interaction values correspond to the local 
maxima of the squared tail norm $\|\psi_{\rm tail}(t)\|^2$. 
Note that for $U$ close to $2.25$ the local maxima visible in Fig.~\ref{fig15} 
of the length scale $\xi_{\rm IPR}$ (computed without the 20\% center box) 
correspond actually to $U=2.3$ with slightly larger values than for $U=2.25$.}
\begin{center}
\begin{tabular}{|r|l|r|r|r|r|r|}
\hline
$\alpha$ & $U$ & $t$ & $\xi_{\rm IPR}$ & 
$\langle H\rangle$ & $\delta^2E$ & $\|\psi_{\rm tail}(t)\|^{2^{\phantom 1}}$ \\
\hline
\hline
$\alpha_1$ & 2.25 & 5120 & 32.27 & -4.744 & 0.475 & 0.00884 \\ 
\hline
$\alpha_1$ & 2.25 & 10240 & 79.48 & -4.828 & 0.127 & 0.107 \\ 
\hline
$\alpha_1$ & 3.6 & 5120 & 101.12 & -0.893 & 0.159 & 0.0449 \\ 
\hline
$\alpha_2$ & 2.25 & 5120 & 30.18 & -4.717 & 0.587 & 0.00562 \\ 
\hline
$\alpha_2$ & 2.25 & 10240 & 64.58 & -4.826 & 0.140 & 0.0709 \\ 
\hline
$\alpha_2$ & 3.6 & 5120 & 45.45 & -0.878 & 0.215 & 0.0188 \\ 
\hline
\end{tabular}
\label{table4}
\end{center}
\end{table}

Table~\ref{table4} summarizes the results of the quantities $\xi_{\rm IPR}$, 
$\langle H\rangle$, $\delta^2E$ and $\|\psi_{\rm tail}(t)\|^2$ (see 
Sec.~\ref{sec4} for the precise definition of them) at 
the two peak values $U=2.25$ and $U=3.6$. 
The values of $\xi_{\rm IPR}$ in Table~\ref{table4} for $U=2.25$ and 
$t=10240$ do actually 
not exactly correspond to the first local maximum 
visible in Fig.~\ref{fig15} 
because $\xi_{\rm IPR}$ is maximal at $U=2.3$ while the value of 
$U=2.25$ corresponds to the local maximum of $\|\psi_{\rm tail}(t)\|^2$. 
However, 
detailed eigenvector calculation for these two interaction values confirm 
that globally the value $U=2.25$ is slightly more optimal than $U=2.3$ with 
stronger delocalization. 

In Table~\ref{table4} we provide for the case $U=2.25$ also the results 
for the two iteration times $t=5120$ and $t=10240$. Obviously, 
$\xi_{\rm IPR}$ and $\|\psi_{\rm tail}(t)\|^2$ are considerably increased 
at $t=10240$ but already at $t=5120$ the strong delocalization FIKS effect is 
visible. The average energy value of the tail state is rather sharp with 
a modest variance $\delta^2E$ for all cases, 
but also with an additional significant decrease 
of $\delta^2E$ between $t=5120$ and $t=10240$ (for $U=2.25$). 

\begin{figure}
\begin{center}
\includegraphics[width=0.48\textwidth]{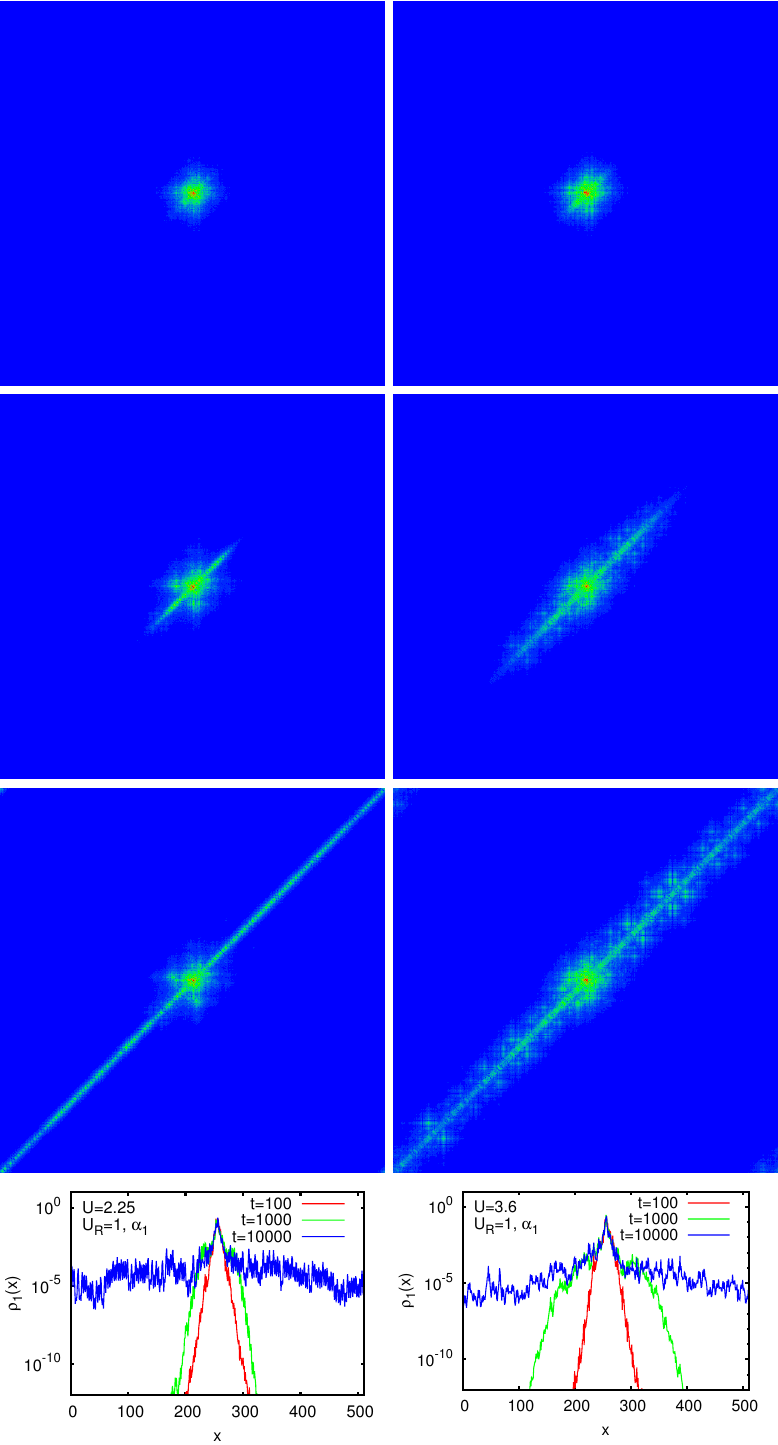}
\caption{Density plot (three top rows of panels) 
of time evolution two-particle states for system size $N=512$, the 
case $\alpha_1$, interaction range $U_R=1$, 
interaction strength $U=2.25$ (left column) or $U=3.6$ (right column), 
iteration times $t=100$ (first row), $t=1000$ (second row) 
and $t=10000$ (third row); panels show the whole system
range $(0 \leq x_, x_2 < 512)$. The fourth row of panels 
shows the one-particle density 
$\rho_1(x)$ in a semi-logarithmic representation for the same 
states as in the three top rows of panels. 
The initial state at $t=0$ is localized with 
both particles in the center position $x_0=N/2$.
}
\label{fig16}
\end{center}
\end{figure}

Globally Fig.~\ref{fig15} and Table~\ref{table4} show that the FIKS 
effect is stronger for $U=2.25$ but at this value 
it requires a longer iteration time to be clearly visible. This observation 
is also confirmed by Fig.~\ref{fig16} which shows 
for $\alpha_1$ and both interaction values $U=2.25$ and $U=3.6$ 
the density plots and the one-particle density of three time evolution 
states at $t=100$, $t=1000$ and $t=10000$. In both cases the state is 
clearly localized at the beginning at $t=100$ and it is delocalized over 
the full system size at $t=10000$ (with a small weight and along the diagonal 
$x_1\approx x_2$ as discussed in Sec.~\ref{sec4}). 
However, for the intermediate time $t=1000$ the state for $U=2.25$ is 
considerably less delocalized than the state for $U=3.6$ at the same 
iteration time clearly confirming the slower delocalization speed for 
$U=2.25$. Thus the velocity of FIKS pairs is smaller at $U=2.25$
than at $U=3.6$ but the weight of FIKS pairs in the initial state is larger
at $U=2.25$. 
Apart from this the delocalized tails of the state at $t=10000$ 
appear somewhat ``stronger'' or ``thicker'' for $U=2.25$ explaining the 
larger values of $\xi_{\rm IPR}$ (for $\alpha_1$). 
Note that Fig.~\ref{fig16} shows the full time evolution states while 
Fig.~\ref{fig2} in Sec.~\ref{sec4}, with the golden ratio value 
for $\alpha/(2\pi$), 
shows only a zoomed range for the right delocalized branche between the right 
border of the 20\% center box and the right border of the full system.

\begin{figure}
\begin{center}
\includegraphics[width=0.48\textwidth]{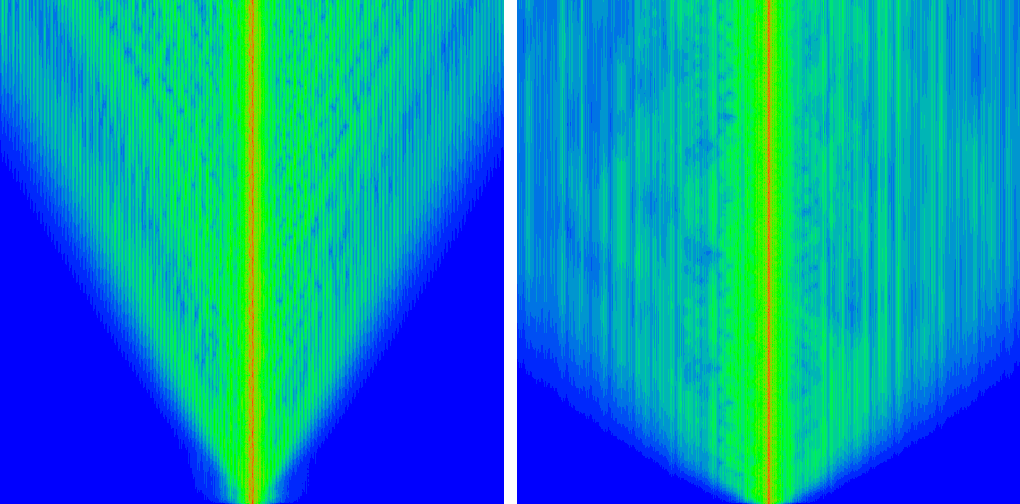}
\caption{Density plot for the time dependence of one-particle density from 
the time evolution state with $x$-position ($0\le x<512$) corresponding to 
the horizontal axis and time $t$ ($0\le t \le 10240$) corresponding to 
the vertical axis. Here $U=2.25$ (left panel),  
$U=3.6$ (right panel) and $\alpha=\alpha_1$, $U_R=1$.
}
\label{fig17}
\end{center}
\end{figure}

Fig.~\ref{fig17} shows the time evolution of the one-particle density 
(for $\alpha_1$) 
with the $x$-dependence corresponding to the horizontal axis and with 
the $t$-dependence ($0\le t\le 10240$) corresponding to the vertical axis. 
This figure provides clear and additional confirmation that the delocalization 
effect is stronger and slower for $U=2.25$ than for $U=3.6$.
It also confirms the linear (ballistic) increase of the delocalized part 
of the state with time (see also Fig.~\ref{fig3}). We mention that 
the other value $\alpha_2$ provides very similar figures as Figs.~\ref{fig15} 
and \ref{fig16} with a slightly reduced delocalization effect for both 
interaction values. 

\begin{figure}
\begin{center}
\includegraphics[width=0.48\textwidth]{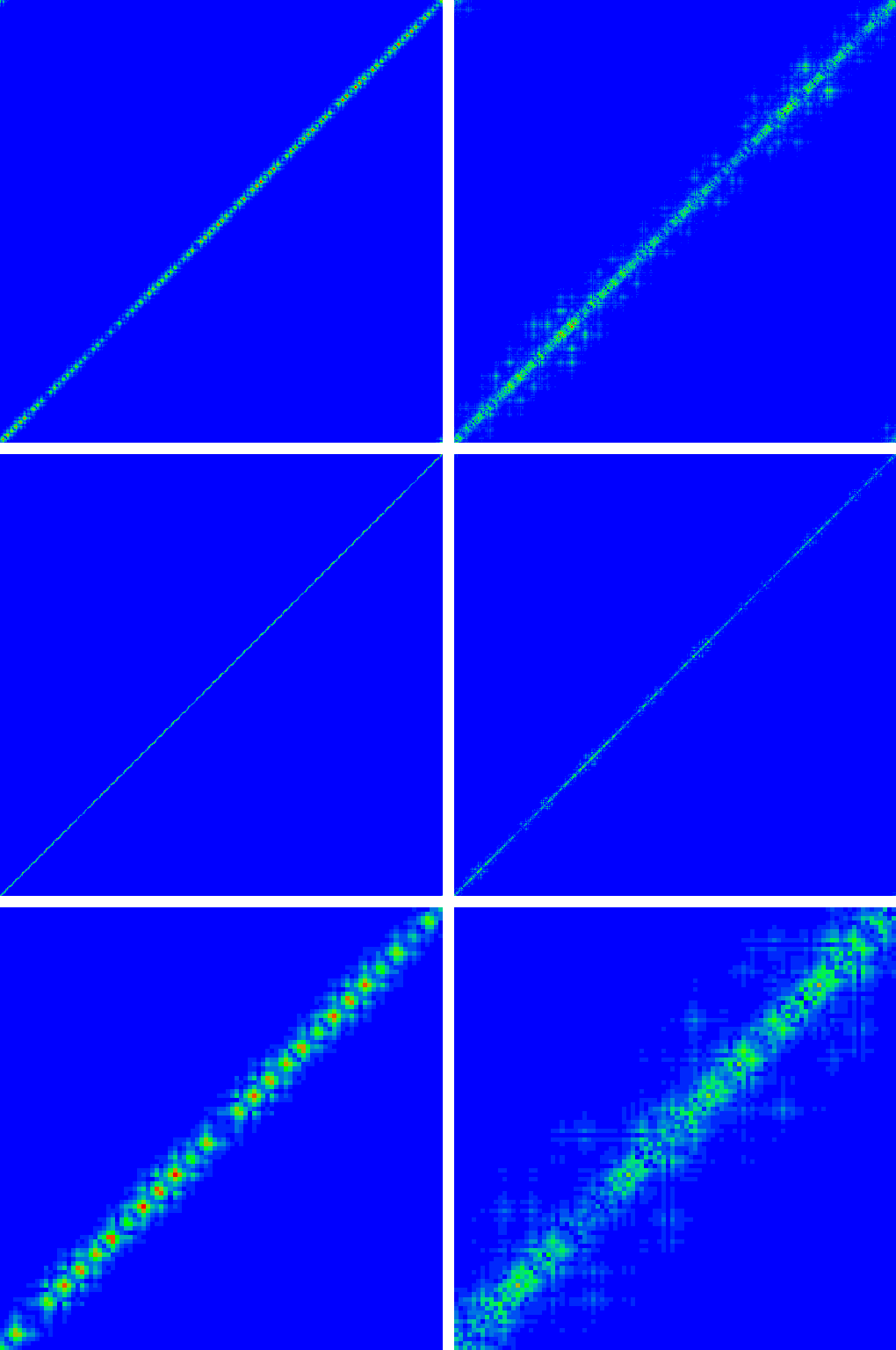}
\caption{Density plot of FIKS eigenstates 
for  rational approximations of $\alpha_2/(2\pi)$ and
$U=2.25$ (left column), $U=3.6$ (right column), 
$N=369$ (top panels), $N=1605$ (center and bottom panels);
$U_R=1$. 
The corresponding energy eigenvalues and values for both types 
of inverse participation ratios are: 
{\em Top left:} $E=-4.85051$, $\xi_E=98.462$, $\xi_x=118.308$.
{\em Top right:} $E=-0.92196$, $\xi_E=113.232$, $\xi_x=108.389$.
{\em Center left:} $E=-4.84994$, $\xi_E=428.375$, $\xi_x=566.237$.
{\em Center right:} $E=-0.92198$, $\xi_E=309.040$, $\xi_x=260.125$.
Bottom panels show a zoomed region of size $100\times 100$ 
with left bottom corner at position $x_1=x_2=350$  
of the center panels. 
}
\label{fig18}
\end{center}
\end{figure}

Following the procedure described in the beginning of Sec.~\ref{sec5} 
we have also computed eigenstates using the Arnoldi Green function method 
with the average energy values $\langle H\rangle$ of Table~\ref{table4} 
as initial Green's function energy for the smallest system size. 
The Green function energies are refined for larger system
sizes using the energy eigenvalue of a well delocalized eigenstate of the 
last smaller system size. Following the spirit of the previous explications 
[see text between Eqs. (\ref{eq_duality2}) 
and (\ref{eq_momentum})]
we choose rational approximations of $\alpha_1/(2\pi)$ 
and $\alpha_2/(2\pi)$ using 
their continued fraction expansions (\ref{eq_alpha1}) and (\ref{eq_alpha2}) 
which provide suitable system sizes given as the denomators of the 
rational approximations. Using a minimal (maximal) system system size 
$\sim 40$ ($\sim 10000$) this provides for $\alpha_1$ the values 
$N=43$, 111, 154, 265, 684, 949, 1633, 4215, 5848, 10063
and for $\alpha_2$ the values $N=43$, 369, 412, 1193, 1605, 
2798, 7201, 9999. Note that the system size $369$ corresponds to the 
rational approximation $\alpha_2/(2\pi)\approx 266/369$ used in the 
experiments of Ref. \cite{bloch}. For each system size we use the 
corresponding rational approximation of $\alpha_j/(2\pi)$ ($j=1,\,2$) 
and $\beta=(\sqrt{5}-1)/2)$ to determine numerically certain eigenstates 
by the Green function Arnoldi method. 

In Fig.~\ref{fig18} we show selected strongly delocalized eigenstates 
for $\alpha_2$ and the two interaction values $U=2.25$ and $U=3.6$ and the 
system sizes $N=369$ and $N=1605$. All eigenstates provide 
nice FIKS pairs with a quite specific particular pattern on 
the diagonal $x_1\approx x_2$ which correponds, for each of the two 
interaction values, rather well to the pattern 
of (the delocalized tails of) the time evolution states for $t=10000$ 
visible in Fig.~\ref{fig16}. For $N=1605$ the pattern for $U=2.25$ 
seems be to considerably more compact than the pattern for $U=3.6$ 
which is also confirmed by a considerably larger value of $\xi_x$. 
The eigenstates for the 
$\alpha_1$ case are very similar for comparable system sizes. 

\begin{figure}
\begin{center}
\includegraphics[width=0.48\textwidth]{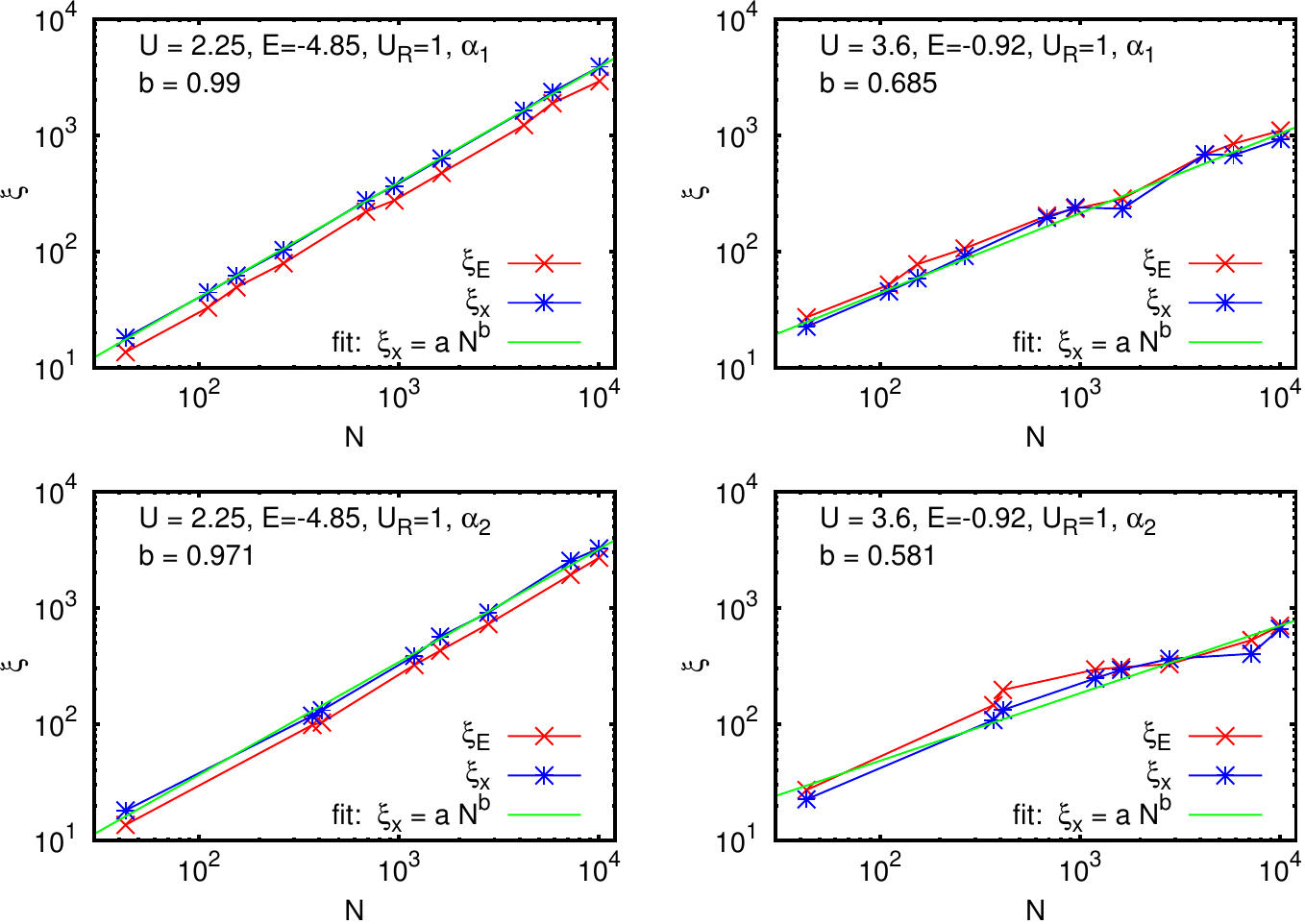}
\caption{Largest inverse participation ratio (for given values of $N$
and approximate energy) of FIKS eigenstates versus system size $N$ in a 
double logarithmic scale for $U_R=1$, for rational approximations 
of $\alpha_1/(2\pi)$ (top panels) or $\alpha_2/(2\pi)$ (bottom panels) 
and for $U=2.25$, $E\approx -4.85$ (left column) or $U=3.36$, 
$E\approx -0.92$ (right column). 
The used system sizes are 43, 111, 154, 265, 684, 949, 1633, 4215, 5848, 
10063 given by the denominators of the rational approximations of 
$\alpha_1/(2\pi)$ and 43, 369, 412, 1193, 1605, 
2798, 7201, 9999 for the rational approximations for 
$\alpha_2/(2\pi)$. The blue line with stars 
corresponds to the inverse participation ratio 
$\xi_x$ in position representation, the red line 
with crosses to the inverse participation ratio 
$\xi_E$ in energy representation and 
the green line to the power law fit $\xi_x=a\,N^b$ with 
fit values given in Table \ref{table5}. 
Note that for given values of $N$ and approximate energy 
the eigenstates with maximal $\xi_x$ and 
maximal $\xi_E$ may be different. 
}
\label{fig19}
\end{center}
\end{figure}

\begin{table}
\caption{Results of the power law fit $\xi_x=a\,N^b$ 
for the four cases of Fig. \ref{fig19}.}
\begin{center}
\begin{tabular}{|r|r|r|r|l|l|}
\hline
$U$ & $U_R$ & $\alpha$ & $E$ & $a$ & $b$ \\
\hline
\hline
$2.25$ & $ 1$ & $\alpha_1$ & $ -4.85$  & $0.424 \pm 0.013$ & $0.990 \pm 0.004$  \\
\hline
$ 3.6$ & $ 1$ & $\alpha_1$ & $ -0.92$  & $1.878 \pm 0.348$ & $0.685 \pm 0.027$  \\
\hline
$2.25$ & $ 1$ & $\alpha_2$ & $ -4.85$  & $0.418 \pm 0.055$ & $0.971 \pm 0.018$  \\
\hline
$ 3.6$ & $ 1$ & $\alpha_2$ & $ -0.92$  & $3.333 \pm 1.197$ & $0.581 \pm 0.050$  \\
\hline
\end{tabular}
\label{table5}
\end{center}
\end{table}

Fig.~\ref{fig19} shows the size dependence of $\xi_x$ and $\xi_E$ 
for the four cases corresponding to any combination of the two interaction 
and the two flux values. The fit results of the power law fit $\xi_x=a\,N^b$ 
are shown in Table~\ref{table5}. For $U=2.25$ both fits for the two flux 
values are very accurate with exponents $b\approx 1$. For $U=3.6$ the fit 
quality is somewhat reduced and the exponents are quite smaller 
$b\approx 0.7$ for $\alpha_1$ and $b\approx 0.6$ for $\alpha_2$ indicating 
a certain fractal structure of eigenstates. At $N\approx 10000$ the 
maximal values of $\xi_x$ for $U=2.25$ at both flux values are at least 
four times larger than the maximal values of $\xi_x$ for $U=3.6$. 
We also observe that the density of good FIKS pairs for $U=2.25$ 
and both flux values is extremely high. In Secs. \ref{sec5} for the rational 
approximation of the golden ratio for $\alpha/(2\pi)$ only 
the case for $U=7.2$ has a comparable density of good FIKS pairs 
(see bottom panels of Fig.~\ref{fig8}). 

\begin{figure}
\begin{center}
\includegraphics[width=0.48\textwidth]{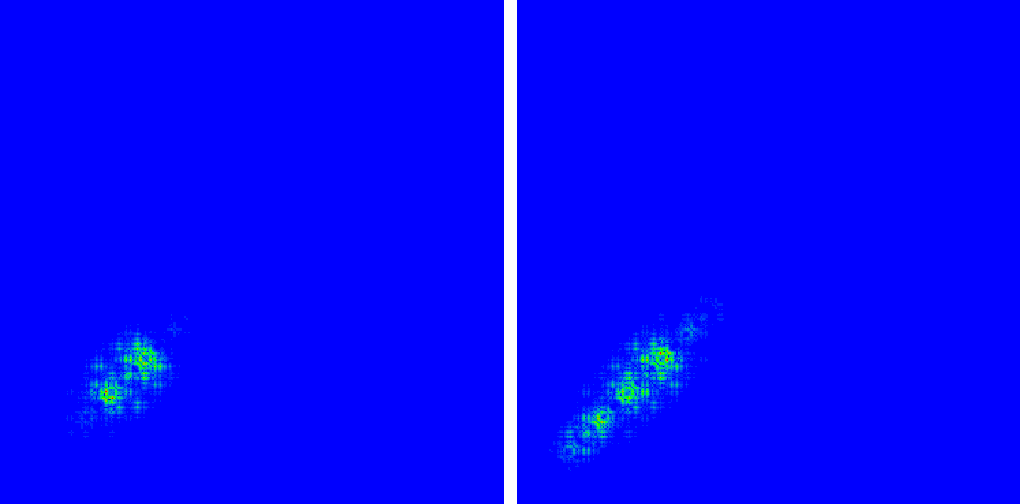}
\caption{Density plot of two selected eigenstates 
for $U=4.5$, $U_R=1$ and for the rational approximations of $\alpha_2/(2\pi)$. 
The corresponding system sizes, energy eigenvalues and values for both types 
of inverse participation ratios are: 
{\em Left:} $N=369$, $E=-2.21758$, $\xi_E=11.378$, $\xi_x=21.029$ 
(2nd largest value of $\xi_E$ and largest value of $\xi_x$ 
for this system size and approximate energy). 
{\em Right:} $N=1605$, $E=-2.21949$, $\xi_E=16.367$, $\xi_x=26.748$ 
(largest value of $\xi_E$ for this system size and approximate energy). 
The left panel shows the full state of size $369\times 369$ 
and the right panel a zoomed region of size $369\times 369$ 
with left bottom corner at position $x_1=x_2=0$ and outside the zoomed 
range no data points different from blue (for zero amplitude) are visible. 
}
\label{fig20}
\end{center}
\end{figure}

We have also tested (for $\alpha_2$) the interaction strength $U=4.5$ 
with approximate energy $E=-3.1$ 
which provided nice FIKS pairs for the golden ratio case 
studied in Sec.~\ref{sec5}. However, here we should not expect 
delocalized FIKS pairs 
since according to Fig.~\ref{fig15} the value of $\xi_{\rm IPR}$ obtained 
from the time evolution state is very small. On the other side, the variance 
length shows some modestly increased values and 
it might be useful 
to verify such cases as well. We applied the standard procedure of 
energy refinement with the  Green function Arnoldi method 
on $U=4.5$ with the initial energy $E=-3.1$ which immediatedly 
selected $E\approx -2.2$ as ``optimal'' energy range (to maximize $\xi_E$). 
Despite some modestly delocalized eigenstates with $\xi_E\sim 15$ and 
$\xi_x\sim 25$ (for the largest considered systems sizes $N=412$, 1193 and 
1605) there are no FIKS pairs with strong delocalization along the diagonal. 
Fig.~\ref{fig20} shows for $\alpha_2$ and $N=369$ or $N=1605$ two such 
modestly delocalized eigenstates which have some ``cigar'' form but 
with a rather short length $\sim 50$-$80$ and a rather elevated width 
$\sim 20$-$30$. 
It seems that the variance length, in contrast to 
$\xi_{\rm IPR}$, does not really allow to 
distinguish between these kind of states and nice FIKS eigenstates. 
Furthermore this example shows that suitable parameters $U$ and $E$ for 
FIKS states depend strongly on the flux parameter $\alpha$, an issue which 
is more systematically studied in the next Section.

\section{Dependence on flux values}
\label{sec9}

A problem with a systematic study of the dependence of the FIKS effect on
different flux values is to select a suitable set of irrational numbers 
of comparable quality and which have roughly the same distance. 
For this we consider at first rational numbers $p/89$ with $44\le p \le 88$ 
where the denominator $89$ has the nice feature of being both a prime 
and a Fibonacci number. We compute for each of these rational numbers 
the canonical variant of its finite continued fraction expansion 
\cite{cf_rational}, reduce the last coefficient by 1 and add an 
infinite sequence of entries of 1. This provides the infinite continued 
fraction expansion of an irrational number which is rather close to the 
initial fraction $p/89$ and which has ``a golden tail'' for the continued 
fraction expansion. It turns out that 
for each value of $p$ the difference between $p/89$ and the corresponding 
irrational number is approximately $5\times 10^{-5}$ therefore 
providing a nice data set of irrational numbers between 0.5 and 1.

In particular for $p=55$, where $55/89$ is a rational 
aproximation of the golden number, we have $55/89=[0;1,1,1,1,1,1,1,1,2]$.
The procedure reduces the last coefficient from 2 to 1 and adds the 
infinite sequence of unit entries just providing exactly the continued 
fraction expansion of the golden number (with all coefficients being unity). 
The golden number is therefore 
one of the data points in the selected set of irrational numbers. 
For $p=64$ we find the irrational value $0.7191011235955056\ldots$ 
which is by construction very close to $64/89$ but also rather close to 
$266/369\approx 0.72087$, which was used in the experiment of 
Ref.~\cite{bloch}, and also to the two irrational numbers 
(\ref{eq_alpha1}) and (\ref{eq_alpha2}) used in the previous Section. 

\begin{figure}
\begin{center}
\includegraphics[width=0.48\textwidth]{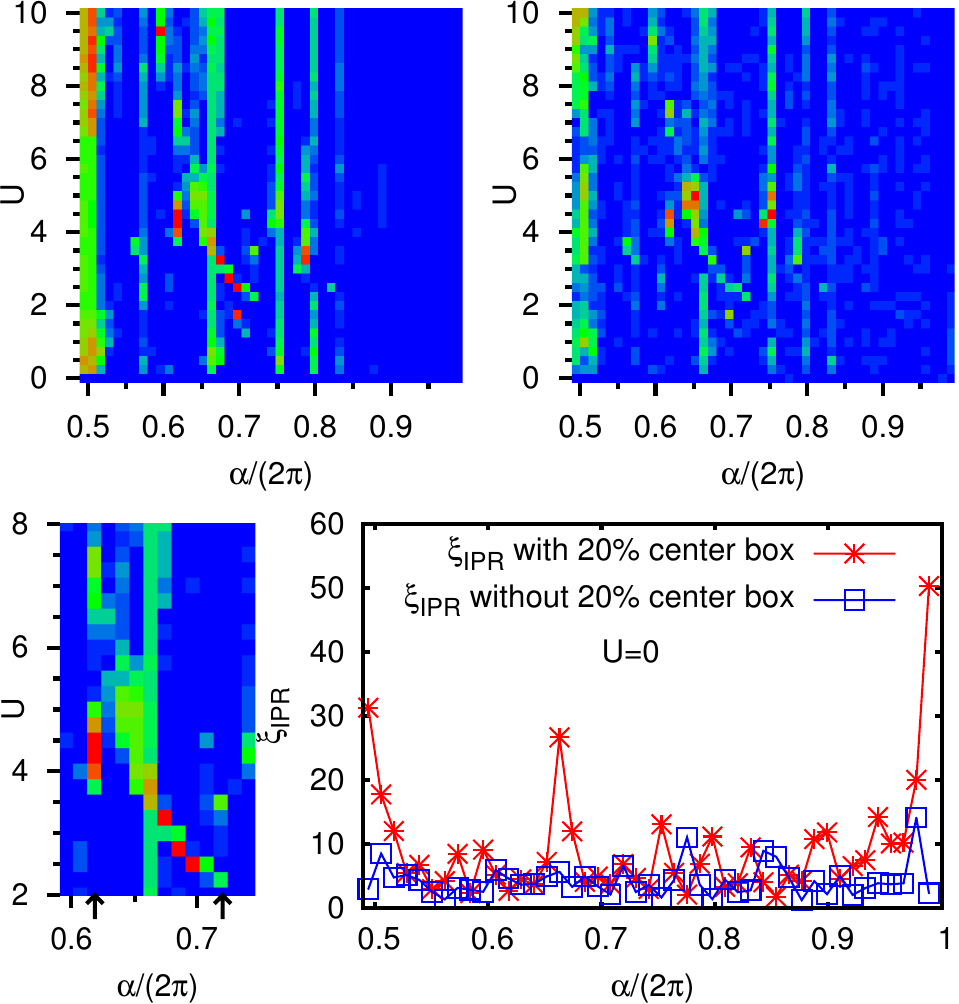}
\caption{{\em Top panels~:} Density plot of the 
squared tail norm $\|\psi_{\rm tail}(t)\|^2$ (left panel) or 
the inverse participation ratio computed without 20\% center box 
(right panel) with horizontal axis representing the parameter 
$\alpha/(2\pi)$ and vertical axis representing the interaction strength $U$ 
using a time evolution state for system size $N=512$, 
iteration time $t=5120$, interaction range $U_R=1$ and a localized 
initial state for $t=0$ with both particles in the center position $x_0=N/2$. 
The bottom left panel shows a zoomed range with $0.6\le \alpha/(2\pi)<0.75$ 
and $2\le U\le 8$ of the top left panel. 
The two arrows indicate the value of the golden 
ratio $\alpha/(2\pi)=(\sqrt{5}-1)/2\approx 0.618$ 
and the value $\alpha/(2\pi)=266/369\approx 0.721$ used in the experiments 
of Ref. \cite{bloch}. 
The bottom right panel shows the inverse participation ratio for $U=0$ 
versus the parameter $\alpha/(2\pi)$ and computed 
with (red crosses) and without (blue squares) the 
20\% center box.
The maximal value is for the squared tail norm $0.12519$ (for 
$\alpha\approx 0.596$ and $U=9.5$) and for the the inverse participation 
ratio without 20\% center box $188.68$ (for 
$\alpha\approx 0.753$ and $U=4.5$). 
The data of this figure are obtained with 
$\beta=(\sqrt{5}-1)/2$ (and not $\beta=0$ as the data of Fig.~\ref{fig1} 
and Table~\ref{table1} in Sec.~\ref{sec4}). 
}
\label{fig21}
\end{center}
\end{figure}

Using these irrationals values for $\alpha/(2\pi)$ and 
$\beta=(\sqrt{5}-1)/2$ we have performed the time evolution analysis 
described in Sec.~\ref{sec4} for system size $N=512$, iteration time 
$t=5120$ and the interaction interval $0\le U\le 10$ 
in steps of $\Delta U=0.25$ providing in total $45\times 21$ data sets. 
The main results of this analysis 
are shown in Fig.~\ref{fig21} containing two density plots 
in $\alpha/(2\pi)$-$U$ plane for the squared tailed norm 
$\|\psi_{\rm tail}(t)\|^2$ and the inverse participation ratio 
$\xi_{\rm IPR}$ (without 20\% center box) both providing the most 
reliable measure of delocalization in the 
framework of the time evolution analysis 
(the variance length provides a considerable amount 
of fluctuation peaks also when the other two quantities are 
very small as can be seen in Figs.~\ref{fig1} and \ref{fig15}).

Concerning the density plots of a quantity $\rho$ we mention that we 
apply the attribution of the different color codes to uniform slices of 
$\rho^{r}$ with $r\le 1$ being some exponent, of typical choice 
$1/4$ or sometimes $1/8$,
to increase the visibility of small values of $\rho$. In Fig.~\ref{fig21} 
we used for the density plot of the squared tail norm the standard 
choice $r=1/4$ due to the large ratio $\sim 10^{12}$ 
between maximum and minimum values but for $\xi_{\rm IPR}$ where this ratio 
is $\sim 10^2$ we chose exceptionnally $r=1$. For these plot parameters 
the density plots for these two quantities provide rather coherent 
and similar results for parameter regions with strong delocalization. 
All raw data of Fig.~\ref{fig21} are available for download 
at \cite{webpage}.

The density plots of Fig.~\ref{fig21} show that for values of 
$\alpha/(2\pi)$ close to the simple fractions $1/2$, $2/3$, $3/4$ 
and even $4/5$ 
there is a certain rather uniform 
delocalization effect for nearly all interaction values $U>0$. 
We attribute this observation to a strong enhancement of 
the one particle location length 
even in absence of interaction for these flux values as can be seen 
in the bottom right panel of Fig.~\ref{fig21} which compares 
the two variants of the 
inverse participation ratio computed with or 
without the 20\% center box for 
vanishing interaction strength $U=0$. 
The first variant of $\xi_{\rm IPR}$ 
measures rather directly the effective one-particle localization length and 
is quite enhanced for the above simple fractions if compared to the 
standard value $\ell=1/\log(\lambda)\approx 4.48$ for $\lambda=2.5$ 
(for irrational values of $\alpha/(2\pi)$ and infinite system size) 
\cite{aubry}. 
It seems that for the irrational values close to simple fractions the 
system size $N=512$ is still too small to see this standard value 
and one observes 
an effective enhanced one-particle localization length. 
We have verified this 
also by direct diagonalization for some example cases. 

Apart from the simple fractions there are certain combinations of 
$\alpha/(2\pi)$ and $U$ with a strong FIKS effect and a non-enhanced 
one-particle localization length. For exemple for the golden ratio case 
one recovers the peaks at $U=4.5$ and $U=7.25$ (being close to $7.2$ 
found in Sec. \ref{sec4}) and also for $\alpha/(2\pi)$ 
close to the 
value of $266/369$ of Ref. \cite{bloch} there are two modest peaks of green 
color at $U=2.25$ and $U=3.5$ (being close to $3.6$ found in the previous
Section) as can be seen from the zoomed density plot of the squared 
tail norm (bottom left panel in Fig.~\ref{fig15}). We remind that 
the value $U=2.25$ also required longer iteration times 
($t=10240$ instead of $t=5120$) to be more clearly visible thus 
explaining the green (instead of red) color for this data point 
since in Fig.~\ref{fig15} we have $t=5120$. 

Other examples are $\alpha\approx 0.596$ and $U=9.5$ (with maximal 
value of the squared tail norm of all data sets), 
$\alpha\approx 0.753$ and $U=4.5$ 
(with maximal value of $\xi_{\rm IPR}$ without 20\% center box) 
and $\alpha\approx 0.697$ with two interaction values $U=1.75$ 
and $U=2.5$. We also computed some eigenvectors by the Green 
function Arnoldi method for these four cases which 
clearly confirms the existence of FIKS eigenstates in each case. 
For example the strongest delocalized eigenstate for $\alpha\approx 0.596$, 
$U=9.5$ and $N=1533$ corresponds to $E=4.72729$, $\xi_E=426.076$, 
$\xi_x=324.511$ and for $\alpha\approx 0.753$, $U=4.5$ and $N=1837$ 
to $E=-0.68824$, $\xi_E=3618.270$, $\xi_x=955.650$.

We mention that, for the golden ratio value, 
the data of Fig.~\ref{fig21} are not perfectly 
identical/coherent to the data of Fig.~\ref{fig1} and Table~\ref{table1} 
due to the different phase offset $\beta=0$ used for the latter.

\section{Discussion}
\label{sec10}

The results presented in this work clearly show the
appearance of completely delocalized FIKS pairs 
induced by interaction in the non-interacting localized phase
of the Harper model 
when all one-particle eigenstates are exponentially localized.
The number of sites (states) $\xi$ populated by FIKS pairs
grows with the system size
approximately like a power law $\xi \propto N^b$
with the exponent being approximately in the range $0.7 \leq b \leq 1$.
We assume that the actual value of $b$ may depend on the energy range
and interaction strength. It is possible that
for $b<1$ we have some multi-fractal structure of FIKS eigenstates.
In spite of a significant numerical progress and large system sizes
studied here (we note that the {\it total Hilbert space} of the TIP problem
 is $N_H = N^2 \approx  10^8$ at maximal $N=10946$)
there are still many
open aspects in this interesting problem of interplay of interactions,
localization and 
quasiperiodicity.
Below we list the main of them.

{\it Physical origin of FIKS pairs.}
We see rather subtle and complex conditions for appearance 
of FIKS pairs. Their regions of existence are rather narrow 
on the energy interval, flux and in the range of interactions 
(see e.g. Figs.~\ref{fig8},\ref{fig21}).
However, at optimal parameters 
we may have up to $12 \%$ of states from the initial 
configuration with particles on the same or nearby site
being projected on FIKS pairs.
Thus the optimal conditions  
and the physical understanding of the FIKS effect should be clarified.
If the energy eigenvalue equation of the original Hamiltonian 
(\ref{eq_h1})-{(\ref{U_operator}) is rewritten in the
basis of non-interacting eigenstates then it gets the form \cite{dlsharper}
\begin{eqnarray}
\nonumber
(\epsilon_{m_1}+\epsilon_{m_2})\chi_{m_1, m_2} &+& 
U \sum_{{m^{'}_1}, {m^{'}_2}} Q_{m_1, m_2, {m^{'}_1}, {m^{'}_2}}
 \chi_{{m^{'}_1}, {m^{'}_2}} \\
  &=& E\chi_{m_{1}, m_{2}}
\label{tipdls1}
\end{eqnarray}
where $\chi_{m_{1}, m_{2}}$ are eigenfunctions of the TIP problem 
in the basis of the non-interacting 
product states $|\phi_{m_1},\phi_{m_2}\!\!>$ introduced in Appendix 
\ref{appendix2}. 
Note that the (second variant) of the Green function Arnoldi method 
computes rather directly  $\chi_{m_{1}, m_{2}}$ and that $\xi_E$ is 
the inverse participation ratio in this energy representation. 
The transition matrix elements produced
by the interaction are (for the Hubbard interaction case)
\begin{equation}
Q_{m_1, m_2, {m^{'}_1}, {m^{'}_2}} = 
\sum_x \phi^*_{m_1}(x)\,\phi^*_{m_2}(x)\,\phi_{m^{'}_1}(x)\,\phi_{m^{'}_2}(x)\,
\label{tipdls2}
\end{equation}
with $\phi_m(x)=<\!x|\phi_m\!>$ being the one-particle eigenfunctions 
of (\ref{eq_h1}) with the one-particle energies $\epsilon_{m}$.

We know that one-particle energies of the Harper model at $\lambda >2$
have gaps and localized eigenstates. We can assume that the sum of 
TIP energies also has gaps (or quasi-gaps) and thus there are some 
narrow FIKS bands with TIP energy width $\lambda_{\rm eff}$.
On the other side the  interaction
generates some transition matrix elements between
these band states with a certain typical transition amplitude  
$t_{\rm eff} \propto U$. Since the energy inside the FIKS band 
oscillates quasiperiodically
with the distance along the lattice we can have approximately the situation of
the original Aubry-Andr\'e model so that the delocalization transition
will take place as soon as $\lambda_{\rm eff} < 2 t_{\rm eff}$.
We think that this is the physical mechanism of 
TIP delocalization in the Harper model.
However, the concrete verification of this mechanism is not so simple: 
the matrix elements are also oscillating with the lattice distance and 
there are quite a several of them
(and not only two as in the Harper model),
 there are also energy shifts produced 
by interaction (the diagonal terms) and probably 
these shifts are at the origin of narrow
regions of interaction where the FIKS pairs appear.

There are some indications from 
the kicked Harper model \cite{lima,ketzmerick1999,prosenprl,artuso},
that coupling transitions between a large number of sites
leads to new effects and even ballistic delocalized states.
Such ballistic states appear in the regime when
the classical dynamics is chaotic and diffusive 
and from the analogy with the quantum 
Chirikov standard map \cite{chirikovscholar} one would expect to find
only pure point spectrum of exponentially localized sates.
Indeed, there are only two transition elements between sites in the Harper model
while in the kicked Harper model there are several of them.
The results presented here also indicate that the interactions 
with a longer range have a larger fraction of FIKS pairs.
Thus for $U_R=5$, which has an optimal 
interaction range, comparable 
with the one-particle localization length, we obtain 
a rather large weight of FIKS pairs of about $10\%$ in energy
and $10\%$ in the interaction range $0<U<20$ (see 
Figs.~\ref{fig1},\ref{fig8},\ref{fig11}).
These fractions exceed significantly the typical interaction 
and energy ranges
for FIKS pairs with the Hubbard interaction.

We assume that the spectrum of FIKS pairs has a structure
similar to the spectrum of the delocalized phase in 
the Aubry-Andr\'e model at $\lambda <2$,
being close to the ballistic spectrum.
Indeed, in the time evolution of 
wave packet (see e.g. Fig.~\ref{fig3})
we see the lines with a constant slope 
corresponding to a ballistic propagation with a constant velocity.
The maximal velocity is $v_p \approx x_{\rm max}/t_{\rm min} \approx 0.2$
being smaller then the maximal velocity $v_p =1$ for one particle 
at $\lambda=0$.  
It is clear that much more further
work should be done to obtain 
a deeper physical understanding of the FIKS effect
in the Harper model.

{\it Mathematical aspects.} The question about the exact spectral structure
of FIKS pairs is difficult to answer only on the basis of numerical simulations
since the system size remains always finite and subtle 
fractal properties of the spectrum require more rigorous treatment.
There are significant mathematical advancements 
in the analysis of quasiperiodic Schr\"odinger operators
reported in \cite{lana1,lana2,lana3}. We hope that the results
presented here will stimulate mathematicians to the analysis of 
properties of the FIKS phase. 
 
{\it FIKS pairs in cold atom experiments}.
The results presented in Sec.~\ref{sec8} show that
the FIKS pairs exists at the irrational flux value
$\alpha/(2\pi) \approx 532/738$ realized in the recent experiments
\cite{bloch}. However, the initial state prepared 
in \cite{bloch} had approximately one atom per each second 
site thus being rather far from the initial configuration considered here.
We think that an initial state with all atoms located
in the center of the lattice will be much more favorable 
for the observation of FIKS pairs. Indeed, such a state is rather 
similar to the initial state considered in our paper (two particles
on same on nearby sites) and thus we expect that in the experiments
one will see ballistic propagating FIKS pairs
on the tails of probability distribution
like it is well seen in Figs.~\ref{fig16},~\ref{fig17}.

We note that the initial state with all atoms in 
the center of the lattice had been used in cold atoms experiments
in the regime of the Aubry-Andr\'e model
\cite{modugno}. In these experiments a subdiffusive
delocalization of wave packet has been observed being similar 
to the numerical studies of the nonlinear Schr\"odinger equation on the 
disordered lattice. Indeed, in the center of the packet with many atoms
the Gross-Pitaevskii description can be more adequate comparing to the 
TIP case considered here. However, on the tails of probability distribution
on larger distances from the center there are only a few atoms
and only FIKS pairs can reach such far away distances.
Thus it is rather possible that  the probability
tails will contain mainly FIKS pairs. In fact the experimental
data in \cite{modugno} (Fig.~3a there) have a plateau
of probability at large  distances. However, at present 
it is not clear if this is an effect of fluctuations 
and experimental imperfections or a hidden effect of FIKS pairs.
We think that the present techniques
of experiments with cold atoms in quasiperiodic lattices
allow to detect experimentally the FIKS pairs 
discussed in this work.

{\it FIKS pairs for charge-density wave and high $T_c$ materials}.
We can expect that at finite electron density in a 1D potential
at certain conditions the main part of electrons below the Fermi energy
will remain well localized creating an incommensurate
quasiperiodic potential for a small fraction of electrons in 
a vicinity of the Fermi level. The FIKS pairs can emerge for this 
fraction of electrons.
Such situations can appear in the regime
of charge-density wave in organic superconductors and conductors
at incommensurate electron density created by doping 
(see e.g. \cite{lebed}). In such a regime it is possible that 
the FIKS pairs will give a significant contribution to 
conductivity in such materials. The proximity between
the charge-density wave regime and high $T_c$ superconductivity
in cuprates \cite{kivelson1,kivelson2}
also indicates a possibility that FIKS pairs
can play a role in these systems.
However, a more detailed analysis of 
finite density systems is required for the solid state systems.

We think that the various aspects of possible implications
of FIKS pairs in various mathematical and physicals problems
demonstrate the importance of further investigations of this
striking phenomenon.

This work was granted access to the HPC resources of 
CALMIP (Toulouse) under the allocation 2015-P0110. 

\appendix

\section{Description of the Arnoldi method}
\label{appendix1}
 For both  Lanczos and Arnoldi  methods 
one chooses some initial vector $|\zeta_1\!>$, which should ideally contain 
many eigenvector contributions, and determines a set of orthonormal vectors 
$|\zeta_1\!>\,\ldots,\,|\zeta_{n_A}\!>$, where we call $n_A$ the 
Arnoldi dimension, using Gram-Schmidt 
orthogonalization on the vector $H|\zeta_k\!>$ with respect to 
$|\zeta_1\!>\,\ldots,\,|\zeta_{k}\!>$ to obtain $|\zeta_{k+1}\!>$. 
This scheme has to be done for $k=1,\,\ldots,\,n_A$ and it also provides 
an approximate representation matrix of ``modest'' 
size $n_A\times n_A$ of $H$ on the {\em Krylov subspace} generated by these 
vectors. The largest eigenvalues of this representation matrix, also 
called {\em Ritz eigenvalues}, are typically very accurate approximate 
approximations of the largest eigenvalues of $H$ and the method also allows 
to determine (approximate) eigenvectors. It requires 
that the product of $H$ to an arbitrary vector can be computed efficiently, 
typically for sparse matrices $H$ but, as we will see in the next Section, 
even non-sparse matrices such as resolvent operators can be used provided an 
efficient algorithm for the matrix vector product is available. 

In its basic 
variant the Arnoldi method provides only the eigenvalues and eigenvectors 
for the largest energies (in module) at the boundary of the band which is not 
at all interesting and in our case it is indeed necessary to be able 
to determine accurately the eigenvalues close to a given arbitrary energy. 

The standard method to determine numerically a modest number of eigenvalues 
localized in a certain arbitrary but small region of the eigenvalue space 
for generic large sparse matrices is the 
{\em implicitly restarted Arnoldi method}. In this method the initial vector 
is iteratively refined by removing eigenvector contributions whose 
eigenvalues are outside the energy interval of interest 
using a subtle procedure based on shifted QR-steps \cite{stewart}. 
Using this algorithm we have been able to determine eigenvalues and 
eigenvectors 
for system sizes up to $N=700$-$1000$ but the computation time 
is very considerable due to the large number of iterations to achieve 
convergence of eigenvectors. Furthermore, in order to limit the computational 
time to a reasonable amount one has to accept eigenvalues of modest 
quality with $\delta^2 E(\psi)=10^{-12}$-$10^{-8}$ where the quantity 
\begin{equation}
\label{deltaE2}
\delta^2 E(\psi)=<\!\psi|\,(E-H)^2\,|\psi\!>
\end{equation}
measures the quality of an approximate eigenvector $|\psi\!>$ with 
an approximate eigenvalue $E=<\!\psi|\,H\,|\psi\!>$. Writing 
$|\psi\!>=|\psi_{\rm exact}\!>+\varepsilon |\delta\psi\!>$ 
with $H|\psi_{\rm exact}\!>=E_{\rm exact}|\psi_{\rm exact}\!>$ 
and $\|\delta\psi\|=1$ one finds easily that 
$<\!\psi_{\rm exact}|\delta\psi\!>={\cal O}(\varepsilon^2)$ (due to 
normalization of $|\psi\!>$ and $|\psi_{\rm exact}\!>$) and therefore 
$E=E_{\rm exact}+{\cal O}(\varepsilon^2)$ and 
$\delta^2 E(\psi)={\cal O}(\varepsilon^2)$. Therefore a value 
of $\delta^2 E(\psi)=10^{-8}$ implies $\varepsilon\sim 10^{-4}$.

\section{Details of the Green function Arnoldi method}
\label{appendix2}

In this appendix we provide some of the details concerning the 
Green function Arnoldi method. 
For the sake of simplicity, we will omit (most of) the details concerning the 
(anti-)symmetrization of two-partic\-le states for bosons (fermions) 
and the corresponding matrix operators acting on them. 
These details are of course important and must be dealt 
with care and precision when implementing the algorithm. 
For example, the efficient algorithm for the position-energy transformation 
(see below) requires a temporary extension of (anti-)symmetrized states 
of the boson (fermion) space of dimension $N_2$ to states in the general 
non-symmetrized two-particle space of dimension 
$N^2$ and a corresponding reduction afterwards. 
However, the details for this kind of extensions or reductions with 
eventual $\sqrt{2}$ factors etc. are based on the application 
of basic text book quantum mechanics and would only obscure the following 
description. 

Our algorithm exploits the fact that the 
interaction operator $\hat U$ acts only on a small number of sites 
$(2U_R-1)\,N\ll N^2$ \cite{U_R} given by the set 
\begin{equation}
\label{eq_setS}
S=\Bigl\{(x_1,x_2)\ \Big|\ |x_1-x_2|<U_R\Bigr\}
\end{equation}
(see again \cite{U_boundary}). Let us 
denote by
\begin{equation}
\label{eq_Pdef}
P=\sum_{(x_1,x_2)\in S} |x_1,x_2\!><\!x_1,x_2|.
\end{equation}
the projector on the sites belonging to the set $S$. Obviously $P$ commutes 
with the interaction operator $\hat U$ given in (\ref{U_operator}) 
and we have 
$P\hat UP=P\hat U=\hat UP=\hat U$. 
For the case of the Hubbard interaction with $U_R=1$ we even have 
$\hat U=UP$ where $U$ is the interaction strength and corresponds to 
the situation considered in  \cite{vonoppen,frahm_tip_green}. However for 
$U_R>1$ and $w>0$ we note that the operators $\hat U$ and $P$ are not 
proportional (but of course they still commute). We denote by 
$H_0=h^{(1)}+h^{(2)}$ the Hamiltonian in absence interaction and by 
$G_0=(E-H_0)^{-1}$ the Green function or resolvent of $H_0$. 
Furthermore we denote by $\bar G_0=PG_0P$ the projected resolvent (for $U=0$) 
which is a non-trivial (non-zero) operator only 
with respect to its diagonal block 
associated to the subspace corresponding to the set $S$. 

In this case we can state the following ``magic'' exact formula
(\ref{eq_Green})
which is the basic ingredient of our numerical approach. 
This formula can be obtained from a perturbative expansion of $G$ with the 
interaction as perturbation and an exact resummation of all terms except 
the first one. It is also possible to provide an algebraic direct proof 
without use of an expansion and we insist on the fact that 
(\ref{eq_Green}) is exact and not approximate. 
Details for both derivations are given in Appendix \ref{appendix3}. 

The key for an efficient determination of $G|\varphi\!>$ using 
(\ref{eq_Green}) is the observation that the operator 
$({\bf 1}-\hat U\bar G_0)^{-1}\hat U$ applied to any vector 
provides only non-zero contributions 
on the subspace associated to the set $S$ and 
the matrix inverse is done for a matrix of size $U_RN\ll N_2$ [or 
$(U_R-1)N\ll N_2$ for the fermion case] \cite{G0} once 
$\bar G_0$ has been determined. This approach generalizes an idea already 
used in 
\cite{vonoppen,frahm_tip_green} where (for the case of Hubbard interaction) 
the projected resolvent (for arbitrary $U$) 
$\bar G=PGP=\bar G_0\,({\bf 1}-\hat U\bar G_0)^{-1}$ 
was calculated to determine 
the localization properties of two interacting particles in one dimension 
from $\bar G$ (we remind that in \cite{vonoppen,frahm_tip_green} a disorder 
and not quasiperiodic potential was studied). 

The numerical algorithm to determine efficiently $G|\varphi\!>$ is composed 
of two parts. The first part is to calculate $\bar G_0$ and the matrix inverse 
$({\bf 1}-\hat U\bar G_0)^{-1}$ which needs to be done only once if the 
value of $E$ is not changed. The second part is to evaluate efficiently 
the successive matrix vector products (with $G_0$, $\hat U$, 
$({\bf 1}-\hat U\bar G_0)^{-1}$ etc.) accordingly to the formula 
(\ref{eq_Green}). 

For both parts we need first to diagonalize the one-particle Hamiltonian $h$ 
resulting in eigenvectors $|\phi_\nu\!>$ and eigenvalues 
$\epsilon_\nu$ which can be done with complexity ${\cal O}(N^3)$ 
(or even better using inverse vector iteration for the eigenvectors). Then 
the resolvant $G_0$ can be determined from 
\begin{eqnarray}
\nonumber
<\!x_1,x_2|&G_0&|y_1,y_2\!>=
\sum_{\nu,\mu} \frac{\phi_\nu(x_1)\,\phi_\mu(x_2)\,
\phi_\mu(y_2)\,\phi_\nu(y_1)}{E-\epsilon_\nu-\epsilon_\mu}\\
\label{eq_proj_resolv0}
&=&\sum_\nu \phi_\nu(x_1)\,g(E-\epsilon_\nu;x_2,y_2)\,
\phi_\nu(y_1),\\
\label{eq_onepart_green}
g(E;x,y)&=&\sum_\mu\frac{\phi_\mu(x)\,\phi_\mu(y)}
{E-\epsilon_\mu}=<\!x|(E-h)^{-1}|y\!>
\end{eqnarray}
where $g(E;x,y)$ is the one-particle Green function and 
$\phi_\nu(x)=<\!x|\phi_\nu\!>$. 

We use (\ref{eq_proj_resolv0}) to determine the {\em projected} resolvent 
$\bar G_0$, i.~e. for $(x_1,x_2),\,(y_1,y_2)\in S$. This requires only 
${\cal O}(N^3\,U_R^2)$ operations in total since for each value of 
$\nu$ we can determine the one-particle Green function as 
inverse of a tridiagonal matrix (with periodic boundary conditions) 
with ${\cal O}(N^2)$ operations using a smart formulation of Gauss algorithm.
Then, still for the same value of $\nu$, we have to update the sums for all 
possible values $(x_1,x_2),\,(y_1,y_2)\in S$ which costs 
${\cal O}(N^2\,U_R^2)$ operations 
which is dominant (or comparable if $U_R=1$) 
to the complexity of the one-particle Green function evaluation. 
The sum/loop over $\nu$ leads then to a further factor of $N$ giving 
${\cal O}(N^3\,U_R^2)$ operations. The subsequent matrix inverse to 
determine $({\bf 1}-\hat U\bar G_0)^{-1}$ requires ${\cal O}(N^3\,U_R^3)$ 
operations. We mention that for the Hubbard interaction case $U_R=1$ 
this algorithm to determine $\bar G_0$ and the inverse was already 
implemented and explained in Ref. \cite{frahm_tip_green}. 

For the second part of the algorithm we still need an efficient method 
to evaluate $G_0|\varphi\!>$ for a given vector $|\varphi\!>$. This 
can actually be done by a transformation from position to energy 
representation, i.~e. an expansion of $|\varphi\!>$ using the 
eigenvectors of $H_0$ given as product states 
$|\phi_\nu,\phi_\mu\!>$. This transformation can be done with 
essentially ${\cal O}(N^3)$ operations using the 
trick to transform first the coordinate of the first particle and 
then in a separate subsequent step the coordinate of the second particle. 
Each one-particle transformation requires ${\cal O}(N^2)$ operations but it 
has to be done for $N$ possible positions of the other particle and the 
transformation for the other particle gives a further factor of 2 resulting in 
$\sim 2N^3$ addition and multiplication operations for one two-particle 
transformation. 
The transformation back into position representation can be done similarly. 

Since $G_0$ is diagonal in the energy representation (with eigenvalues 
$(E-\epsilon_\nu-\epsilon_\mu)^{-1}$) the product $G_0|\varphi\!>$ 
in this representation only requires ${\cal O}(N^2)$ operations. 
Once this is done the resulting vector is transformed back into position 
representation (also with ${\cal O}(N^3)$ operations). 
Then the product of the matrix $({\bf 1}-\hat U\bar G_0)^{-1} \hat U $ 
to a vector in position representation only requires 
${\cal O}(N^2 U_R^2)$ operations (provided that the matrix inverse is 
calculated and stored only once in advance for a fixed value of $E$). Finally 
a further double-transformation-multiplication step with $G_0$ is 
necessary. 
Combing all this it is possible to evaluate $G|\varphi\!>$ by 
(\ref{eq_Green}) by ${\cal O}(N^3)$ operations (but with a rather big 
prefactor) where the most complex part consists of the 
two position-energy transformations and the two 
inverse energy-position transformations. 

In summary we have described an algorithm to determine $G|\varphi\!>$ 
by ${\cal O}(N^3\,U_R^3)$ operations for the initial preparation for 
a given energy $E$ and ${\cal O}(N^3)$ operations for each product 
(i.~e. $G$ applied to several different vectors) 
provided the initial value of $E$ is not changed. 
In terms of the matrix size $N_2\approx N^2/2$ this implies a complexity of 
${\cal O}(N_2^{3/2})$ operations which is more expensive than the 
product $H|\varphi\!>$ with ${\cal O}(N_2)$ operations but still much better 
than the naive matrix vector multiplication with ${\cal O}(N_2^{2})$ 
operations. 

The position-energy transformation can be furthermore optimized 
for larger system sizes using that the one-particle eigenfunctions 
$\phi_\nu(x)$ are localized around some position $x_{\rm max}$ 
with localization length $\ell$. In this case the ratio 
$|\phi_\nu(x)/\phi_\nu(x_{\rm max})|$ is below $10^{-17}$ 
(the numerical rounding error for standard double precision numbers) 
for $|x-x_{\rm max}|>c$ with the constant 
$c=17\,\log(10)\,\ell\approx 175$ if we replace 
the value $\ell\approx 4.48$ for $\lambda=2.5$. The positions $x$ 
fulfilling this condition can be safely excluded in the multiple sums 
for the position-energy transformation therefore reducing the complexity 
to ${\cal O}(c\,N^2)$. 

This first variant of the algorithm combined with the (simple) Arnoldi method 
for $G$ is already very efficient and very superior to the implicitly 
restarted Arnoldi meth\-od applied to $H$ and produces for a sufficiently 
large value of the Arnoldi dimension $n_A$ easily more than $50\%-70\%$ 
of numerically accurate eigenvalues close to the energy $E$ appearing 
in the Green function (from all $n_A$ Ritz eigenvalues produced by the 
Arnoldi method). For example for the Hubbard case with $U=7.8$ and 
$E= -2.78$ we have been able, on a machine with 
64 GB of RAM memory, to increase the system size up to $N=4181$ (which is a 
Fibonacci number) and to choose the Arnoldi dimension $n_A=900$ 
and about 620 out of 900 obtained eigenvalues have a quality 
with $\delta^2 E(\psi)<10^{-20}$ \cite{quality}. 
Furthermore most of the important parts of the algorithm can be quite well 
parallelized for multiple core machines. 

As start vector for the Arnoldi iteration 
we choose a vector proportional to the projection 
$P\sum_{x_1,x_2}|x_1,x_2\!>$, i.~e. a vector 
with uniform identical values for the sites in the set $S$ 
where the interaction acts. In this way we avoid (most of) the many useless 
contributions from eigenstates which are essentially localized product states 
$|\phi_\nu,\phi_\mu\!>$ with both particles localized very far away 
such that the interaction has no effect on them. With this start vector 
we capture all well ``delocalized'' states with energies close to the 
value of $E$. The Arnoldi method still provides a considerable number of 
eigenstates being similar to 
strongly localized product states where the distance between particles 
is ``modest'', i.~e. sufficiently large that the product states are indeed 
relatively good eigenstates of $H$ but also sufficiently small 
that the initial vector has small contributions of these states which will be 
amplified by the Green function Arnoldi method if the eigenvalue of 
the product state is sufficiently close to $E$. 

For small values of $U_R$ the memory requirement of the Arnoldi method is 
determined by the number 
$n_A$ of iteration vectors which need to be stored and the size of these 
vectors $N_2\approx N^2/2$ which provides the essential limitation of this 
method concerning the choice of $n_A$ and $N$. For larger values of 
$U_R$, e.~g. $U_R=20$ the largest value we have considered, the requirement 
to store multiple matrices of size $U_RN\times U_RN$ is also important 
(or even dominant for the second variant described below). 

However, this first variant of the Green function Ar\-noldi method, 
which works with vectors stored in the position representation, can be 
considerable improved by using vectors stored in the non-interaction 
energy representation using an expansion in terms of the non-interacting 
product states $|\phi_\nu,\phi_\mu\!>$. This modification allows for 
several improvements. 

First, the number of the rather expensive energy-po\-si\-tion 
(or inverse position-energy) transformation steps is reduced 
from four to two when evaluating $G|\varphi\!>$ since, according to 
the above description of the algorithm, the 
first energy-position and the last inverse position-energy transformation 
can be avoided if the vector $|\varphi\!>$ is by default already available 
(or needed) in 
energy representation (instead of position representation). 

Second, in this modified variant it is natural 
to choose a somewhat different start vector, i.~e. a vector given as sum 
of product states with maximal positions in the set $S$ which is qualitatively 
similar to the other initial vector used for the first variant but still 
different due to the finite one-particle localization length. The important 
point is that the new initial vector contains less contributions from 
useless products states. For given values of $n_A$ and $N$ this improves 
considerably the quality of the eigenvectors by reducing  the value of 
the quantity (\ref{deltaE2}) and one obtains 
more nicely ``delocalized'' states (with eigenvalues a bit further away from 
$E$) and less useless product states. 

The third improvement concerns the possibility to reduce considerably the 
dimension of the Hilbert space in energy representation from 
$N_2\approx N^2/2$ to $cN$ (with $c\approx 175$ for $\lambda=2.5$)
since one can simply remove all product states with maximal positions 
further away than $c$ because these states do not feel the interaction at all 
(i.~e. with interaction coupling matrix elements smaller than $10^{-17}$). 
This reduces the amount of memory usage and also computation time for 
the Arnoldi iterations by a factor $2c/N$ which becomes quite small for 
large system sizes ($N>1000$). Especially the reduced memory 
requirement allows to perform computations with larger values of $N$ and 
$n_A$, for example for $U_R=1$ we have been able to choose a system size 
$N=10946$ 
with Arnoldi dimension $n_A=3000$ (on a machine with 64 GB of RAM memory). 
For the case $N=4181$ and $n_A=900$, the maximum possible size for the 
first variant with 64 GB, 
the computation time for the second variant of the method is reduced by a 
factor of ten if compared to the first variant.

The overall complexity of the Green function Arnoldi method for small 
systems ($N\le c$) is given by 
$C_1(U_RN)^3+C_2 N^3\,n_A+C_3N^2\,n_A^2$ with three terms representing 
the initial preparation part (first term with the constant $C_1\sim 1$), 
the Green function vector multiplications (second term with the constant 
$C_2\sim 5$) and the Gram-Schmidt orthogonalization scheme (third term 
with the constant $C_3\sim 1$). For larger systems $N\gg c=175$ we have 
to replace in the second and third term a factor of $N$ by $c$ resulting in
$C_1(U_RN)^3+C_2\,c\, N^2\,n_A+C_3\,c\,N\,n_A^2$. 
If one choose typically $n_A\sim N$ the second and third term have 
comparable complexity $\sim c\,N^3$ 
but in practice the second term is dominant due to 
a considerably larger value of the constant $C_2$. Therefore it is not 
interesting to use the Lanczos method (instead of the full Arnoldi iterations) 
because this would only remove in the last, non-dominant, term 
one factor of $N$. The memory requirements (in units of size of double 
precision numbers) scale with 
$C_4\,(U_RN)^2+C_5\,c\, N\,n_A$ with $C_4\sim 5 C_5$ and $C_5\sim 1$ because 
one has to store several copies of matrices of size $(U_RN)\times (U_RN)$ 
and $n_A$ vectors of size $c\,N$ for the Arnoldi iterations.

With increasing values of the interaction range $U_R$ the memory requirement 
and also computation time of the initial preparation part become more 
important or dominant, for $U_R=20$, but even for this extreme case we have 
been able to push the system size up to $N=1597$ and one can (should) 
choose very large values for $n_A$ for the second Arnoldi-iteration 
part to better exploit the 
computational ``investment'' of the preparation part. Even with $n_A=2500$ 
for $N=1597$ the second and third part require only about 
5\% of the computation time while for $U_R=1$ the first preparation part 
is typically negligible (at most 7\% for the largest system size $N=10946$, 
$n_A=3000$ we considered).

We close this Appendix mentioning that the effective algorithm to compute 
arbitrary resolvent vector products can also be used to calculate more 
directly (or improve) individual eigenvectors if the eigenvalue 
(or an approximate eigenvector) is known with sufficient precision by the 
method of inverse vector iteration. We have for example been able to improve 
the modest quality eigenvectors which we had obtained 
by the implicitly restarted Arnoldi method to maximum possible precision only 
using a few number of these iterations. Actually, also a random 
initial vector can be used if a rather good approximate eigenvalue is known. 
However, to achieve a good efficiency for a systematic computation of 
{\em many} eigenvectors with close energies 
the Arnoldi method for the resolvent is the 
best choice to exploit the Green function algorithm. 
The reason is the expensive initial part of the algorithm [the rather 
expensive initial computation of $\bar G_0$ and the matrix inverse in 
(\ref{eq_Green})] which is 
only done once for the Arnoldi method and has to be repeated for any new 
individual eigenvalue when using inverse vector iteration.

\section{Projected Green's function formula}
\label{appendix3}

In this appendix we show the formula (\ref{eq_Green}) where 
$G=(E-H)^{-1}$, $G_0=(E-H_0)^{-1}$, $H=H_0+\hat U$,
$\bar G_0=PG_0P$ and $P=P^2$ is 
a projector such that $\hat U=P\hat U=\hat UP=P\hat UP$, i.~e. $\hat U$ has 
the same eigenvectors as $P$ and only non-vanishing eigenvalues if 
the corresponding eigenvalue of $P$ is unity. 

\subsection{Perturbative expansion of $G$}

The proof of (\ref{eq_Green}) by an expansion in a matrix power series is 
quite illustrative. First we express $G$ as
\begin{eqnarray}
\nonumber
G&=&\Bigl[({\bf 1}-\hat UG_0)(E-H_0)\Bigr]^{-1}\\
\label{eq_app1}
&=&G_0({\bf 1}-\hat U G_0)^{-1}=
G_0\sum_{n=0}^\infty (\hat UG_0)^n \\
\label{eq_app2}
&=&G_0+G_0\left(\sum_{n=0}^\infty (\hat U G_0)^n\right) \hat U G_0
\end{eqnarray}
where we have assumed that the matrix power series converges well which is 
the case for sufficiently large values of $E$ in the 
complex plane. Using the relations between $\hat U$ and $P$ we may 
rewrite the expression (\ref{eq_app2}) as:
\begin{equation}
\label{eq_app3}
G=G_0+G_0\left(\sum_{n=0}^\infty (\hat U PG_0P)^n\right) \hat U G_0
\end{equation}
which becomes after replacing $\bar G_0=PG_0P$ and resumming the 
series (in parentheses) just formula (\ref{eq_Green}). Furthermore applying an 
argument of analytic continuation the validity of (\ref{eq_Green}) is 
extended to all values of $E$ in the complex plane (except the singularities 
of $G$ or $G_0$). This calculation shows the crucial role of the 
relations between the interaction operator $\hat U$ and the projector $P$ and 
which finally allow to reduce the difficulty to determine the 
resolvent $G$ by using a matrix inverse in a subspace of considerably 
smaller dimension which is just the subspace onto which $P$ projects. 

\subsection{Algebraic direct proof}

The expansion in a matrix power series and the argument of analytic 
continuation can be avoided by a direct but somewhat ``less clear'' 
calculation. For this we write:
\begin{eqnarray}
\nonumber
G&=&G(E-H_0)G_0=G(E-H+\hat U)G_0
=G_0+G\hat UG_0\\
\label{eq_app4}
&=&G_0+G_0({\bf 1}-\hat UG_0)^{-1}\hat UG_0\\
\label{eq_app5}
&=&G_0+G_0\hat O\hat U G_0
\end{eqnarray}
where we have used the first identity of (\ref{eq_app1}) to 
obtain (\ref{eq_app4}). The operator $\hat O$ is given by 
$\hat O=(1-PA)^{-1}P$ and $A=\hat UG_0$ and to obtain 
(\ref{eq_app5}) we have used (twice) that $P\hat U=\hat U$. 
We rewrite $\hat O$ in the form
\begin{equation}
\label{eq_app6}
\hat O=({\bf 1}-PA)^{-1}P({\bf 1}-PAP)({\bf 1}-PAP)^{-1}
\end{equation}
and since $P({\bf 1}-PAP)=({\bf 1}-PA)P$ we obtain the expression
\begin{displaymath}
\hat O=P({\bf 1}-PAP)^{-1}=({\bf 1}-PAP)^{-1}P
=({\bf 1}-\hat U\bar G_0)^{-1}P
\end{displaymath}
which together with (\ref{eq_app5}) (and again $P\hat U=\hat U$) 
provides the formula (\ref{eq_Green}).



\begin{thebibliography}{99}
\bibitem{U_boundary} In view of the periodic boundary conditions 
the condition $|x_1-x_2|<U_R$ is understood to be true also for the case 
$N-|x_1-x_2|<U_R$, i.~e. if $x_1$ is close to one boundary and $x_2$ 
to the other boundary.
\bibitem{arnoldi_comment} The Lanczos method is theoretically 
more efficient than the Arnoldi method since it requires only 
orthogonalization with respect to two and not all previous orthonormal 
vectors. However, this 
mathematical property is numerically problematic and may lead to the 
appearance of additional incorrect eigenvalues also know as ``ghosts''. 
In this work, we used the less problematic but somewhat more 
``expensive'' Arnoldi method, since the gain of efficiency of the Lanczos 
method is finally not relevant due to other parts in our numerical approach 
with a higher complexity. In view of this we speak in this work only of 
the Arnoldi method.
\bibitem{U_R} The expression $(2U_R-1)N$ for the number of sites in the 
set $S$ does not take into account the reduction of dimension 
due to (anti-)symmetrization.
Of course for the boson (fermion) case there are only $U_RN$ 
[$(U_R-1)N$] physical (anti-)symmetrized states associated to the set $S$.
\bibitem{G0} 
The projected resolvent $\bar G_0$ takes a block diagonal 
form for the boson and fermion cases and when expressed in a basis 
of (anti-)symmetrized states. Therefore the matrix inverse 
involving $\bar G_0$ can be done for a matrix size $U_RN$ 
[$(U_R-1)N$] for bosons (fermions) and does not require a matrix size 
$(2U_R-1)N$. To keep the discussion simple 
the details of these complications are omitted 
in Appendix \ref{appendix2}.
\bibitem{quality} In all our programs, based on different variants or 
algorithms, the quality of obtained eigenvectors was checked 
independently by calculating the quantity (\ref{deltaE2}) using 
the direct matrix vector multiplication of the sparse matrix $H$ 
with the numerically obtained eigenvector.
\bibitem{cf_rational} For each rational number there are two possible finite 
continued fraction expansions, one where the last coefficient is $\ge 2$
and the other one with one additional coefficient which is $1$ and 
where the second last coefficient is reduced by $1$. We choose 
the first variant which is also called the canonical representation.
\end{thebibliography}
\end{document}